\documentclass{aa}  

\usepackage{graphicx}
\usepackage{url}            
\usepackage{txfonts}
\usepackage{natbib}
\bibpunct{(}{)}{;}{a}{}{,}  
\usepackage{color}
\usepackage[percent]{overpic} 

\def\OII{[O\,\textsc{ii}]}

\def\microns{$\textrm{$\mu$m}$}                             
\def\num{\hbox{N$^{\underline{o}}$}}                        

\begin{document}

   \title{Galaxy population properties of the massive X-ray luminous galaxy cluster XDCP\,J0044.0-2033 at $z\!=\!1.58$\thanks{Based on observations under programme ID 084.A-0844, 087.A-0351, and 089.A-0419 collected at the European Organisation for Astronomical Research in the Southern Hemisphere, Chile.} }

   \subtitle{Red-sequence formation, massive galaxy assembly, and central star formation activity}

   \author{R. Fassbender
          \inst{\ref{OAR},\ref{MPE}}
          \and
             A. Nastasi
          \inst{\ref{Orsay},\ref{MPE}}
          \and
          J.S. Santos
          \inst{\ref{Arcetri},\ref{ESAC}}
          \and          
         C. Lidman
          \inst{\ref{AAO}}
           \and        
          M. Verdugo
          \inst{\ref{Vienna}}
          \and
          Y. Koyama
          \inst{\ref{Durham},\ref{Japan_NAO}}        
           \and
          P. Rosati
          \inst{\ref{Ferrara},\ref{ESO}}   
          \and                 
          D. Pierini 
          \inst{\ref{MPE}}
          \and
          N. Padilla
          \inst{\ref{Catolica}}
          \and 
          A.D. Romeo
          \inst{\ref{Catania}}
          \and
          N. Menci
          \inst{\ref{OAR}}
          \and
          A. Bongiorno
          \inst{\ref{OAR}}
          \and
          M. Castellano
          \inst{\ref{OAR}}
                    \and           
           P. Cerulo   
           \inst{\ref{Swinburne}}          
          \and
          A. Fontana
          \inst{\ref{OAR}}
          \and  
          A. Galametz
          \inst{\ref{OAR}}
          \and
          A. Grazian
          \inst{\ref{OAR}}
          \and
          A. Lamastra
          \inst{\ref{OAR}}
         \and                 
          L. Pentericci
          \inst{\ref{OAR}}
          \and
          V. Sommariva
          \inst{\ref{OAR}}
          \and
         V. Strazzullo  
           \inst{\ref{Saclay}}
                    \and
          R. \v{S}uhada
          \inst{\ref{USM}}
           \and                 
          P. Tozzi
          \inst{\ref{Arcetri}}
          }

  \institute{INAF-Osservatorio Astronomico di Roma (OAR), Via Frascati 33, 00040, Monteporzio Catone, Italy \\
           \email{rene.fassbender@oa-roma.inaf.it} \label{OAR}
          \and
          Max-Planck-Institut f\"ur extraterrestrische Physik (MPE), Postfach 1312, Giessenbachstr.,  85741 Garching, Germany   \label{MPE}
         \and
        Institut d'Astrophysique Spatiale, B\^{a}timent 12,  Universit\'{e} Paris-Sud, 91405 Orsay, France    \label{Orsay}
          \and
         INAF-Osservatorio Astrofisico di Arcetri, Largo Enrico Fermi 5,  50125 Firenze, Italy  \label{Arcetri}
         \and
        European Space Astronomy Centre (ESAC), 7828691 Villanueva de la Canada, Madrid, Spain \label{ESAC}
         \and
         Australian Astronomical Observatory, PO BOX 915, North Ryde 1670, Australia    \label{AAO}      
            \and
        University of Vienna, Department of Astronomy, T\"urkenschanzstra\ss e 17, 1180, Vienna, Austria \label{Vienna} 
           \and
        Department of Physics, Durham University, South Road, Durham DH1 3LE, UK \label{Durham}
          \and
         National Astronomical Observatory of Japan, Mitaka, Tokyo 181-8588, Japan \label{Japan_NAO}
          \and
        Universit\`{a} degli Studi di Ferrara, Via Savonarola 9,  44121 Ferrara , Italy \label{Ferrara}
          \and
         European Southern Observatory (ESO), Karl-Scharzschild-Str.~2, 85748 Garching, Germany \label{ESO}
         \and
         Departamento de Astronom\'ia y Astrof\'isica, Pontificia Universidad Cat\'olica de Chile, Casilla 306, Santiago 22, Chile  \label{Catolica}
           \and
           INAF-Osservatorio Astrofisico, Via S.Sofia 78, 95123 Catania, Italy   \label{Catania}        
          \and
         CEA Saclay, Orme des Merisiers, 91191 Gif sur Yvette, France  \label{Saclay}
           \and
Centre for Astrophysics and Supercomputing, Swinburne University of Technology, PO Box 218, Hawthorn, VIC 3122, Australia\label{Swinburne}
           \and
          Department of Physics, Ludwig-Maximilians Universit\"at M\"unchen, Scheinerstr. 1, 81679  Munich, Germany \label{USM}
             }

   \date{Received Apr 3, 2014; accepted May 3, 2014}

 
  \abstract
   {Recent observational progress has enabled the detection of galaxy clusters and groups out to very high redshifts and 
  for the first time allows detailed studies of galaxy population properties in these densest environments in what was formerly known as the `redshift desert' at $z\!>\!1.5$. }
   {We aim to investigate various galaxy population properties of the massive X-ray luminous galaxy cluster XDCP\,J0044.0-2033 at $z\!=\!1.58$, which 
constitutes the most extreme currently known matter-density peak at this redshift.}
   {We analyzed deep VLT/HAWK-I near-infrared data with an image quality of 0.5\arcsec \ and  limiting Vega magnitudes (50\% completeness) of 24.2 in J- and 22.8 in the Ks band, complemented by similarly deep Subaru imaging in i and V,   {\it Spitzer} observations at 4.5\microns, and new spectroscopic observations with VLT/FORS\,2.}
   {We detect a cluster-associated excess population of about 90 galaxies, most of them located within the inner 30\arcsec\,(250\,kpc) 
  of  the X-ray centroid, which follows a centrally peaked, compact  NFW  galaxy surface-density profile  with a concentration of $c_{200}\!\simeq\!10$. Based on the {\it Spitzer}  4.5\,\microns \ imaging data, we measure a total enclosed stellar mass  of  $M*_{500}\!\simeq\!(6.3 \pm 1.6)\!\times\!10^{12}\,\mathrm{M_{\sun}}$  
  and a resulting stellar mass fraction of  f$_{*,500}=M_{*,500}/M_{500}  = (3.3\pm1.4)$\%, consistent with local values. The total J- and Ks-band galaxy luminosity functions of the core region yield characteristic magnitudes J* and Ks* consistent with expectations from simple  $z_{\mathrm{f}}\!=\!3$ burst models. However, a detailed look at the morphologies and color distributions of the spectroscopically confirmed members reveals that the most massive galaxies are undergoing a very active mass-assembly epoch through merging processes. Consequently, 
  the bright end of the cluster red sequence is not in place, while a red-locus population is present at intermediate magnitudes [Ks*,Ks*+1.6], which is then sharply truncated at magnitudes fainter than Ks*+1.6. The dominant cluster-core population comprises post-quenched galaxies transitioning toward the red sequence at intermediate magnitudes, while additionally a significant blue-cloud population of faint star-forming galaxies is present even in the densest central regions. 
 Based on a  color-color selection performed to separate different cluster galaxy types, we find 
  that the blue star-forming population is concentrated in clumpy structures and dominates in particular at and beyond the  R$_{500}$ radius. 
On the other hand, the fraction of post-starburst galaxies steadily
increases toward the center, while the red-locus population and red-sequence transition galaxies seem to reach their peak fractions already at intermediate cluster-centric radii of about $r\!\sim\!200$\,kpc.      
%
   }
   {Our observations support the scenario in which  
   the dominant effect 
   of the dense $z\!\simeq\!1.6$ cluster environment is an accelerated mass-assembly timescale ($\sim$1\,Gyr or shorter) 
   through merging activity that is responsible for 
    driving core galaxies across the mass-quenching threshold of  $log(M_*/M_{\sun})\!\simeq\!10.4$. 
    Beyond this mass limit, star formation is suppressed on timescales of $\sim$1\,Gyr, while the direct environmental quenching process seems to be subdominant and is acting on significantly longer timescales ($\sim$2-3\,Gyr).
   }




\maketitle
%
\section{Introduction}
\label{c1_Intro}

The discovery of clusters of galaxies at redshifts $1.5\!<\!z\!\la\!2$ over the past few years with infrared  \citep[e.g.,][]{Gobat2011a,Zeimann2012a,Stanford2012a,Muzzin2013a,Papovich2010a,Andreon2014a} or X-ray selection techniques \citep[e.g.,][]{Fassbender2011a,Santos2011a,Tozzi2013a,Tanaka2013a} has enabled a  new view on galaxy evolution studies 
in what was formerly known as the `redshift desert'   with large ensembles of co-eval galaxies in the densest environments.
However, owing to the detection challenge and the relative rareness of these systems, detailed studies of the galaxy population properties in the densest $z\!>\!1.5$  environments have so far only been available for a small number of galaxy groups with total masses of  
$M_{200}\!<\!10^{14}\,\mathrm{M_{\sun}}$ \cite[e.g.,][]{Rudnick2012a,Papovich2012a,Strazzullo2013a,Newman2014a}.


While the galaxy populations in the centers of massive galaxy clusters show  surprisingly little change in their overall appearance and properties all the way out to  $z\!\simeq\!1.4$ \cite[e.g.,][]{Mei2009a,Strazzullo2010a}, it has now been firmly established 
that the $z\!\simeq\!1.5$ regime is 
a crucial transition epoch beyond which the predecessors  of the homogeneous  `red and dead'  galaxy population of lower-$z$ systems undergo drastic changes as the star formation activity is switched on in many systems all the way to the densest core regions  \cite[e.g.,][]{Tran2010a,Hayashi2010a,Hilton2010a,Fassbender2011a,Bayliss2013a,Romeo2008a}.

However, observations of 
high star formation rates in cluster cores
and active mass-assembly activity  
\cite[e.g.,][]{Lotz2013a} in massive group galaxies and the associated gradual disappearance of a well-defined red sequence on one hand are currently confronted with recent reports on seemingly evolved galaxy populations in some systems at  $z\!\simeq\!1.6$ 
\citep[e.g.,][]{Tanaka2013a} and beyond \citep{Andreon2014a}. Current observations of the general evolutionary state of galaxies in  $z\!\ga\!1.5$ group and cluster environments therefore do not provide a homogeneous picture yet owing to the still limited statistics of investigated systems and, in particular, the lack of mass leverage into the more massive cluster regime.  

Clearly, the influence and effects of the  total cluster mass $M_{200}$ on the overall galaxy evolution process at $z\!\ga\!1.5$ is one missing key aspect to be investigated further. Moreover, at lookback times of around 9.5\,Gyr and beyond, the cluster formation and galaxy evolution timescales become similar to the cluster 
ages since gravitational collapse. 
Therefore the individual mass accretion histories of the cluster can be expected to play a significant role in the evolutionary state of the overall system and its galaxy populations. Improved sample statistics and a wide mass-baseline  at $z\!\ga\!1.5$ are therefore key to any detailed understanding of the early formation and evolution process of group and  cluster galaxies. 
 


In this study we aim to investigate the galaxy population properties of the massive cluster XDCP\,J0044.0-2033 at $z\!=\!1.58$ \citep{Santos2011a}  using new deep near-infrared (NIR) observations in the J- and Ks bands along 
other complementary data sets. 
The cluster was originally discovered based on its extended X-ray emission (centroid coordinates: RA=00:44:05.2, DEC=$-$20:33:59.7) within the framework of the  XMM-{\it Newton} Distant Cluster Project \citep[XDCP,][]{Fassbender2011c}, 
which implies that the selection is unbiased with respect to the underlying galaxy population of the system.
Based on the discovery X-ray data set, the cluster has a measured bolometric luminosity of  
L$^{\mathrm{bol}}_{\mathrm{X,500}}\!\simeq\!6\times10^{44}$\,erg\,s$^{-1}$ and a derived luminosity-based total system mass estimate  \citep{Reichert2011a}
of M$_{\mathrm{200}}\!\simeq\!3\times10^{14}$\,M$_{\sun}$, making it the most massive X-ray selected cluster at  $z\!>\!1.5$ currently known. 
This paper is the first in a series of upcoming studies with the aim to provide a very detailed multiwavelength characterization of the system.
These works will~include, for example, updated cluster mass measurements  based on new observations with {\it Chandra}  (Tozzi et al., subm.), the total cluster star formation activity derived from {\it Herschel} data (Santos et al., in prep.), and studies  based on {\it Spitzer} observations  and additional spectroscopic and optical imaging coverage. 

The structure of this paper is as follows: in Sect.\,\ref{c2_DataSets} we introduce the imaging and spectroscopic data sets used for this study and discuss the data reduction; 
in  Sect.\,\ref{c3_Results} we present the results of various aspects of the 
galaxy population properties of the system. Section\,\ref{c4_GlobalView} provides a global view of the cluster, Sect.\,\ref{c5_Discussion}
discusses the findings and compares them with galaxy evolution models, and we conclude in
Sect.\,\ref{c6_Conclusions}.
Throughout this paper we assume a standard $\Lambda$CDM cosmology with
$\Omega_\mathrm{m}\!=\!0.3$, $\Omega_{\Lambda}\!=\!0.7$, and $h\!=H_0/(100\,\mathrm{km\,s}^{-1}\,\mathrm{Mpc^{-1}})\!=\!0.7$. 
At a redshift of $z\!=\!1.58$, this implies a lookback time of 9.45\,Gyr, an age of the Universe of 4.0\,Gyr, and an angular scale of 8.47\,kpc/\arcsec \ very close to the absolute maximum of the angular diameter distance D$_A$.
We use the notation with
subscript $X_{500}$ ($X_{200}$) 
to refer to physical quantities $X$ measured inside a radius, for which the mean enclosed total
density of the cluster is 500 (200) times the critical energy density of the Universe $\rho_{\mathrm{cr}}(z)$ at the
cluster redshift. All reported 
 magnitudes are given in the Vega system.




\section{Data sets and data reduction}
\label{c2_DataSets}

The data sets for this work include deep NIR imaging observations in the J- and Ks band with VLT/HAWK-I presented in Sect.\,\ref{s2_HawkI}, complemented by new spectroscopic data (Sect.\,\ref{s2_Spectroscopy}), archival Subaru imaging in the V- and i band (Sect.\,\ref{s2_Subaru}), and mid-infared   observations  with {\it Spitzer} (Sect.\,\ref{s2_Spitzer}).

\subsection{Near-infrared observations with VLT/HAWK-I}
\label{s2_HawkI}


The cluster field of XDCP\,J0044.0-2033 was observed with VLT/HAWK-I  \citep{Kissler2008a} on 27 June 2011 and 08 August 2011 with the J- and Ks-band filters under excellent observing conditions with 0.5\arcsec \ seeing  (program ID 087.A-0351). In total, the field was observed for a net exposure time of 80\,min in J (40 positions each with 12$\times$10sec) and 36\,min in Ks (36 positions each with 6$\times$10sec) with slight telescope dithering offsets between observing positions. The focal plane of HAWK-I consists of four  2048$\times$2048 pixel detectors with 15\arcsec gaps between them, providing a field of view (FoV) for each quadrant of  3.6\arcmin \ on the side with a pixel scale of 0.106\arcsec/pixel.  The cluster center was placed well inside one quadrant (Q3) to provide a homogeneous deep exposure time in the cluster region without interference of the detector gaps. For this study we used the other three quadrants as control fields to accurately determine the background source density at the full depth of the observations.

The data reduction was performed in standard NIR manner, closely following the processing steps described in \citet{Lidman2008a}. In brief, these steps include dark-frame subtraction, flatfielding, background subtraction with object masking, astrometric calibration, and the final seeing-weighted image combination using {\em SWarp} \citep{Bertin2002a}. The seeing 
of the resulting deep combined images was measured to be 0.51\arcsec \ (FWHM) in Ks and 0.53\arcsec \ in the J band. The photometric zero points for both bands were determined to an accuracy of about 0.02\,mag (0.05\,mag) in J (Ks) using 2MASS point sources \citep{Skrutskie2006a} in the FoV  as well as dedicated standard star observations. 


The photometry was performed with  {\em SExtractor}  \citep{Bertin1996a} in dual-image mode. 
Since the depth of the J- and Ks-band image relative to the expected magnitudes of the red cluster galaxy population was designed to be comparable, 
 we co-added the final image stacks, which 
resulted in a noise-weighted combined J+Ks detection image with the maximum achievable depth. This combined frame was then used as object-detection image for the first {\em SExtractor} pass, whereas the final photometry was measured in the original individual bands in J and Ks. To accurately determine object colors the Ks image quality was matched to the slightly different seeing conditions in the J image.    
The magnitudes of the resulting photometric catalogs were corrected for galactic extinction. We used  MAG\_AUTO for the total object magnitudes and 1\arcsec \ aperture magnitudes ($\sim$2$\times$seeing) to determine the J$-$Ks object colors.   
Finally, we applied the small color term correction to transform the HAWK-I into the 2MASS filter system following \citet{Lidman2008a}.

\begin{figure}[t]
    \centering
    \includegraphics[width=\linewidth, clip]{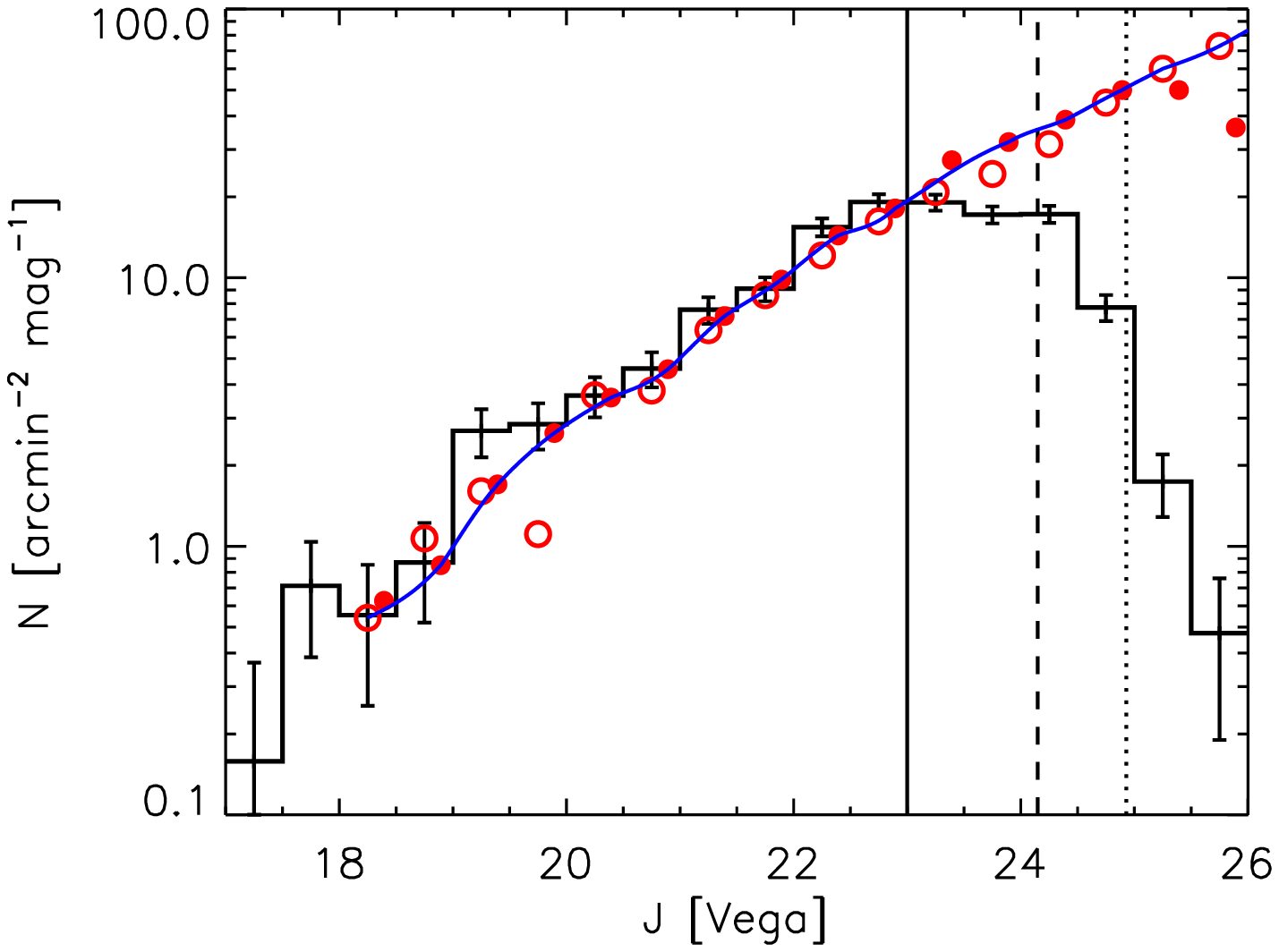}    
    \includegraphics[width=\linewidth, clip]{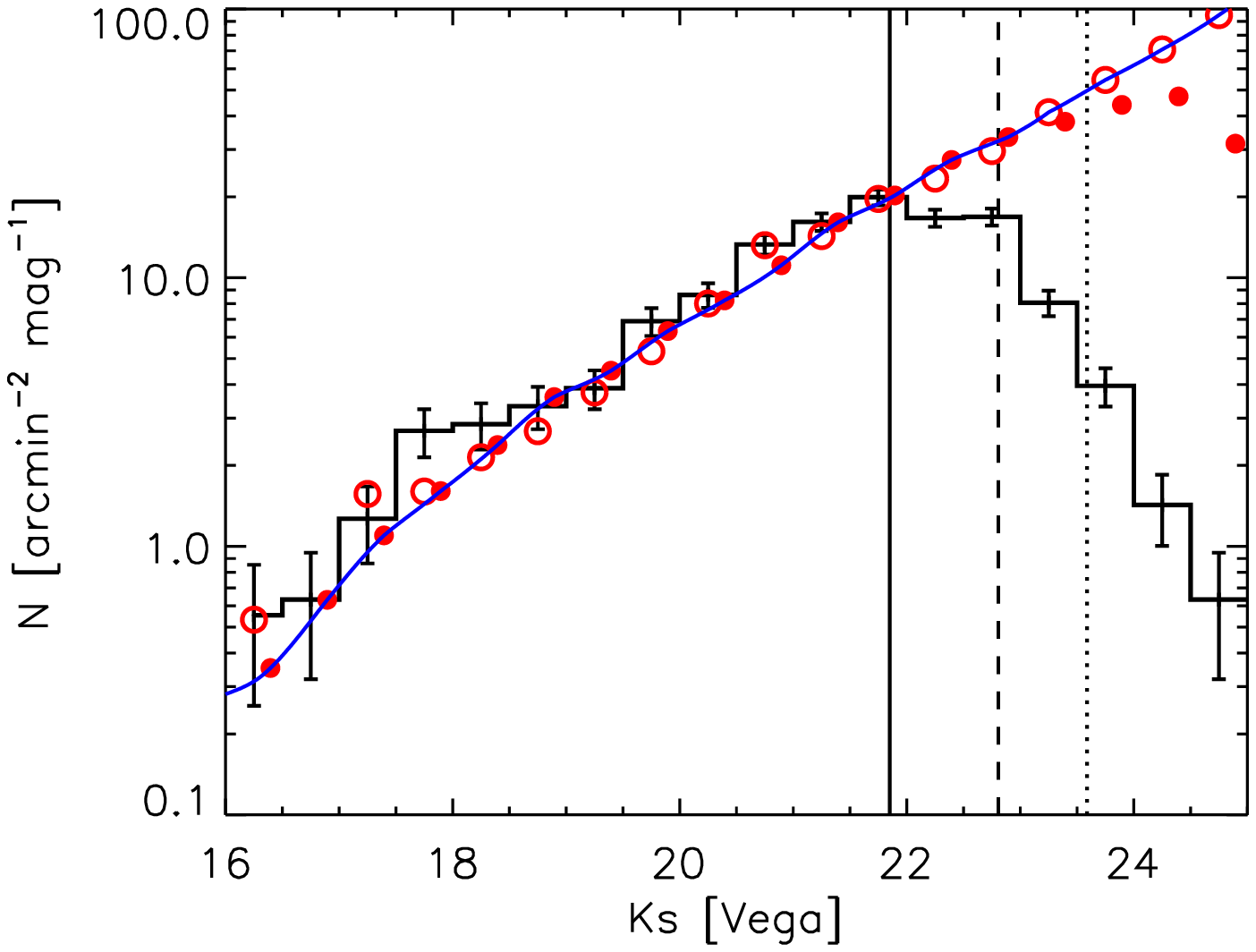}      
      \caption{Differential galaxy number counts in the J ({\it top panel}) and Ks-band ({\it bottom panel}). The black histograms show the measured differential galaxy counts in the 
  background regions with Poisson errors. Open circles show literature number counts taken from  \citet{Maihara2001a} while filled circles are counts from larger fields from  \citet{Windhorst2011a} in J and  \citet{Galametz2013a} in Ks. The solid lines show the combined reference counts used to determine the completeness levels as a function of magnitude. The 100\%/50\%/10\% completeness levels are indicated by solid/dashed/dotted vertical lines. The slight excess in some bins can be attributed to the large-scale structure environment of XDCP\,J0044.0-2033 and the nearby cluster A2813 at $z\!=\!0.292$. 
} 
         \label{fig_NumberCounts}
\end{figure}

Figure\,\ref{fig_NumberCounts} shows the measured differential galaxy number counts in three dedicated background regions, that
is,~in the three quadrants that do not contain 
the cluster center. These background regions with a combined sky area of 25.3 square arcmin
were selected as areas with the full homogeneous exposure time and  without contaminating bright stars. The measured galaxy counts (black histograms) are compared with published literature counts \citep{Maihara2001a,Windhorst2011a,Galametz2013a} in the same filters to evaluate the completeness levels of the observations as a function of limiting magnitudes. These limiting magnitudes for completeness levels of  
100\%/50\%/10\% (vertical lines) were determined to be 23.0/24.2/24.9\,mag (21.9/22.8/23.6\,mag) in the J (Ks) band.

In some of the bins, in particular for brighter magnitudes, a slight excess of observed counts on the 1-2$\sigma$ level with respect the reference counts is observed. This is to be expected given the proximity of the observed HAWK-I field to the massive low-$z$ cluster A2813 at $z\!=\!0.292$,  which was the target of the original XMM-{\it Newton} archival field. The mass of A2813 of $M_{200}\!\simeq\!7\!\times\!10^{14}\,\mathrm{M_{\sun}}$ translates into a cluster radius of $R_{200}\!\simeq\!1.7\,\mathrm{Mpc}\!\simeq\!6.6\arcmin$  \citep{Zhang2008a}. The distant cluster XDCP\,J0044.0-2033 is located approximately 10.5\arcmin \ to the northeast of A2813, implying that the 
 HAWK-I background fields are at cluster-centric distances of 1.5-2\,$R_{200}$ away from the A2813 center, that is, still in the large-scale-structure (LSS) environment for which a cluster-associated excess of foreground galaxies can be expected. 
Moreover, the background-field locations also are at similar cluster-centric distances in units of  $R_{200}$ from 
 XDCP\,J0044.0-2033 itself, implying that the LSS of the system under investigation might contribute in a non-negligible way to the observed background counts. The appropriate background reference counts, that is,~measured in background fields versus literature counts, to be used thus depend on the specific applications discussed in  Sect.\,\ref{c3_Results}.

%
%


\begin{figure*}[t]
    \centering
    \includegraphics[width=0.476\linewidth, clip]{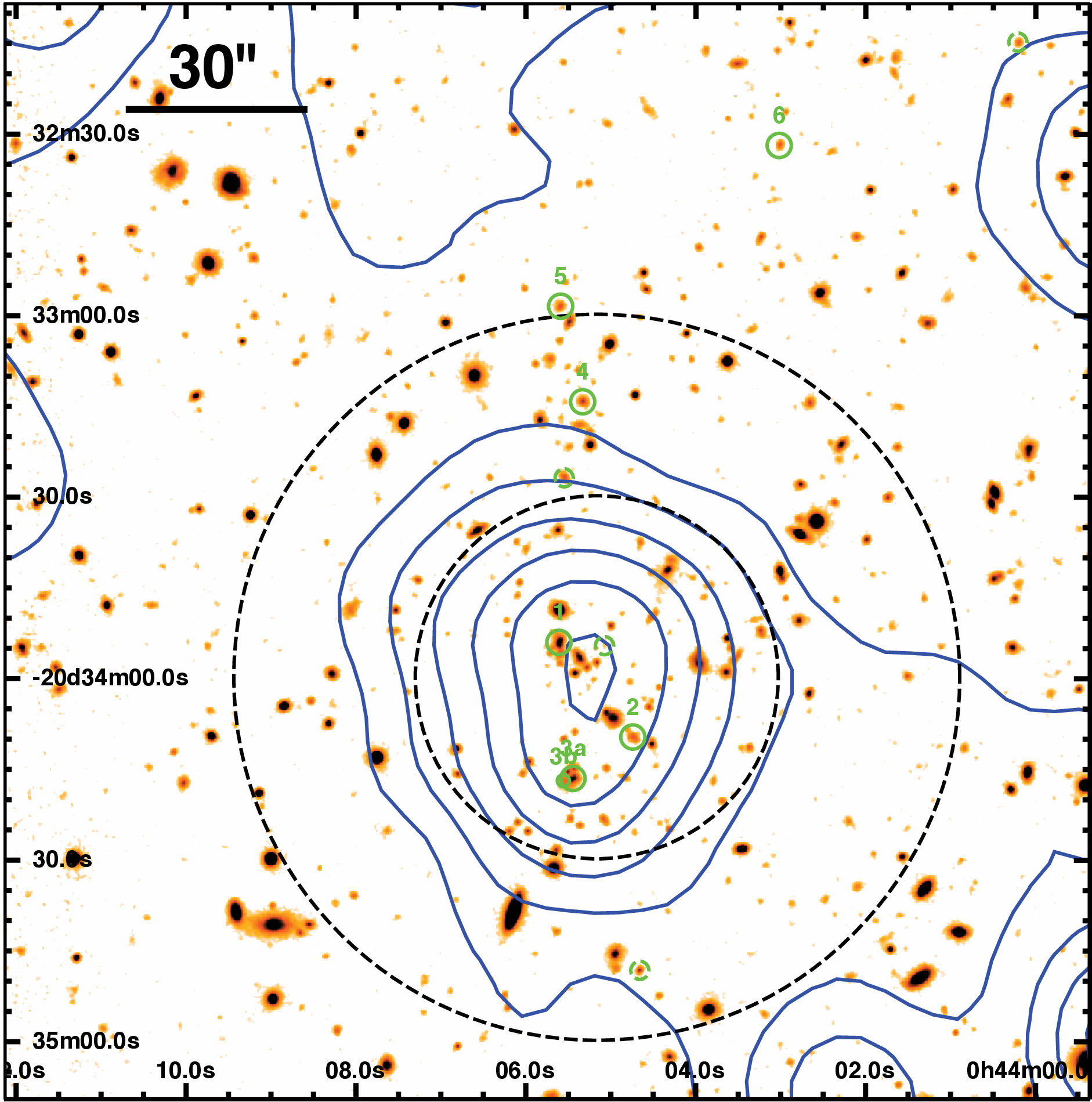}       
    \includegraphics[width=0.485\linewidth, clip]{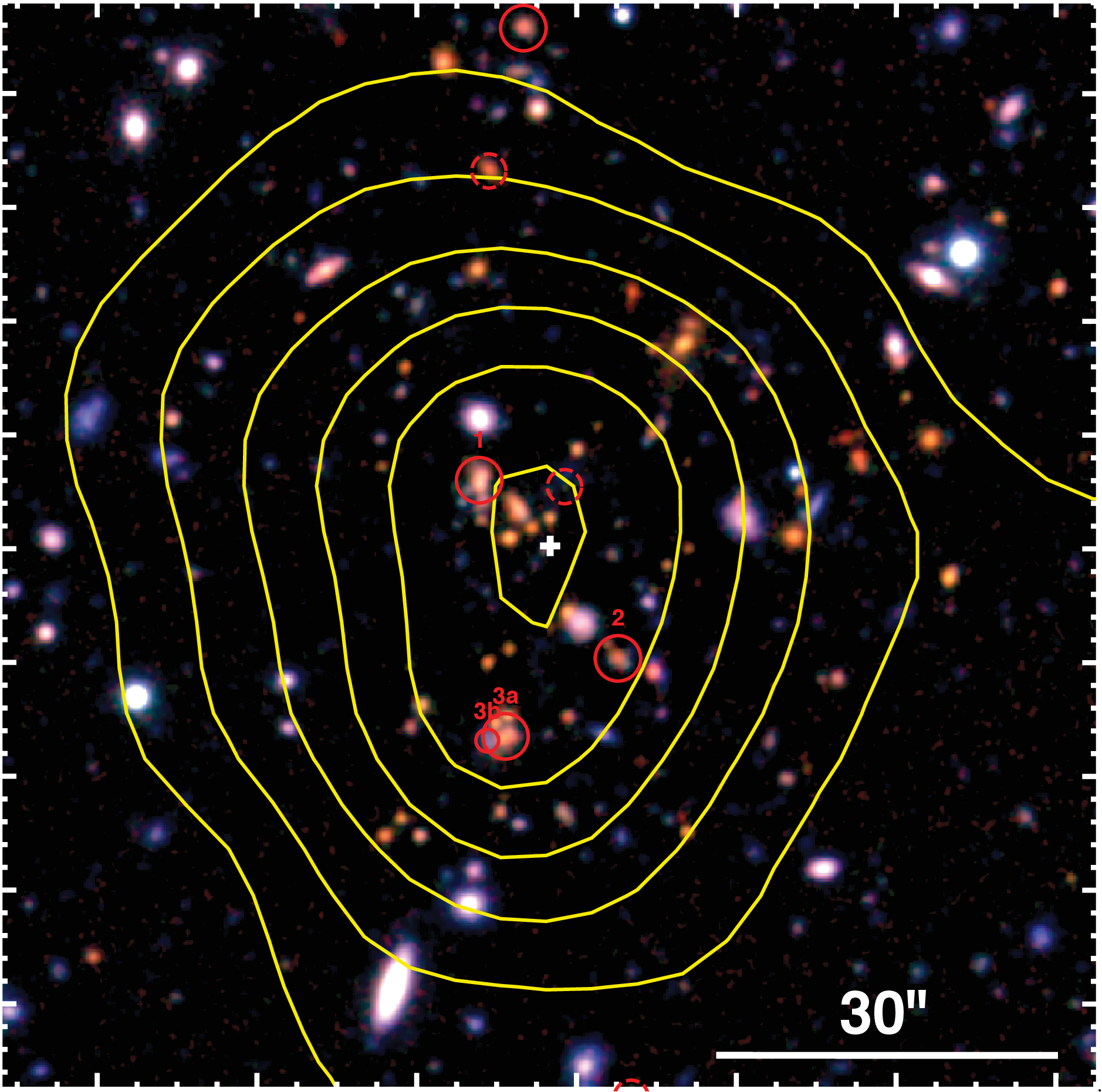}    
      \caption{Environment of the galaxy cluster XDCP\,J0044.0-2033 at $z\!=\!1.58$. Secure spectroscopic cluster members are marked by solid small circles with their object ID number on top, the positions of tentative system members are indicated by small dashed circles. The logarithmically spaced contours show the spatial distribution of the X-ray emission detected with XMM-{\it Newton}. 
      {\it Left panel}:  3\arcmin$\times$\,3\arcmin \ overview image of the combined J+Ks detection band including all currently known secure and tentative cluster member galaxies. The large dashed circles indicate the  30\arcsec \,and 60\arcsec\,($\simeq$R$_{500}$) radii about the X-ray centroid position. {\it Right panel}: color composite image centered on the X-ray centroid (white cross) made up of the Ks (red channel), J (green), and  i-band (blue) with dimensions 1.6\arcmin$\times$\,1.6\arcmin.
} 
         \label{fig_OpticalOverview_X0044}
\end{figure*}

\subsection{Optical spectroscopy with VLT/FORS\,2}
\label{s2_Spectroscopy}

The original discovery data set of XDCP\,J0044.0-2033 presented in \citet{Santos2011a} included only three secure spectroscopic member galaxies of the cluster. To increase the number of spectroscopically confirmed members, we initiated a new optical spectroscopic campaign with VLT/FORS\,2 for a new deep MXU (mask exchange unit)  observation using a customized mask of targeted color-selected candidate member galaxies (program ID: 089.A-0419). However, during the recent observing season only one-third of the scheduled observations was completed, resulting in a net on-target exposure time of 1.45\,hours in good subarcsec observing conditions executed on 1 and 2 January 2013.  

The data were reduced in standard manner using the new publicly  available FORS\,2 pipeline F-VIPGI described in detail in
\citet{Nastasi2013a}. The limited exposure time allowed only the secure redshift determination for member galaxies with significant 
\OII-emission line. This way, we identified four new secure emission-line cluster members listed in 
Table\,\ref{tab_specmembers}  with their identification numbers (2, 3a, 5, and 6) together with the redshifts, their cluster-centric distance, total Ks magnitude, and J$-$Ks and i$-$Ks colors. Together with the previously known members (IDs 1, 3b, and 4) this increases the number of known secure spectroscopic members to seven, leaving the previously determined median system redshift unchanged at $z_{\mathrm{med}}\!=\!1.579$. 
Additionally, four new tentative cluster members were identified, whose redshifts await a final confirmation once the full observational data set is available. The spatial distribution of all secure and tentative spectroscopic cluster member galaxies are indicated in Fig.\,\ref{fig_OpticalOverview_X0044}. More details on the spectroscopic properties of confirmed cluster member galaxies will be presented in a forthcoming publication for a  combined analysis of all new spectroscopic data sets (Nastasi et al., in prep.).

The double object with IDs 3a and 3b is of particular interest. From its NIR morphology it is identified as a single object (3a) by {\em SExtractor}, see also the upper right panel of Fig.\,\ref{fig_PostageStamps} for a close-up view. However, the subcomponent 3b has a significantly bluer color than the central part of the main component 3a, which 
establishes their 
nature as two distinct objects, as can be seen in Fig.\,\ref{fig_Videt_CoreView}. This 
is now confirmed by the  new  spectroscopic results, which yield two distinct spectra and redshifts with  $z\!=\!1.5699$ for the main component 3a and $z\!=\!1.5716$ for the bluer subcomponent 3b. With a projected separation of about 1.6\arcsec$\simeq$13\,kpc and a restframe velocity offset of $\Delta v\!\simeq\!200$\,km/s we can establish this double system as an ongoing merging event of the blue satellite galaxy (3b) with the main red galaxy (3a), which is the second-ranked member galaxy in NIR luminosity among the secure members.


\subsection{Subaru V- and i-band data }
\label{s2_Subaru}

The field of the low-$z$ cluster A2813 has archival coverage of good quality wide-field imaging data in the V- and i bands with Subaru/Suprime-Cam originally intended for a weak-lensing study of  A2813. These data were taken on 6 November 2010 in subarcsec observing conditions for a total exposure time of 28\,min (7$\times$240\,sec) in V and 36\,min (9$\times$240\,sec) in i. These archival data complement our dedicated follow-up programs of XDCP\,J0044.0-2033 in the near- and mid-infrared  toward the optical bands with similar depth and image quality.

The Subaru data were reduced in standard manner using the SDFRED2 software \citep{Ouchi2004a}. 
The resulting image quality of the final stacked images was measured to be 0.77\arcsec \ (FWHM) in V and   0.72\arcsec \ in the i
band. Photometric zero points were obtained from observations of the standard-star field SA95 \citep{Landolt1992a} during the same night. The optical Subaru images were co-aligned with the HAWK-I data and re-gridded to the same pixel scale. 
Photometric catalogs were obtained in the same manner as described in Sect.\,\ref{s2_HawkI} with appropriate seeing adjustments for color measurements in i$-$Ks.   
The limiting magnitudes for the i-band image were determined to be 27.0/28.0/28.5\,mag (Vega) for completeness levels of  100\%/50\%/10\%.

\subsection{Mid-infrared observations with {\it Spitzer}/IRAC }
\label{s2_Spitzer}

{\it Spitzer}/IRAC observations in the 3.6$\mu$m and 4.5$\mu$m channels were obtained for XDCP\,J0044.0-2033 on  24 August 2011 as part of program 80136. 
The total integration time in each channel was 270\,s tailored to reach a depth for the detection of individual galaxies of up to  m*+2 at $z\!\sim\!1.6$.

We retrieved processed (post-BCD) images for both channels  from the {\it Spitzer} Heritage Archive. Tests were performed with the APEX/MOPEX pipeline to assess the quality of the reduction, which was found to be sufficient for our science purposes  without additional processing steps. The photometry of individual objects was performed with  {\em SExtractor} in both channels, which for this work was used only to mask bright foreground galaxies.


\begin{table*}[t]    
\caption{Spectroscopically confirmed, secure member galaxies of the cluster XDCP\,J0044.0-2033.
The table lists the object IDs, the galaxy coordinates, the cluster-centric distances d$_{\mathrm{cen}}$ relative to the XMM-{\it Newton} determined centroid of the extended X-ray emission, the total Ks-band magnitude,  the J$-$Ks color, and the i$-$Ks  color within 1\arcsec \ apertures.
Typical measurement uncertainties for the individual galaxy redshifts are $\sigma_{z}\!\sim\!0.0005$. The Ks magnitude of ID\,3b is given for a 1\arcsec \ aperture since there is no clear separation from the main component 3a (see upper right panel of Fig.\,\ref{fig_PostageStamps}). All galaxies but ID\,1 exhibit a detected \OII \ emission line.
}
 \label{tab_specmembers}

\centering
\begin{tabular}{ c c c c c c c c}
\hline \hline

ID & RA     & DEC     &   $z_{\mathrm{spec}}$   & d$_{\mathrm{cen}}$   & Ks  & J$-$Ks  & i$-$Ks \\

   &  J2000 &  J2000  &  &   arcsec   & Vega mag  & Vega mag   & Vega mag      \\

\hline
1 &     11.02336 &      -20.56509 & 1.5797  &   8.2 &   17.96 &       1.98 & 4.21      \\      
2 &     11.01973 &      -20.56936 & 1.5936  &   11.7 &  19.52 &       2.08 & 4.23      \\ 
3a & 11.02269 & -20.57125 & 1.5699  &   17.3 &  18.06 & 2.33 & 4.86   \\     
3b & 11.02316 &         -20.57135 & 1.5716       &      18.0 &       21.04  &        1.38 & 2.98      \\         
4 &     11.02220 &      -20.55395 & 1.5787  &   45.5 &  19.25 &       2.10 & 4.27      \\ 
5 &     11.02333 &      -20.54957 & 1.5778  &   61.6 &  19.58 &       1.92 & 4.20      \\     
6 &     11.01255 &      -20.54218 & 1.5986  &   92.9 &  20.09 &       1.82 & 3.71      \\

\hline
\hline
\end{tabular}
\end{table*}


For the current work we only aim at obtaining a measurement of the integrated 4.5$\mu$m MIR flux 
within a given projected distance from the cluster center
independent of the detection and proper deblending of individual galaxies, for which we applied the following approach: we masked out foreground sources brighter than the most luminous cluster galaxies with m(Vega)$\le$15\,mag (see left panel of Fig.\,\ref{fig_Spitzer_View}). We then measured the average integrated background surface brightness for the 4.5$\mu$m channel in a dozen independent background regions outside the cluster volume, most of which are located in the parallel field observations obtained simultaneously during the channel 1 on-target integration. This determination of the integrated 4.5$\mu$m background surface brightness yielded a field value of  0.1378$\pm$0.0028\,MJy/sr
with a standard deviation representative for square arcmin scales. This background value was subsequently subtracted from the integrated  4.5$\mu$m flux measured inside the analysis region 
 of XDCP\,J0044.0-2033.

\section{Galaxy population properties of XDCP\,J0044.0-2033}
\label{c3_Results}

Figure\,\ref{fig_OpticalOverview_X0044} provides an overview of the cluster and its environment together with the spatial distribution of the detected XMM-{\it Newton} X-ray emission and the location of the spectroscopic members. The image in the left panel covers an area of  3\arcmin$\times$\,3\arcmin, while the i+J+Ks band color composite in the right panel shows the central 1.6\arcmin$\times$\,1.6\arcmin \ of the galaxy cluster. From the original 
total mass estimate of M$_{\mathrm{200}}\!\simeq\!3\times10^{14}$\,M$_{\sun}$ we can approximate the characteristic cluster radii to be  R$_{\mathrm{200}}\!\simeq\!760$\,kpc$\simeq\!90\arcsec$ and R$_{\mathrm{500}}\!\simeq\!490$\,kpc$\simeq\!58\arcsec$.

In the following subsections, we investigate 
different aspects of the cluster 
galaxy population properties starting with purely statistical 
measures such as radial surface density profiles (Sect.\,\ref{s3_Profiles}), the stellar mass profile (Sect.\,\ref{s3_MIRproperties}), and NIR luminosity functions  (Sect.\,\ref{s3_NIR_LF}), followed by a closer look at the morphologies of individual galaxies  (Sect.\,\ref{s3_morphologies}), the color-magnitude relation  (Sect.\,\ref{s3_CMR}),
the faint end of the red sequence  (Sect.\,\ref{s3_FaintEnd_RS}), the spatial distribution of color-color selected galaxy types    (Sects.\,\ref{s3_color_color}\,\&\,\ref{s3_SpatialDistribution}), and the formation of the brightest cluster galaxy in the core region  (Sect.\,\ref{s3_BCGobs}).



\begin{figure}[t]
    \centering
    \includegraphics[width=\linewidth, clip]{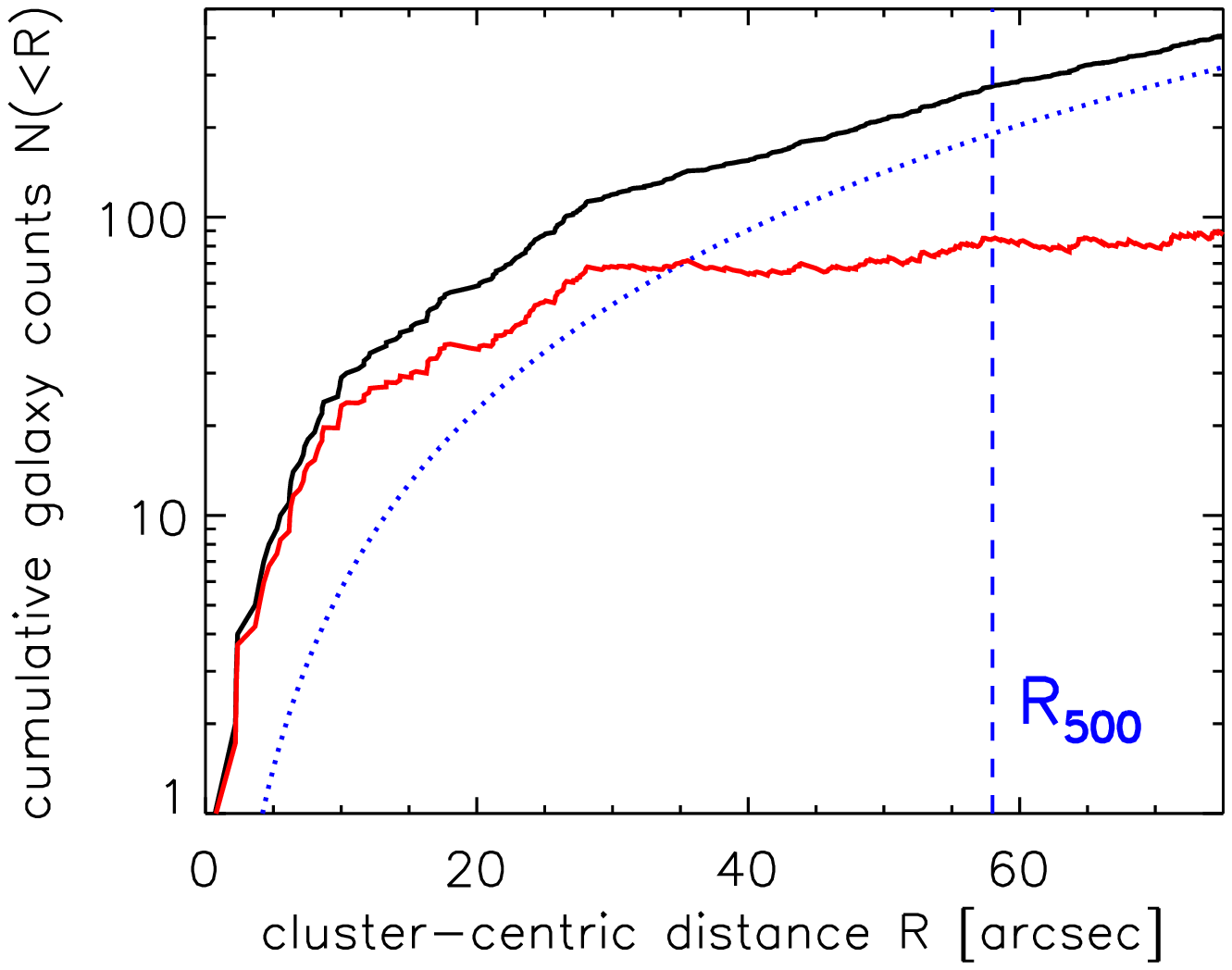}    
    \includegraphics[width=\linewidth, clip]{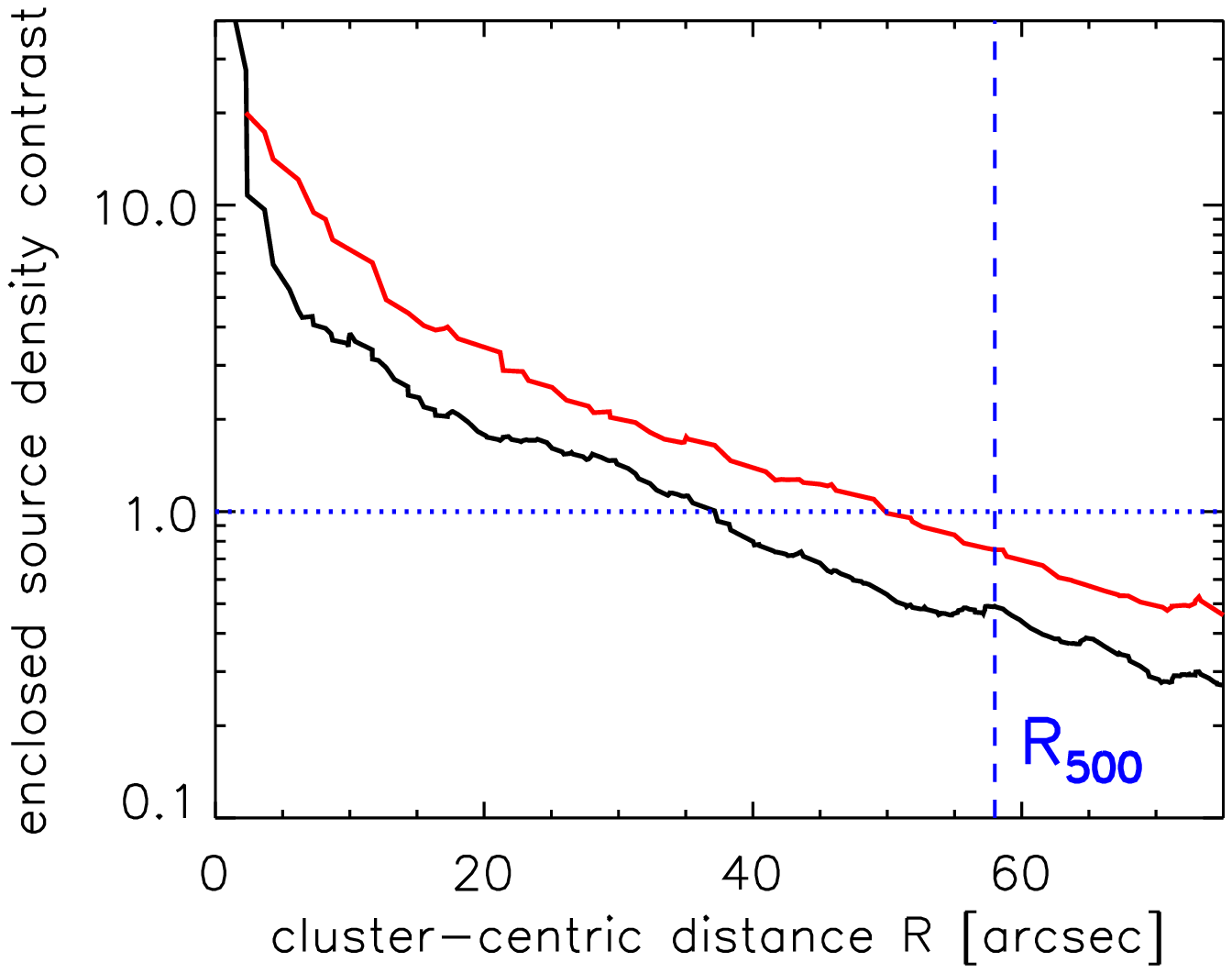}    
      \caption{Enclosed galaxy count properties within the projected cluster-centric distance $R$ from the X-ray centroid based on all HAWK-I detected sources. The dashed vertical line indicates the  R$_{\mathrm{500}}$ radius of the cluster.
      {\it Top panel}: Cumulative radial galaxy count profile $N(<R)$ of enclosed sources within 
      $R$  for all galaxies (black solid line) and background-subtracted net cluster member counts (red line). The average background counts are shown by the blue dotted line. A total of about 90 excess galaxies associated with the cluster are detected.
 {\it Bottom panel}: Enclosed galaxy density contrast ($[N(<R)-B(<R)]/B(<R)$) within radius $R$ for all galaxies with 17$\le$Ks$\le$22 (black line) and brighter galaxies with   17$\le$Ks$\le$20 (top red line). At the intersection point with the horizontal dashed dotted line the enclosed cluster galaxy density equals the background density.     
} 
         \label{fig_RadialProfile}
\end{figure}

\begin{figure}[t]
    \centering
    \includegraphics[width=\linewidth, clip]{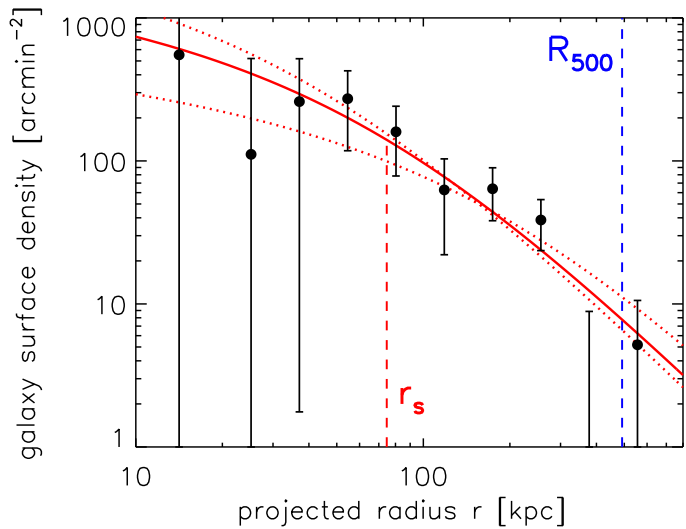}    
      \caption{Projected cluster galaxy surface-density profile (black data points) and best-fitting projected NFW profile (red solid line) with a concentration parameter of $c_{200}\!=\!10.2\pm6.6$ and scale radius $r_s\!=\!R_{200}/c_{200}\!\simeq\!75$\,kpc (dashed red vertical line).     
Dotted red lines depict the profiles with $c_{200}\!\pm\!1\sigma$.} 
         \label{fig_NFW_SD}
\end{figure}

\begin{figure}[h]
    \centering
    \includegraphics[width=\linewidth, clip]{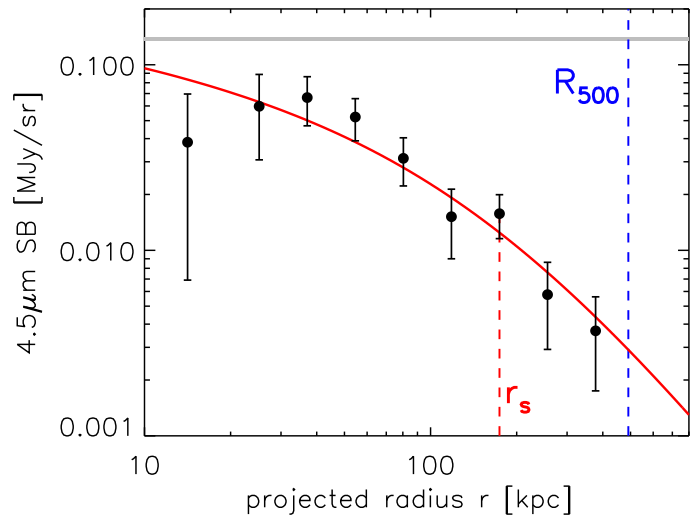}     
      \caption{4.5\microns \ surface-brightness profile (black data points) based on the {\it Spitzer} data shown in Fig.\,\ref{fig_Spitzer_View}. The best-fitting projected NFW profile (red solid line) with a concentration parameter of $c_{200}\!=\!4.4\pm1.6$ is overplotted in red. The median background level is indicated by the gray horizontal band on top, vertical lines are as in Fig.\,\ref{fig_NFW_SD}. 
      } 
         \label{fig_Spitzer_SB}
\end{figure}

\subsection{Radial galaxy density profiles}
\label{s3_Profiles}

As a first application of the deep NIR HAWK-I data of Sect.\,\ref{s2_HawkI}, we measured the overdensity contrast and radial profile of the galaxy population of  XDCP\,J0044.0-2033. 
To this end, we determined the background galaxy density of all nonstellar sources in the three control quadrants as 68.1$\pm$2.5\,arcmin$^{-2}$, where the background uncertainty is given by Poisson error along
with the expected influence of the LSS contribution of the cluster environments.  The top panel of Fig.\,\ref{fig_RadialProfile} shows the detected cumulative galaxy counts $N(<R)$ within the cluster-centric distance $R$ for all galaxies, the
expected background, and the net excess of cluster galaxies. In total, we detect a cluster associated net excess of about 90 galaxies, approximately 75\% of which are located within 30\,arcsec from the cluster center. The enclosed high-$z$ cluster galaxy population  equals the background galaxy contribution at a cluster-centric distance of 35\,arcsec, outside of which the background is dominant.  

The enclosed density contrast of cluster galaxies with respect to the background   is  shown in the bottom panel of Fig.\,\ref{fig_RadialProfile} for the bulk of the galaxy population and bright galaxies alone.  
At $z\!\sim\!1.6$,  even a rich cluster such as XDCP\,J0044.0-2033  stands out from the background only within about half an arcmin in deep NIR imaging data. Source-density contrasts on the order of 10 or more are only reached within a few arcsec from the X-ray center. When the enclosed density profile is restricted to brighter magnitudes 17$\le$Ks$\le$20 (red line), for instance,~with shallower NIR data, the density contrast is shifted upward  owing to 
a significantly reduced background density ($\simeq$10.5\,arcmin$^{-2}$) and the preferred location of luminous cluster members in the central region.

With the given 
number of detected cluster-associated galaxies for XDCP\,J0044.0-2033 one can attempt to measure the NIR galaxy surface-density  profile of the cluster. This is shown 
in  Fig.\,\ref{fig_NFW_SD} as a function of projected physical cluster-centric radius with Poisson errors following \citet{Gehrels1986a}. The red solid line shows the best fit  projected NFW profile \citep[e.g.,][]{Lokas2001a}
to the source surface density distribution with a formal best-fit solution for the concentration parameter  of $c_{200}\!=\!10.2\pm6.6$. Despite the significant statistical uncertainty in determining the concentration parameter, as can be expected for a single galaxy cluster at  $z\!\sim\!1.6$, the compact distribution of cluster galaxies can be visually confirmed in the color image of  Fig.\,\ref{fig_OpticalOverview_X0044} (right panel), where very few red galaxies are visible beyond R$\ga$30\arcsec, 
in particular toward the South (bottom) and East (left). 

Based on the compact best-fit galaxy surface-density profile with $c_{200}\!\simeq\!10$, constrained mostly within R$_{500}$, with its relatively high concentration parameter compared to local clusters  \citep[e.g.,][]{Budzynski2012a}, estimates for the expected number of detectable galaxies in the cluster outskirts can be obtained.
This suggests 
 that only about 20\% of all galaxies within the fiducial cluster radius R$_{200}$ are 
 located in the cluster outskirts beyond the R$_{500}$ radius, that is,~at cluster-centric distances of 58-90\arcsec.
However, in these outer regions  the Poisson noise of the background counts dominates the modest expected increase of the cumulative cluster-galaxy count function (top panel of Fig.\,\ref{fig_RadialProfile}). 
With the current data, this function can be reliably  traced out to a cluster-centric radius of $\simeq$75\arcsec, beyond which the chip gaps of the HAWK-I detector array influence the source detection. However, the almost flat cumulative profile shape in the outer cluster region
with only a modest increase in excess cluster-counts is fully consistent with the extrapolated expectations from the best-fit NFW surface density function within the uncertainties imposed by the increasing Poisson noise.


\subsection{Mid-infrared view of the cluster and stellar mass
profile}
\label{s3_MIRproperties}

\begin{figure*}[t]
    \centering
    \includegraphics[width=0.885\linewidth, clip]{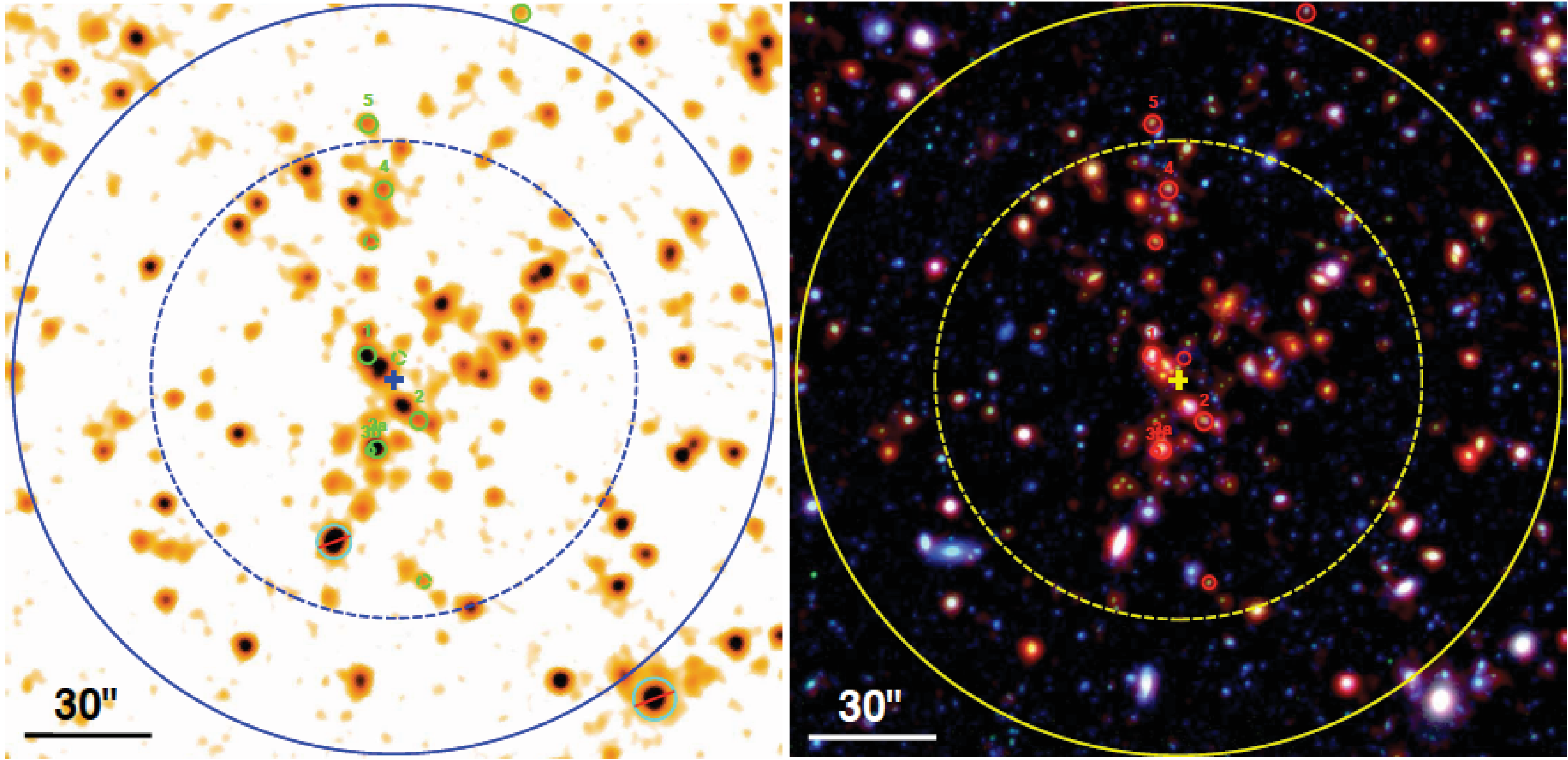} 
      \caption{{\it Left panel}:  {\it Spitzer}/IRAC 4.5\microns \ view of the cluster volume (3\arcmin$\times$\,3\arcmin). The cluster center (central cross),  R$_{500}$ (dashed  blue circle), and R$_{200}$ (solid  blue circle) are indicated, small 
      circles mark spectroscopic members as in Fig.\,\ref{fig_OpticalOverview_X0044}. The crossed-out cyan objects were removed from the analysis. {\it Right panel}: color composite with the same FoV by adding the V-band (blue) and the combined J+Ks image (green) to the   {\it Spitzer} 4.5\microns  \ data (red).} 
         \label{fig_Spitzer_View}
\end{figure*}

\begin{figure}[t]
    \centering
 \includegraphics[width=\linewidth, clip]{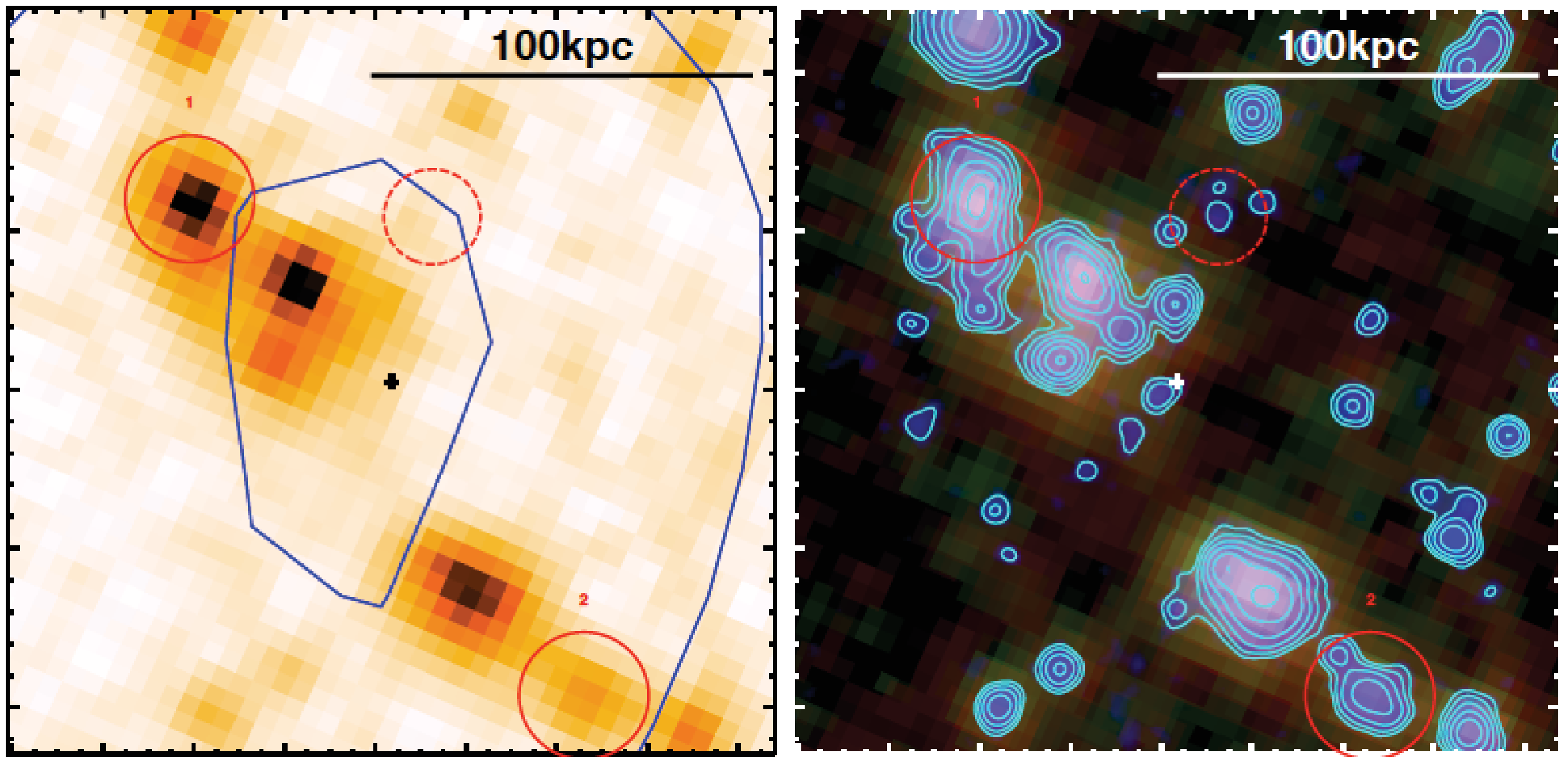} 
      \caption{{\it Spitzer}/IRAC view of the central 100\,kpc of the cluster core of XDCP\,J0044.0-2033.
      Left panel:  4.5\microns \ image with X-ray contours overlaid in blue and symbols as in Fig.\,\ref{fig_Spitzer_View}.
    Right panel: same field-of-view as color image composed of 4.5\microns \ (red), 3.6\microns \ (green), and the 0.5\arcsec-resolution NIR JKs detection image (blue) with cyan NIR isophotal contours derived from the latter band to visualize the source confusion at the {\it Spitzer} resolution. }
         \label{fig_Spitzer_View_100kpc}
\end{figure}

\begin{figure}[h]
    \centering
    \includegraphics[width=\linewidth, clip]{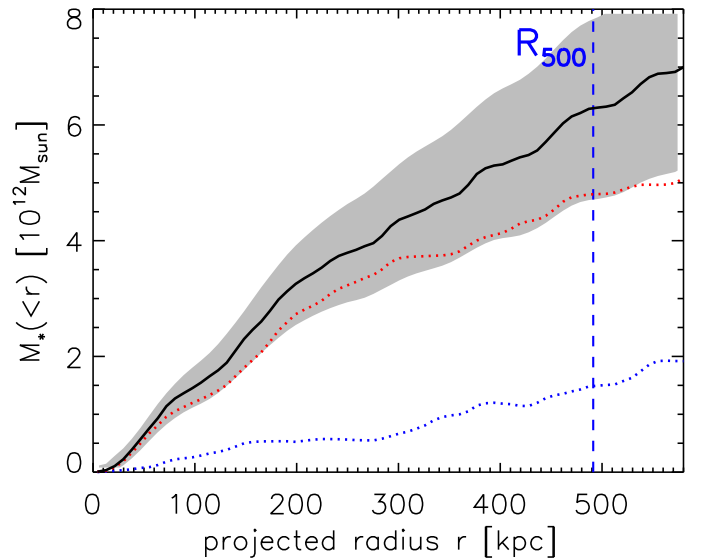}     
      \caption{Cumulative total stellar mass profile (black solid line) for XDCP\,J0044.0-2033 as a function of  projected cluster-centric radius with 1\,$\sigma$ uncertainties indicated by the gray-shaded area. The relative contributions to the total mass profile of the  red-masked objects (red) and the remaining area (blue) are shown by the dotted lines.      
      } 
         \label{fig_Spitzer_StellarMassProfile}
\end{figure}



After constraining the radial number-density profile of the cluster-associated galaxy population  of XDCP\,J0044.0-2033, we analogously examined the stellar mass distribution of the cluster. To this end, we used the mid-infrared (MIR) {\it Spitzer}/IRAC observations introduced in Sect.\,\ref{s2_Spitzer}. At $z\!\sim\!1.6$, the observed 4.5\microns-band traces the restframe NIR peak of stellar light emitted by old stellar populations (see Fig.\,\ref{fig_TemplateFilters})  and is therefore optimally suited to derive integrated quantities of the  stellar light and mass content of high-$z$ clusters \citep[e.g.,][]{Brodwin2006a}.

The {\it Spitzer}/IRAC view of the projected cluster volume of XDCP\,J0044.0-2033 is shown in Fig.\,\ref{fig_Spitzer_View} for the 4.5\microns \ surface brightness distribution (left panel) and as a color-image representation in combination with the NIR and optical data (right panel). Using the determined integrated background value of Sect.\,\ref{s2_Spitzer}, we 
 measured a total net  4.5\microns \ cluster flux  of 1.78$\pm$0.40\,mJy (2.58$\pm$0.62\,mJy) inside the projected  
R$_{500}$  (R$_{200}$) radius shown in Fig.\,\ref{fig_Spitzer_View}, corresponding to a total integrated apparent (Vega) magnitude of  m$^{4.5\mu m}_{500}$=12.45$\pm$0.22\,mag and m$^{4.5\mu m}_{200}$=12.04$\pm$0.24\,mag, respectively.


Figure\,\ref{fig_Spitzer_SB} shows the observed  4.5\microns \ surface brightness profile inside R$_{500}$ .
The best-fitting projected NFW profile to the observed surface density yields a result for the concentration parameter of $c_{200}\!=\!4.4\pm1.6$, lower than the compact profile parameter of the galaxy surface density of Fig.\,\ref{fig_NFW_SD}. 
In contrast to the galaxy number density, the 4.5\microns \ surface brightness, used as proxy for stellar mass, is not centrally peaked, but drops within $\simeq$40\,kpc of the X-ray centroid position.

The central 100\,kpc region of the cluster core  is shown as a zoom in Fig.\,\ref{fig_Spitzer_View_100kpc} and is discussed in more detail in Sect.\,\ref{s3_BCGobs}. The brightest central {\it Spitzer} source is located northeast of the X-ray center (left panel) and is a blend of  at least 
four red galaxies closest to the center (see right panel). The most central galaxies at the current accuracy of the centroid position are faint blue galaxies with little flux in the  {\it Spitzer}  band, but counting toward the galaxy surface density with a resulting higher NFW concentration parameter. However, outside the very core region the NFW profile fits the observed 4.5\microns \ surface brightness profile quite well, which is also reflected in the reduced  $\chi^2$ value of the fit of 0.8.

To convert the observed  4.5\microns \ cluster flux to stellar mass at $z\!=\!1.58$, we evaluated the stellar-mass-to-light-ratio M$_*$/L$_{4.5\mu m}$ based on a grid of stellar population models with different star formation histories and metallicities. 
To account for changes of  M$_*$/L$_{4.5\mu m}$ across the cluster volume owing to the varying underlying galaxy populations, we adopted a two-model approach that accounts for old red galaxies with a solar metallicity  $z_{\mathrm{f}}$=3 model on one hand and bluer galaxies with a younger ($z_{\mathrm{f}}$=2) stellar population on the other.  
To this end, we masked the areas of detected galaxies with red colors of J$-$Ks$\ge$1.84 (see Sects.\,\ref{s3_CMR}\,\&\,\ref{s3_FaintEnd_RS}) and converted the background-subtracted cluster flux with the $z_{\mathrm{f}}$=3 model, while using the 
$z_{\mathrm{f}}$=2 model for the unmasked regions.

Figure\,\ref{fig_Spitzer_StellarMassProfile} shows the resulting cumulative total stellar mass profile enclosed within the projected cluster-centric radius $r$ and the two contributing components for the old galaxy model and the galaxies  with an assumed younger stellar formation epoch. The gray-shaded area shows the 1\,$\sigma$ uncertainties of the total enclosed stellar mass, which are dominated by net cluster flux errors and the remaining model uncertainties of the stellar-mass-to-light ratios.  

We measure a  4.5\microns -based total  integrated  stellar mass of  XDCP\,J0044.0-2033  within the projected R$_{500}$  and  R$_{200}$ radii of  $M*_{500}\!\simeq\!(6.3 \pm 1.6)\!\times\!10^{12}\,\mathrm{M_{\sun}}$  and $M*_{200}\!\simeq\!(9.0 \pm 2.4)\!\times\!10^{12}\,\mathrm{M_{\sun}}$, respectively. The central 100\,kpc  cluster core region alone contains a stellar mass of  $M*_{100\,\mathrm{kpc}}\!\simeq\!(1.5 \pm 0.3)\!\times\!10^{12}\,\mathrm{M_{\sun}}$.
Based on the original 
estimate of the total mass of  XDCP\,J0044.0-2033  with a $\sim$35\% uncertainty, we can obtain a first approximate determination of the stellar mass fraction\footnote{The stellar mass-fraction measurement is to first order independent of the exact value of the total cluster mass, since total mass and stellar mass follow similar radial profiles. However, since the cluster radius changes with total mass, the exact value of $M*_{500}$ does increase for higher estimates of the total cluster mass.}
of a  $z\!\sim\!1.6$ cluster, yielding f$_{*,500}=M_{*,500}/M_{500}  = (3.3\pm1.4)$\%. This value is fully consistent with the best-fit model of stellar mass fractions of low-redshift systems by 
\citet{Giodini2009a}, which predicts  a stellar mass fraction of $\simeq$3\% with 50\% intrinsic scatter
for the mass of this cluster.





\subsection{J- and Ks-band galaxy luminosity functions}
\label{s3_NIR_LF}

\begin{figure}[t]
    \centering
    \includegraphics[width=\linewidth, clip]{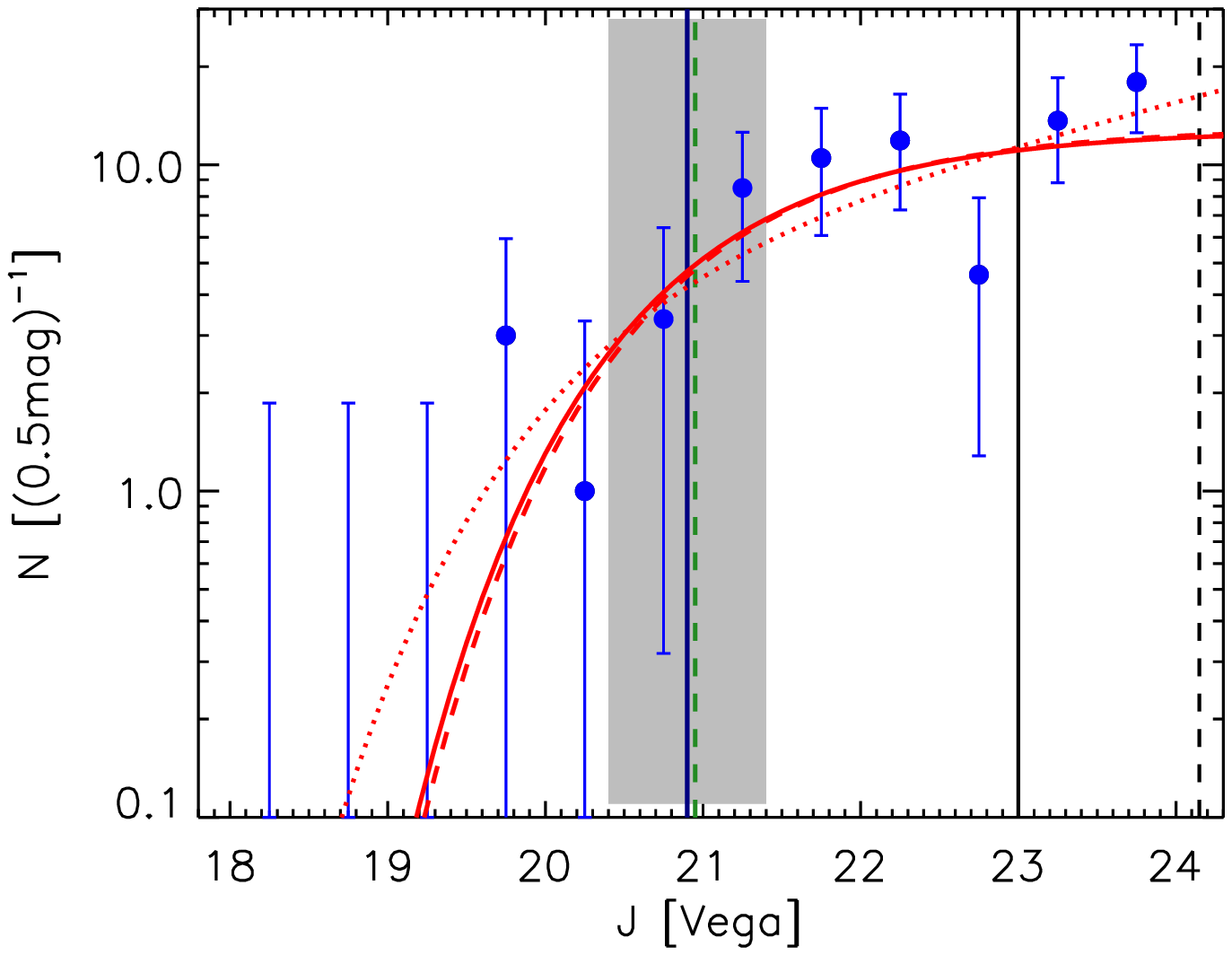}    
    \includegraphics[width=\linewidth, clip]{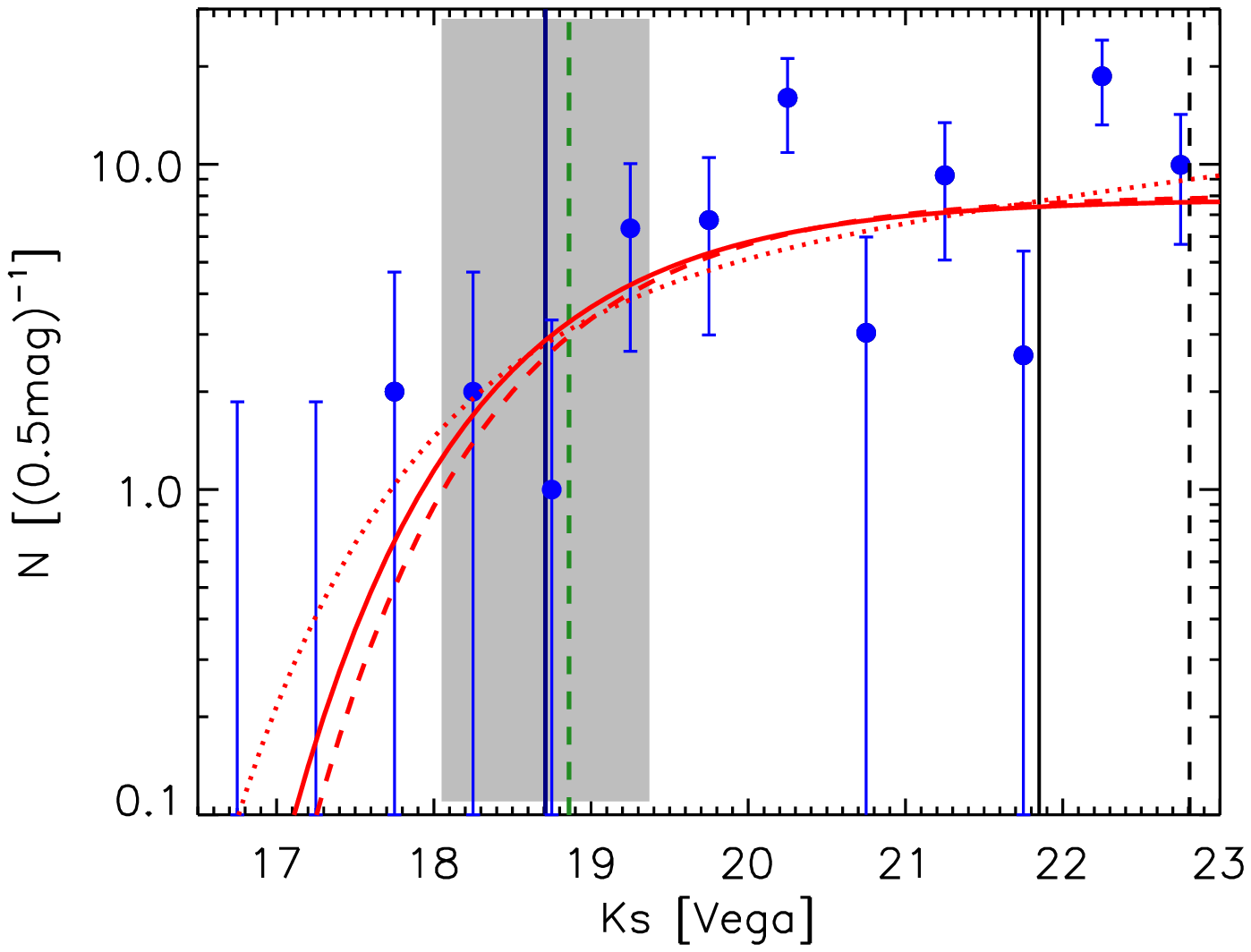} 
      \caption{Observed galaxy luminosity functions (blue data points) in the J- ({\it top}) and Ks-band ({\it bottom}) for XDCP\,J0044.0-2033 extracted within a cluster-centric radius of 30\arcsec\,($\simeq$250\,kpc). The red lines show the best fit Schechter functions for three ($\phi$*, m*, $\alpha$) free parameters (dotted), two free parameters with a fixed faint end slope of $\alpha\!=\!-1$ (solid), and a single parameter normalization fit (dashed) using a SSP model magnitude for m* (green dashed vertical line). The dark blue vertical line with the grey shaded area indicates the best fit solution of the characteristic magnitude m* of the two parameter fit with 1$\sigma$ uncertainties. The vertical black (dashed) solid lines on the right indicate the 100\% (50\%) completeness limits. 
} 
         \label{fig_NIR_LFs}
\end{figure}

Next, we derived 
the near-infrared galaxy luminosity (LF) functions of  XDCP\,J0044.0-2033 in the J and Ks bands 
for a first global characterization of the cluster galaxy population properties.
To this end, the signal-to-noise ratio (S/N) of the background-subtracted excess cluster members with respect to the Poisson noise of all galaxies within an encircled region need to be as high as possible. This S/N-optimized cluster-centric  extraction radius is determined to be 30\arcsec \ ($\simeq$250\,kpc at $z\!=\!1.58$) with an
S/N$\simeq$6.2, enclosing approximately 68 cluster galaxies and 52 background sources (see Fig.\,\ref{fig_RadialProfile}).    

To determine the NIR cluster galaxy luminosity functions of XDCP\,J0044.0-2033  we followed the approach adopted by \citet{Strazzullo2006a} for other individual high-$z$ systems at $z\!\simeq\!1.2$. 
We selected 
 all galaxies  in the S/N-optimized extraction region within $R\! \le \!30\arcsec$  using a binning of 0.5\,mag width. We considered 
 all galaxies with magnitudes up to the 50\% completeness limits in J and Ks, 
as determined in Sect.\,\ref{s2_HawkI}, with the appropriate completeness correction factor applied to the two faintest magnitude bins. The expected background contribution to each bin in the analysis region was then estimated based on the literature model counts 
(Fig.\,\ref{fig_NumberCounts}) and subtracted from the measured total counts in a given bin. The 
uncertainties of the net counts were estimated based on the Poisson errors of both the cluster and background counts following  \citet{Gehrels1986a} for the low count regime.

At magnitudes brighter than 19 in Ks and 20.5 in J, the background and cluster counts per bin within the analysis area drop to the
order of unity and below, at which point the statistical background-subtraction approach is quite uncertain. 
To improve the reliability of the LF constraints on the bright end, all eight galaxies in the analysis region with Ks$<$19 were checked individually and assigned a foreground or cluster membership flag. As can be seen in Fig.\,\ref{fig_Videt_CoreView}, three of the galaxies (labels f1-f3) can be readily identified as low-$z$ interlopers and were therefore considered as foreground galaxies. 
Two galaxies (ID 1 and 3a) are spectroscopically confirmed cluster members, and the three remaining ones are flagged as high-probability  cluster member candidates (labels c1-c3) with colors encompassed by the two bright spectroscopic members in a region of the color-magnitude diagram with  very low 
background/foreground contamination (see Sect.\,\ref{s3_CMR}). 

Figure\,\ref{fig_NIR_LFs} shows the  resulting observed net cluster counts 
for XDCP\,J0044.0-2033 per magnitude bin  in  the J- (top panel) and Ks band (bottom panel). To obtain constraints on the characteristic magnitude m* at $z\!\sim\!1.6$ a 
Schechter 
function \citep{Schechter1976a} with the additional free parameters for the normalization $\phi$* and the faint end slope $\alpha$ was fitted to the data. 
However, a three-parameter LF fit  for a single cluster at this redshift  has too many degrees of freedom for useful constraints on each parameter. As a realistic constrained baseline model we therefore used a two-parameter fit with a fixed faint-end slope of $\alpha\!=\!-1$  as found, for example, by~\citet{Strazzullo2006a} and \citet{Mancone2012a}
for their composite cluster LF at $z\!\simeq\!1.2$. 
For consistency checks, we also fitted an LF with a single free parameter for the normalization based on the prediction of the characteristic magnitude m* from a simple stellar population (SSP) evolution model \citep{Fioc1997a}.




Although deriving an LF measurement for a single $z\!\sim\!1.6$ cluster is a challenging task, the observed J-band luminosity function of XDCP\,J0044.0-2033 in the top panel of Fig.\,\ref{fig_NIR_LFs} is surprisingly well defined and is consistently described by all three underlying Schechter fits with reduced $\chi^2$ values of $\sim0.5$.
The two-parameter fit (solid red line) of the luminosity function yields a characteristic magnitude of J*$=$20.90$\pm$0.50  that
agrees very well with the SSP model prediction for a formation redshift of $z_{\mathrm{f}}\!=\!3$ of J*$_{\mathrm{SSP}}$$=$20.95.
The three-parameter  fit is consistent within the expected large uncertainties (J*$=$20.11$\pm$1.60, $\alpha\!=\!-1.30\pm0.37$).


The LF of 
the Ks-band data (bottom panel of Fig.\,\ref{fig_NIR_LFs}) is not quite as well defined as J, with several outliers on the  2$\sigma$ level from the best-fitting LF models and corresponding reduced  $\chi^2$ values of $\sim1.2$.
Nevertheless, the derived characteristic magnitude based on the two-parameter fit of 
Ks*$=$18.71$\pm$0.66 still agrees well with the SSP model value of  Ks*$_{\mathrm{SSP}}$$=$18.86.  The formal three-parameter fit yields Ks*$=$18.17$\pm$1.54, $\alpha\!=\!-1.15\pm0.29$.

Since a single, well-defined J$-$Ks color for the cluster galaxies would only 
result in a shift along the x-axis between the J and Ks data points, one can immediately conclude from the measured magnitude distribution that the cluster galaxies cannot have a single predominant  J$-$Ks color. This aspect is investigated in more detail in Sect.\,\ref{s3_CMR} based on the color-magnitude diagram.





\begin{figure*}[t]
    \centering
\includegraphics[width=0.8\linewidth, clip]{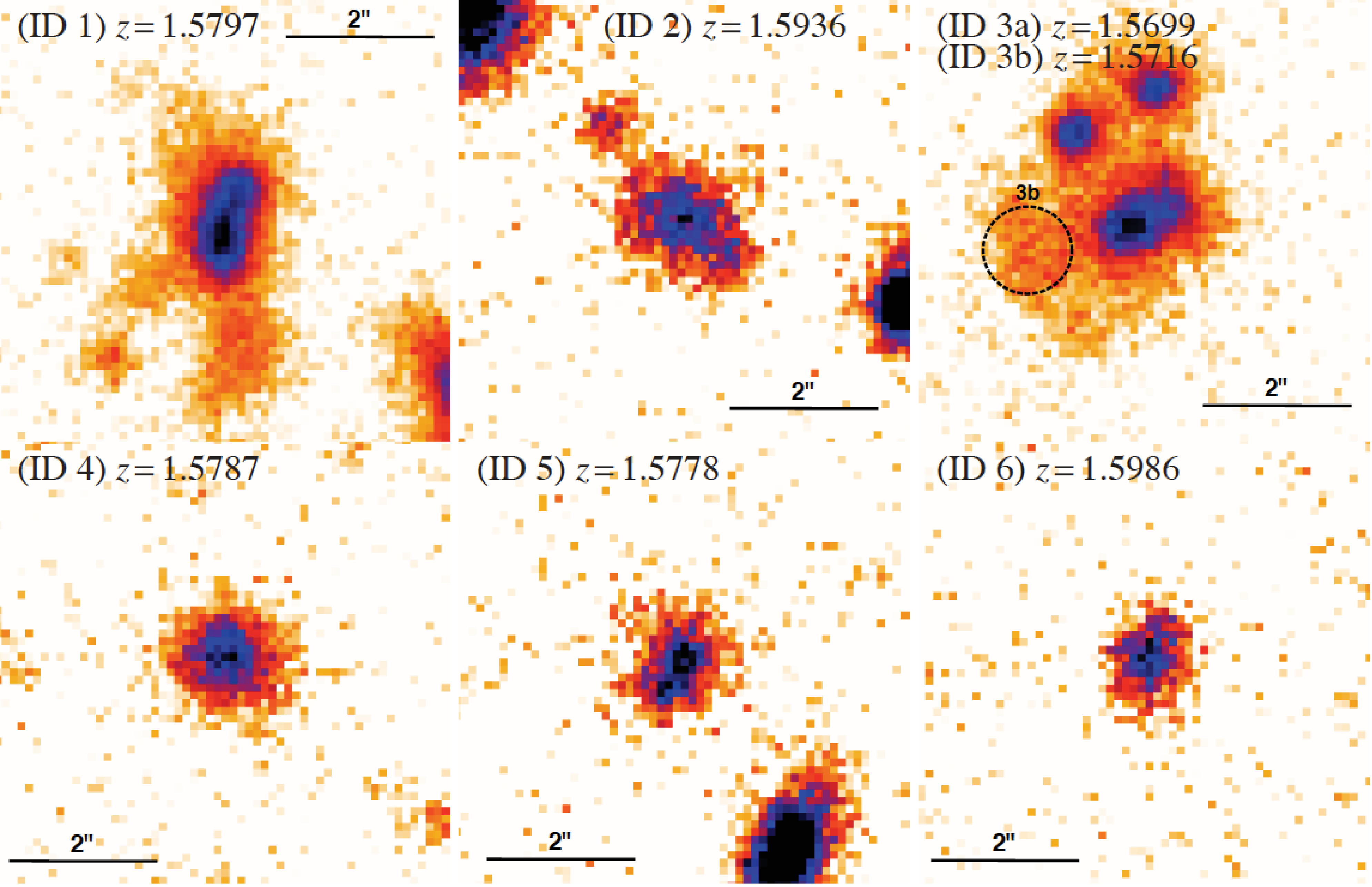} 
%
%
      \caption{Image cutouts of the secure cluster member galaxies based on the combined J+Ks band detection image without any smoothing applied. Each postage-stamp cutout has dimensions 6\arcsec$\times$\,6\arcsec \,($\sim$50\,kpc$\times$50\,kpc) centered on the member galaxy with object ID and redshift stated on top of each panel. The color scale is linear with increasing flux toward darker colors. The position and redshift of the secondary object clump with ID 3b is indicated in the upper right panel.
}
         \label{fig_PostageStamps}
\end{figure*}

\subsection{Morphologies of secure cluster members}
\label{s3_morphologies}

After the purely statistical analyses of the last three sections, we now examine the spectroscopically confirmed secure cluster member galaxies more closely. Although still limited to seven 
secure members, these galaxies represent a test sample to cross-check and confirm the results of the statistical analyses without the uncertainty introduced by the interloping foreground/background population.

Figure\,\ref{fig_PostageStamps} presents 6\arcsec$\times$\,6\arcsec \ postage-stamp cutouts of the secure member galaxies based on the NIR co-added J+Ks detection image. 
The corresponding physical projected dimensions of these 
cutouts are 50.8\,kpc$\times$50.8\,kpc, with a scale of individual pixels of 0.9\,kpc/pixel.  
The seeing-limited resolution scale of about 4.5\,kpc (0.52\arcsec) allows a first-order evaluation of the underlying morphologies.  
The spectroscopic members in the top row of Fig.\,\ref{fig_PostageStamps} are located within the 30\,\arcsec \,($\simeq$250) core region of the cluster with the highest densities, whereas members with IDs\,4-6 in the bottom row are located at projected radii of (0.5-1)\,R$_{200}$, as shown in Figs.\,\ref{fig_OpticalOverview_X0044}\,and\,\ref{fig_Spitzer_View} and Table\,\ref{tab_specmembers}.

The most luminous spectroscopic member of the current sample in the NIR  is ID\,1, which is also a 
strong candidate for the current brightest cluster galaxy (see Sect.\,\ref{s3_BCGobs}). This m*-1 galaxy is located about 70\,kpc in projection NE of the X-ray centroid position. Based on its strongly distorted NIR morphology (top-left panel of Fig.\,\ref{fig_PostageStamps}) in the N-S direction and two further extensions toward the South and SE, it is readily visually classified as system in a state of very dynamic mass-assembly activity with probably several ongoing and recent merger events. Although the color of the main component is bluer than many of the surrounding red galaxies (Figs.\,\ref{fig_OpticalOverview_X0044}\,and\,\ref{fig_Videt_CoreView}), this is the only spectroscopic member without any detected \OII \ line emission in the optical spectrum.  The two S/SE extensions exhibit even bluer colors than the main component.      

The second-most luminous spectroscopic member is the red galaxy with  ID\,3a with its spectroscopically confirmed smaller blue companion system with ID\,3b (see Sect.\,\ref{s2_Spectroscopy}). Based on the NIR morphology, this double system is not separable (top right panel of Fig.\,\ref{fig_PostageStamps}), which is clearly the case when adding the optical bands (see Fig.\,\ref{fig_Videt_CoreView}). Overall, the morphology of the main component seems less distorted than for ID\,1. However, two additional red galaxies are located within 2\arcsec \ in projection toward the North, which could imply imminent additional merging activity.

The core member galaxy with ID\,2 is about 1.5\,mags fainter than the brightest systems, but also shows signs of dynamical assembly activity in the NE-SW directions with another NE companion within  2\arcsec \ in projection.  
Bluer optical emission mainly originates from the SW and NW extensions.  

The three spectroscopic members with IDs\,4-6 in the outer regions of the cluster have NIR magnitudes in the range m*+0.5 to m*+1. Their NIR morphologies 
are compact and more regular than the core galaxies without any detectable companion within 2\arcsec. However, IDs\,5\,and\,6 
do exhibit inhomogeneous color distributions within the galaxies. 

In summary, the distorted NIR morphologies, color inhomogeneities, and close companions within  2\arcsec \ point toward very active mass assembly 
of the most central spectroscopic member galaxies, while the secure members in the cluster outskirts exhibit more regular and compact morphologies.

\subsection{Color-magnitude relation}
\label{s3_CMR}

\begin{figure}[t]
    \centering
    \includegraphics[width=\linewidth, clip]{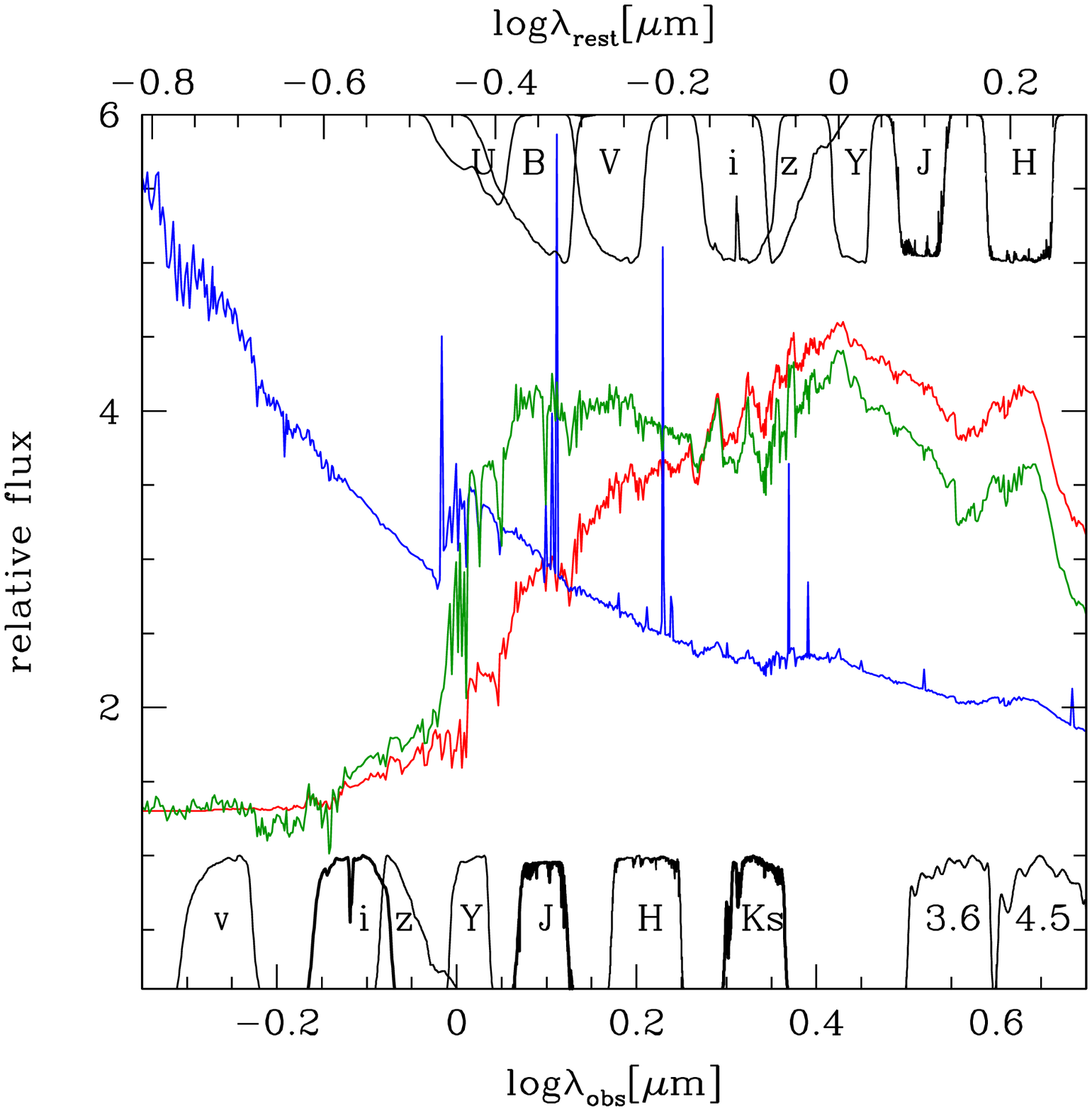}     
      \caption{SED template spectra for an old elliptical galaxy (red), a post-starburst 1\,Gyr after SF (green), and an active starburst (blue) redshifted to observed wavelengths at the cluster redshift of XDCP\,J0044.0-2033 (bottom axis) and in restframe units (top axis). The main filter transmission curves are plotted in black at the bottom of the panel with their names indicated. The corresponding restframe wavelength coverage is shown on top.} 
         \label{fig_TemplateFilters}
\end{figure}

We now proceed to add quantitative color information and investigate the distribution of the cluster galaxy population in the J$-$Ks and  i$-$Ks versus Ks  color-magnitude diagrams (CMDs).  Figure\,\ref{fig_TemplateFilters} provides a general overview of the observational setup for $z\!\simeq\!1.58$ cluster galaxies by illustrating 
redshifted template spectra based on Galaxy Evolutionary Synthesis Models (GALEV\footnote{ \url{http://www.galev.org}})  \citep{Kotulla2009a} and relevant optical and infrared filter bandpasses (bottom) with their cluster-restframe counterparts (top). The three 
template spectra depict the expected observed spectral energy distribution (SED) of an elliptical galaxy with an old passively evolving stellar population, a post-starburst template observed one gigayear after the star formation epoch, and the SED of a galaxy with ongoing starburst activity. 

At the cluster redshift of XDCP\,J0044.0-2033, the 4000\,\AA-break is shifted to the observed Y-band. The deep HAWK-I observations in J and Ks thus probe the SEDs of cluster members redward of the break, corresponding approximately to a restframe B$-$i color. With the available deep Subaru imaging data, we obtain an additional  measurement blueward of the 4000\,\AA-break by including the i-band, 
which probes the restframe ultraviolet part of the spectrum of cluster galaxies.

\begin{figure*}[t]
    \centering
    \includegraphics[width=0.74\linewidth, clip]{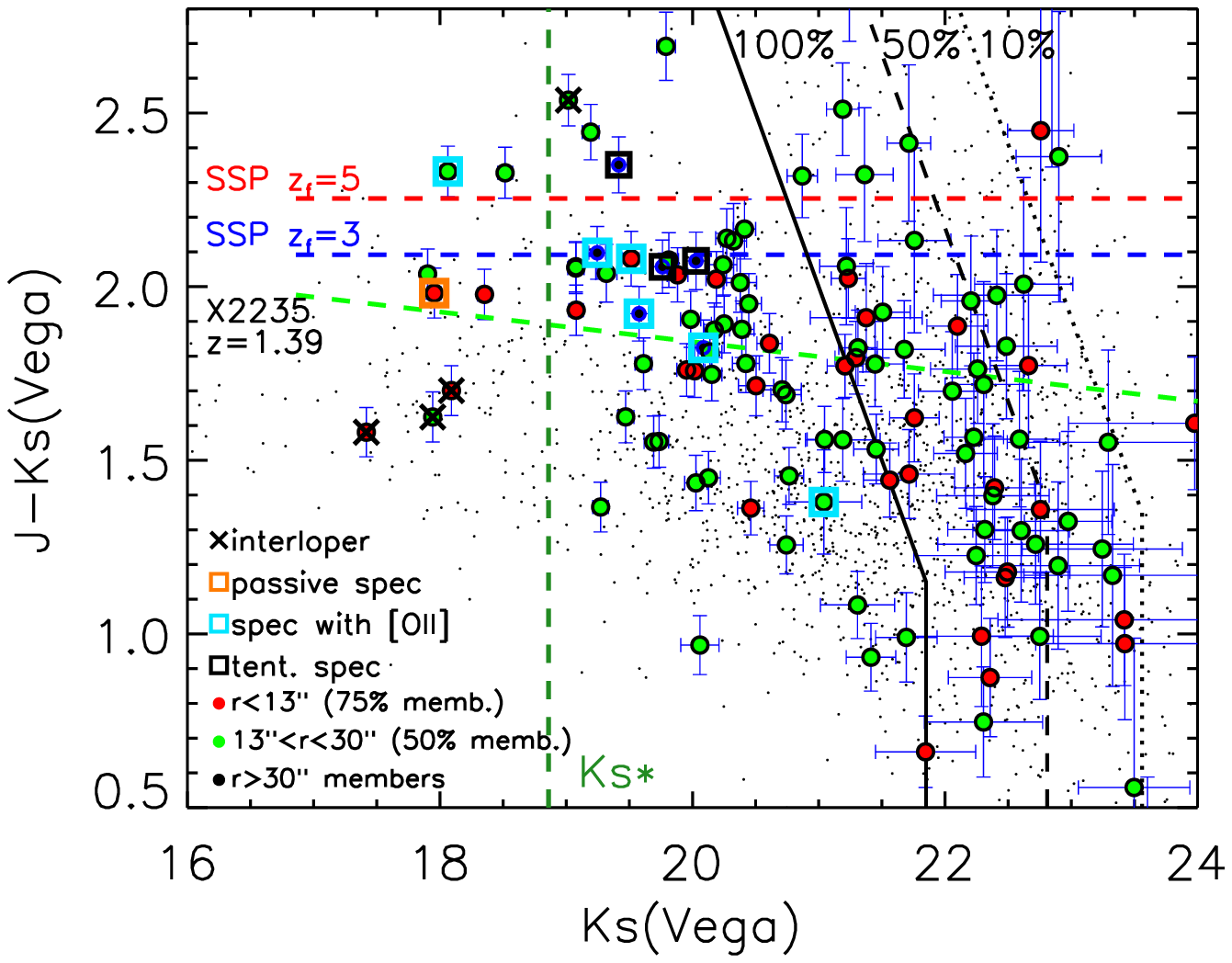}    
     \includegraphics[width=0.74\linewidth, clip]{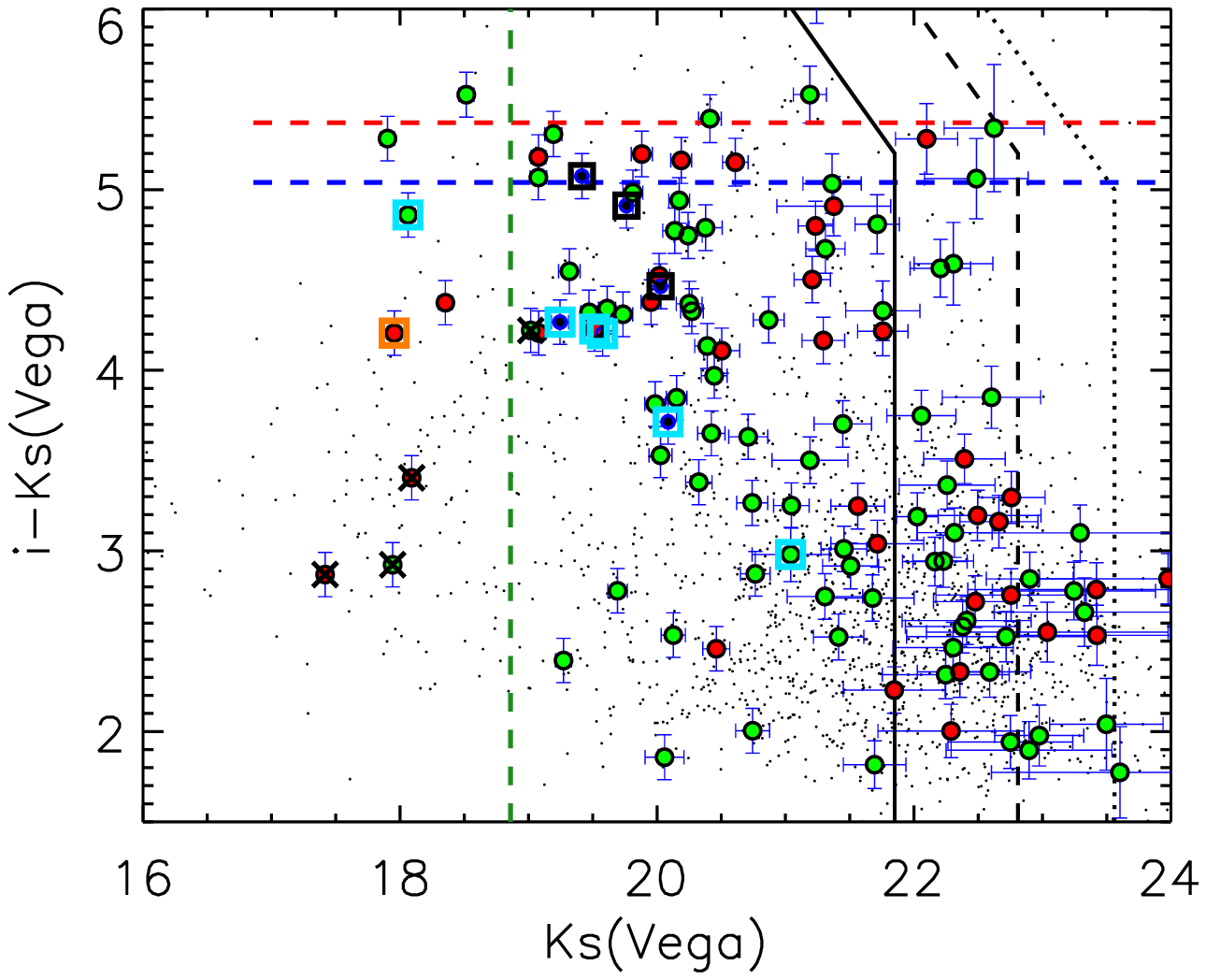}
      \caption{J$-$Ks versus Ks ({\it top}) and i$-$Ks versus Ks ({\it bottom}) CMDs of the inner region of XDCP\,J0044.0-2033 for galaxies at 
 cluster-centric distances $\le$13\arcsec \ (red circles) with a statistical cluster membership of 75\% and  objects within 13-30\arcsec \ (green circles) with a purity of 50\%. Spectroscopic (tentative) members outside 30\arcsec \ are indicated by black circles, small dots mark all other galaxies in the field. Secure (tentative) spectroscopic members are marked by orange/cyan (black) squares with cyan indicating the presence of an \OII-emission line. The 100\%/50\%/10\% completeness limits are shown by the solid/dashed/dotted black lines. The dashed green line in the top panel depicts the red-sequence location of XMMU\,J2235.3-2557 at $z\!=\!1.39$ from \citet{Lidman2008a} as  reference. Horizontal dashed lines mark solar metallicity SSP-model color predictions for formation redshifts of 3 (blue) and 5 (red), the vertical green line shows the corresponding characteristic Ks* magnitude. } 
         \label{fig_CMR} 
\end{figure*}

The top panel of Fig.\,\ref{fig_CMR} shows the NIR HAWK-I  J$-$Ks versus Ks color magnitude diagram for galaxies within 13\arcsec \ of 
the X-ray centroid, at cluster-centric distances of 13-30\arcsec and for all other galaxies at larger distances from the cluster center. The  two CMD analysis radii where chosen to provide a 75\% statistical cluster member purity  for the objects in the very core region based on the overdensity analysis in Sect.\,\ref{s3_Profiles} and a   50\% statistical purity for the galaxies. In total, the central 30\arcsec \ region contains about 120 NIR detected galaxies of which about 70 are statistically member galaxies of  XDCP\,J0044.0-2033. The secure spectroscopic members are indicated by cyan (with detected \OII-emission) and orange  squares (without  \OII-emission), while tentative spectroscopic members are marked by smaller black squares.

The NIR HAWK-I CMD is based on total Ks magnitudes and 1\arcsec-aperture J$-$Ks colors and represents the most accurate photometric data currently available for  XDCP\,J0044.0-2033. Black lines in  Fig.\,\ref{fig_CMR}  indicate the completeness limits of the 0.5\arcsec-seeing 
data as determined in Sect.\,\ref{s2_HawkI} and  the vertical green dashed line marks the position of the characteristic SSP model magnitude at the cluster redshift as discussed and observationally confirmed in Sect.\,\ref{s3_NIR_LF}. As visual reference, the horizontal lines depict the predicted colors of solar metallicity SSP models with  $z_{\mathrm{f}}$=3 and   $z_{\mathrm{f}}$=5, while the slanted light green line shows the best fit observed J$-$Ks  red-sequence location of the cluster  XDCP\,J2235.3-2557 at $z\!=\!1.396$ found by  \citet{Lidman2008a}.


A clear locus of  about 15 galaxies is evident in J$-$Ks color-space at the location of the color prediction of the SSP  $z_{\mathrm{f}}$=3 model in the magnitude range between the characteristic magnitude Ks* and Ks*+1.5. This locus would be classically called the cluster red sequence, which has been shown to be in place in massive clusters at least up to  $z\!\simeq\!1.4$ \citep[e.g.,][]{Strazzullo2010a} and is composed of predominantly passive galaxies with early-type morphologies. However, identifying a well-defined red sequence is 
not straightforward for XDCP\,J0044.0-2033 because the spectroscopic members in this locus show indications of active star formation based on their detected  \OII-emission and deformed morphologies due to active merging activity, as discussed in Sect.\,\ref{s3_morphologies}. Given the difficulties of separating passive early-type galaxies with high confidence  based on the currently limited spectroscopic coverage, we refrain from a quantitative red-sequence fit for the mean color, slope, and scatter at this point. However, the part of the red sequence that appears to be already in place at  $z\!\simeq\!1.58$ is fully consistent with the simple stellar-population 
prediction for a formation redshift $z_{\mathrm{f}}$=3 with a negligible slope. 

The situation for distinguishing red-sequence galaxies from non-red-sequence objects is even more difficult at the bright end of the galaxy population with magnitudes brighter than the characteristic magnitude Ks*.  The BCG candidate (ID\,1), which is also the 
only spectroscopic member without a clearly detected  \OII-emission line, is bluer than the discussed locus in J$-$Ks color-space and the  expectations from lower redshift clusters. The bluer color of this galaxy  and the other bright central galaxy (labeled c1) compared with the red-locus population is readily visible in the color-image representation of  Fig.\,\ref{fig_Videt_CoreView}. The central galaxy region of the spectroscopic member with ID\,3a, on the other hand, shows a  J$-$Ks color that is significantly redder than the red locus, although the spectrum exhibits detected   \OII-emission. The region in   J$-$Ks color-space between the SSP models for  $z_{\mathrm{f}}$=3 (blue) and  $z_{\mathrm{f}}$=5 (red), where one would naively expect the bright end of the red-sequence galaxy population to be located, is completely devoid of objects within the 30\arcsec \ analysis region from the X-ray centroid. Based on  this observation, we can conclude that the bright end of the red sequence is not yet in place for  XDCP\,J0044.0-2033  for galaxies brighter than the characteristic magnitude Ks*.

The underlying physical reasons for the significant color deviations of the bright cluster galaxies from simple model expectations are of prime interest within the overall galaxy evolution framework. In particular for galaxies with colors above the red-locus (e.g.,~ID\,3a), the detailed interplay of ongoing star formation activity and possible significant dust  attenuation \citep[e.g.,][]{Pierini2004a,Pierini2005a}
remains to be investigated in depth to derive firm conclusions. To this end, we initiated spectroscopic NIR integral-field observations with VLT/KMOS of bright cluster galaxies, which will be discussed in forthcoming publications (Fassbender et al., in prep.).

\begin{figure}[t]
    \centering
    \includegraphics[width=\linewidth, clip]{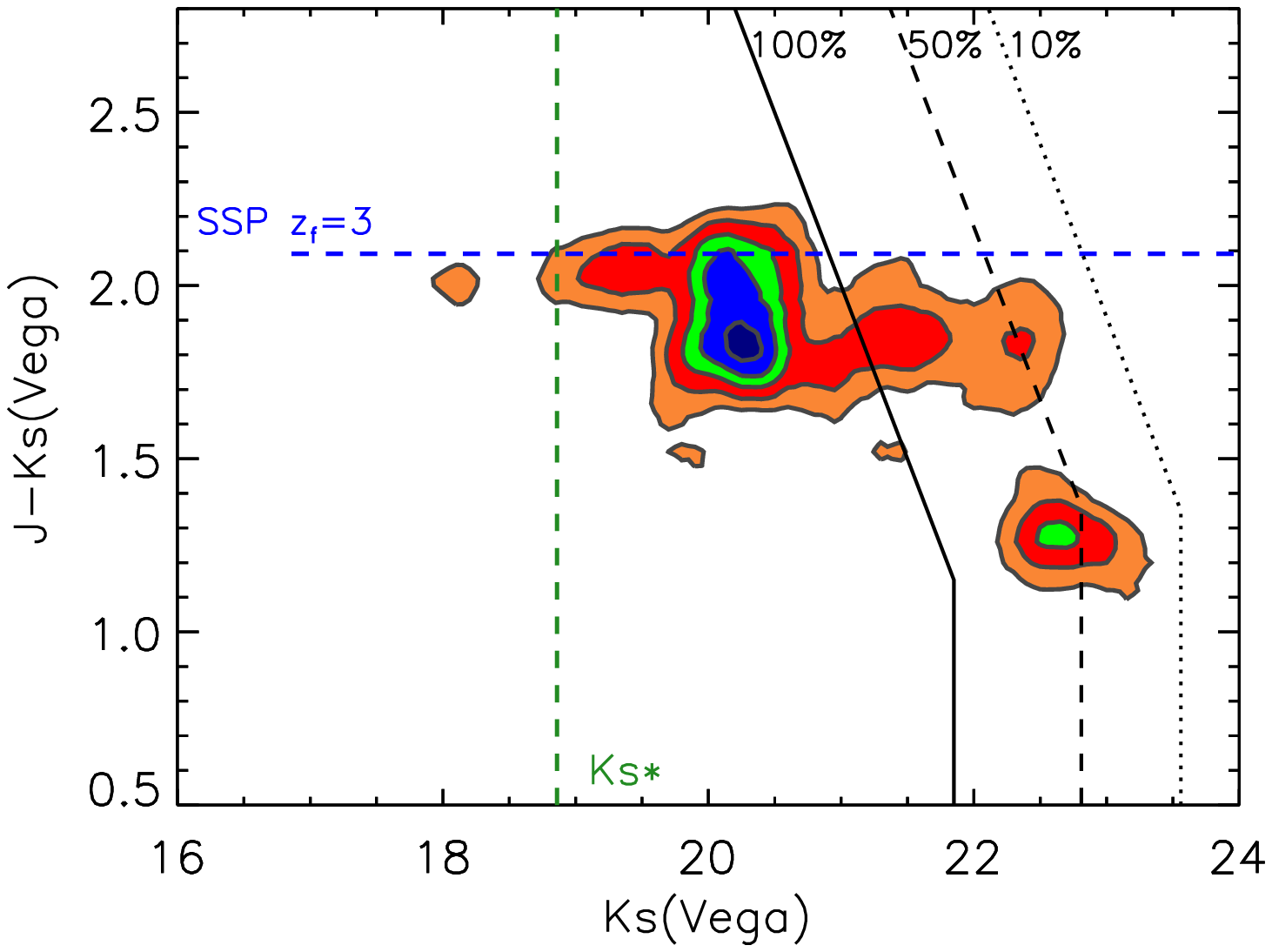}
    \includegraphics[width=\linewidth, clip]{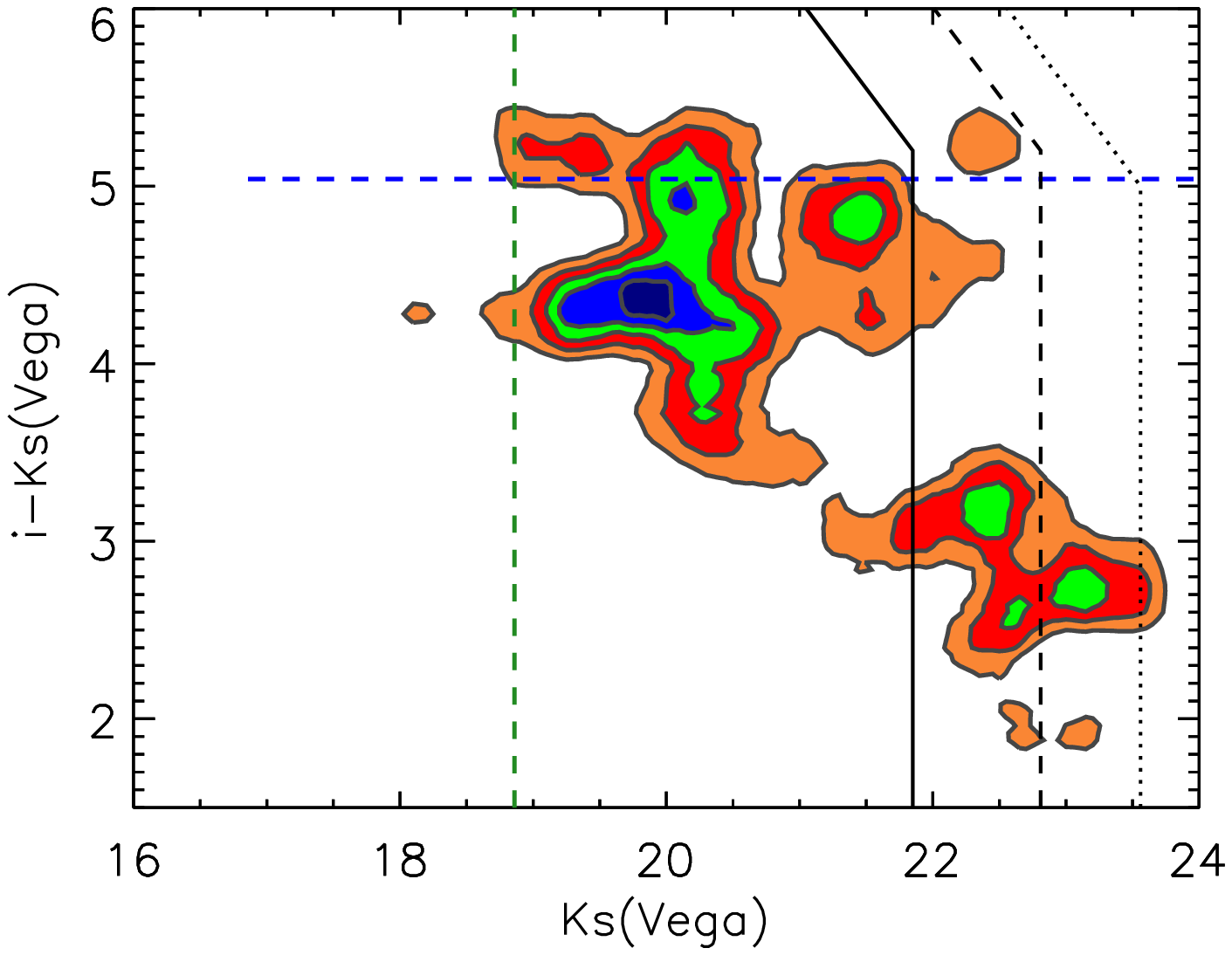}    
      \caption{Background-subtracted densities of cluster galaxies of XDCP\,J0044.0-2033  in the J$-$Ks versus Ks ({\it top})  and i$-$Ks versus Ks ({\it bottom}) color magnitude diagrams based on the data shown in Fig.\,\ref{fig_CMR}. The cluster region was constrained to a radius of 30\arcsec \ from the X-ray centroid. The contours and the corresponding filled regions correspond to detected cluster galaxy densities of 3 (orange), 5 (red), 7 (green), 9 (blue), 11 (dark blue) in units of  galaxies per arcmin$^{2}$ per mag in Ks per 0.2 mag in J$-$Ks for the top panel and densities of 2.3 (orange), 3.6 (red), 4.9 (green), 6.2 (blue), 7.5 (dark blue)  galaxies per arcmin$^{2}$ per mag in Ks per 0.4 mag in i$-$Ks for the bottom panel.
The sharp truncation of the cluster red-sequence at Ks$\sim$20.5 and the significant population of faint blue galaxies in the lower right corner are clearly visible. Lines have the same meaning as in Fig.\,\ref{fig_CMR}.
      } 
         \label{fig_CMRdensity} 
\end{figure}

The situation at the faint end of the red-locus population is equally interesting and is discussed in detail in 
Sect.\,\ref{s3_FaintEnd_RS}. For now, we examine the bulk of the galaxies within the  30\arcsec \ analysis region with colors significantly bluer than the red-locus population. In these bluer and fainter parts of the CMD the density of the foreground/background populations increases as well, therefore it is not immediately evident in which regions of color-magnitude-space the bulk of the cluster galaxies are located. To overcome this problem and separate the 
cluster contribution we created a background-subtracted version of the CMD in Fig.\,\ref{fig_CMRdensity} (top panel) that illustrates the net cluster galaxy density contours with 
levels of 3-11 net cluster galaxies per square arcmin per magnitude in Ks per 0.2 mag in  J$-$Ks color.

These background-subtracted density contours in the top panel of Fig.\,\ref{fig_CMRdensity}   show  that at least three more components of the cluster galaxy population can be distinguished in addition to the already discussed red-locus population (i): (ii) a
`red-sequence transition population'  marking the highest density region in the J$-$Ks CMD,  (iii) a tail extending to fainter Ks magnitudes with a quasi-constant color offset to the red locus, and (iv) a population of faint and very blue cluster galaxies in the lower right part of the CMD. 


The lower panels of Figs.\,\ref{fig_CMR}\,and\,\ref{fig_CMRdensity} show the analogous color-magnitude and density diagrams for the i$-$Ks color, which now brackets the 4000\,\AA-break and therefore
is more sensitive to ongoing star formation activity. Owing to the 40\% increased seeing  in the i band compared to Ks, the absolute seeing-matched  i$-$Ks  photometric color uncertainties   are correspondingly larger than for J$-$Ks, which is partially compensated for by the wider spread along the color axis due to the additional additive  i$-$J contribution to the total  i$-$Ks color.

As expected, all spectroscopic members with detected   \OII-emission are now shifted down in  i$-$Ks farther away from the expected color of passive SSP model galaxies.  In effect, the locus of red galaxies in  J$-$Ks is now split into two branches, one in the region of the SSP model expectations, and a second one almost one magnitude bluer. 
This bluer second branch includes three spectroscopic members with detected   \OII  \ plus the  BCG candidate galaxy, as clearly visible in the density plot of Fig.\,\ref{fig_CMRdensity} (lower panel). This means that the locus of red galaxies consistent with old passive stellar populations is more sparsely populated than the  J$-$Ks color, which is less sensitive to residual star formation. 
The  cluster galaxy population of faint blue objects in the lower right corner
is again clearly visible in the i-Ks CMD; it is  now distributed over a wider color spread, as are the other components identified in the J$-$Ks. These different components of the cluster galaxy population are investigated in depth in 
color-color space in Sects.\,\ref{s3_color_color}\,and \ref{s3_SpatialDistribution}, following the examination of the faint end of  red-locus galaxies in the next section.

\enlargethispage{2ex}
\subsection{Faint end of the red sequence}
\label{s3_FaintEnd_RS}




To study  the faint end of the red sequence, we focus on the  J$-$Ks versus Ks CMD to minimize the potential budget of systematics at faint magnitudes due to,
for instance, the different instruments, PSF matching, and the sparser population of the red galaxy locus. However, the general conclusion is independent of the color used and is also evident in i$-$Ks.

\begin{figure}[t]
    \centering
    \begin{overpic}[width=\linewidth, clip]{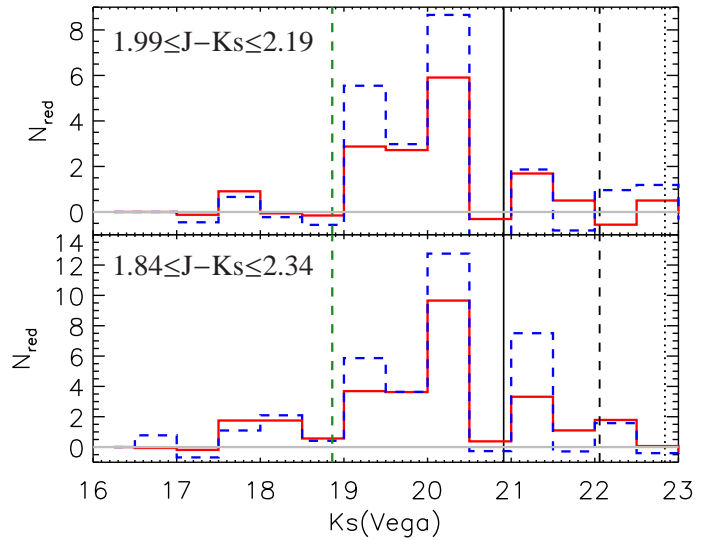}
\put(15,72){\large 1.99$\le$J$-$Ks$\le$2.19}
\put(15,39){\large 1.84$\le$J$-$Ks$\le$2.34}
  \end{overpic}     
      \caption{Background-subtracted cluster counts of red galaxies in J$-$Ks about the color of the  SSP $z_f\!=\!3$ model with a J$-$Ks color widths of $\pm$0.1\,mag (top panel) and  $\pm$0.25\,mag (bottom panel) for the R$\le$30\arcsec \ core region (red solid lines) and the full projected region within R$_{500}$ (blue dashed line)
The characteristic magnitude Ks* at the cluster redshift is indicated by the green dashed line on the left, the solid/dashed/dotted vertical lines on the right mark the 100\%/50\%/10\% completeness levels for the selected color interval.
The sharp cut-off in the red galaxy counts at K$\simeq$20.5\,mag ($\simeq$Ks*+1.6) is clearly visible for all histograms.
      } 
         \label{fig_RS_counts} 
\end{figure}

Figure\,\ref{fig_RS_counts} shows the observed histogram of the number of background-subtracted red galaxies as a function of Ks magnitude for a narrow ($\pm$0.1\,mag) selected color interval about the SSP   $z_{\mathrm{f}}$=3  model color (top panel) and a wider ($\pm$0.25\,mag) color selection (bottom panel). The red lines show the red galaxy counts for the CMD analysis region within    
30\arcsec, while the blue dashed histogram illustrates the integrated net counts out to the R$_{500}$ radius. In all cases, independent of color cut and selection aperture, the red galaxy population exhibits a very sharp truncation at Ks$\simeq$20.5\,mag or about Ks*+1.6 after reaching the maximum density in the magnitude interval 20$\le$Ks$\le$20.5.
This truncation magnitude of   Ks$\simeq$20.5\,mag is about half a magnitude brighter than the 100\% completeness limit and even 1.5\,mag away from the 50\% limit of the HAWK-I data.

\newpage

\clearpage

There are several potential systematic effects that might bias the conclusion of a sharp truncation of the red-locus cluster population at  Ks$\ga$20.5\,mag. First, the determined limiting depth of the data might be overestimated so that the apparent suppression at faint magnitudes might be due to incompleteness. That this is not the case was demonstrated with the galaxy number counts in Sect.\,\ref{s2_HawkI}. 
Moreover, this can be 
visually checked by confirming the clear detections of red-locus galaxies in Fig.\,\ref{fig_Videt_CoreView} (top) at the truncation limit (label `rs1', Ks$\simeq$20.4) and beyond  (label `rs2', Ks$\simeq$21.2). Second, increasing color errors at faint magnitudes might depopulate the red locus by scattering objects away from the true underlying color. While this is an unavoidable observational bias, we showed in the lower panel of  Fig\,\ref{fig_RS_counts} that a wider color-cut does not change the overall picture. And third, one could invoke that faint red-locus galaxies possibly
have larger effective radii than other cluster galaxies and might
therefore fall below the surface brightness limit for a detection. However, such a scenario is neither supported by the compact nature  of the detected red-locus galaxies (e.g.,~IDs\,4\,\&\,5 in Fig.\,\ref{fig_PostageStamps} or Fig.\,\ref{fig_Videt_CoreView}) nor by other observational high-$z$ studies of passive cluster galaxies  \citep[e.g.,][]{Strazzullo2013a}.
In summary, none of the potential systematic biases are likely to influence the apparent lack of red-locus galaxies in the Ks-magnitude interval of interest [20.5,21.5] in a significant way. 

We conclude that the population of red-locus galaxies of XDCP\,J0044.0-2033 is strongly suppressed beyond the surprisingly sharp truncation magnitude of Ks$\simeq$20.5\,mag. This cut-off magnitude in observed red galaxy counts is directly related to the  turn-over point of the quiescent galaxy luminosity function, as found previously in a high-$z$ group environment at similar redshift \citep{Rudnick2012a}. 




\subsection{Cluster galaxies in color-color space}
\label{s3_color_color}


\begin{figure}[t]
    \centering
    \includegraphics[width=\linewidth, clip]{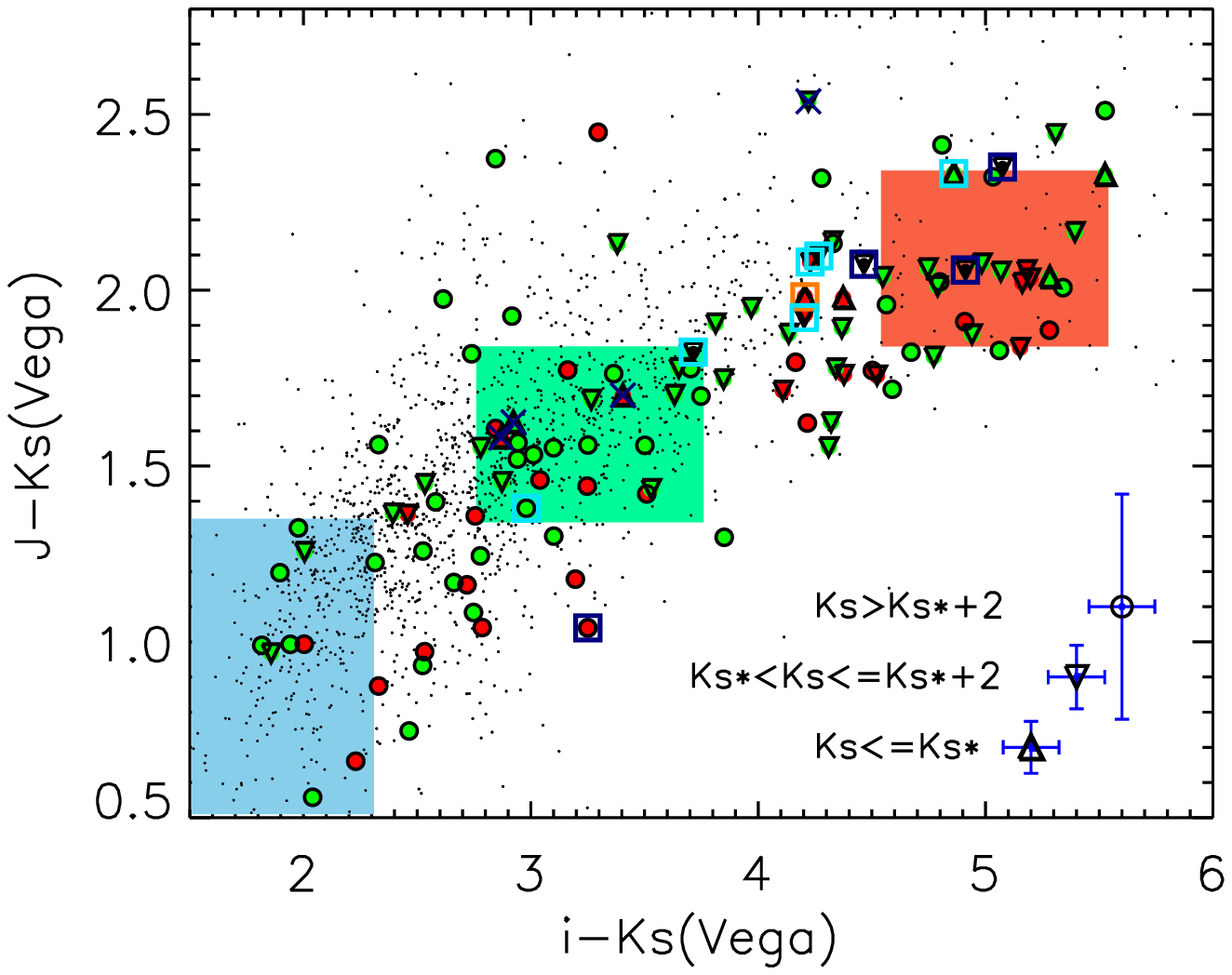}
 \begin{overpic}[width=0.99\linewidth, clip]{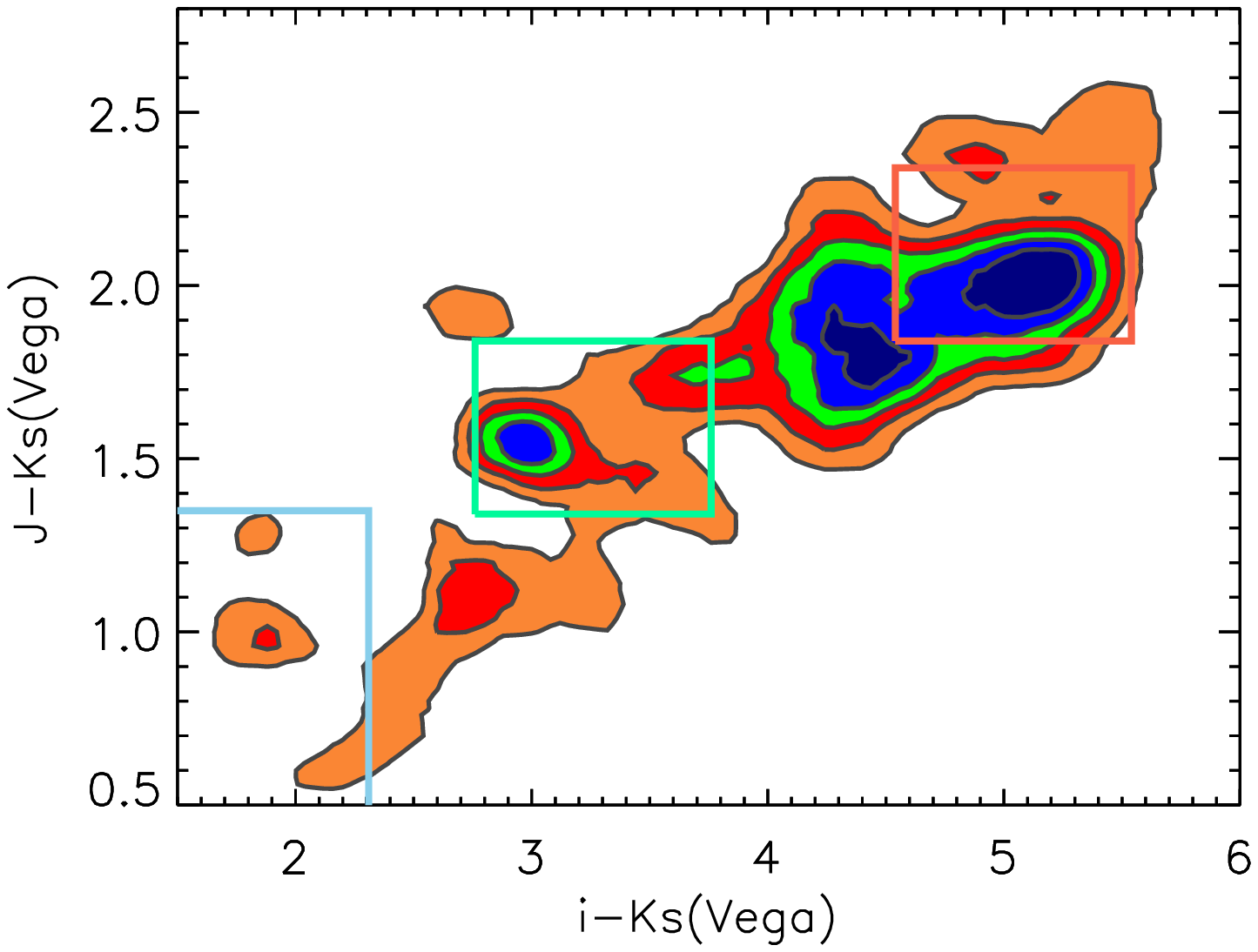} 
 \put(81,54.5){\Large \textcolor{white}{1}}
 \put(64.5,47){\Large \textcolor{white}{2}}
 \put(43.5,39){\Large \textcolor{white}{3}}
  \put(21.5,16){\Large \textcolor{black}{4}}
 \end{overpic}  
      \caption{{\it Top panel}: Observed J$-$Ks versus i$-$Ks color-color-diagram of the galaxies shown in Fig.\,\ref{fig_CMR} with the same color coding. The Ks magnitude of the galaxies is indicated by the outer shape:
 triangles for galaxies brighter than Ks* at $z\!=\!1.58$ , inverted triangles for magnitudes between Ks* and Ks*+2 , and circles for galaxies fainter than Ks*+2. 
       The median color errors for the three magnitude bins are shown in the legend in the lower right corner. The red, green, and blue boxes show  the  immediate color-color environment of galaxies with a SED consistent with 
  the template spectra at the cluster redshift  (Fig.\,\ref{fig_TemplateFilters}) for a passive galaxy (red), a post-starburst template (green), and active starbursts (blue).
{\it Bottom panel}: Background-subtracted cluster galaxy densities in color-color space with density levels of
1 (orange), 2.2 (red), 3.4 (green), 4.6 (blue), 5.8 (dark blue) core cluster galaxies   per arcmin$^{2}$ per 0.4 mag in i$-$Ks per 0.2 mag in J$-$Ks. Numbers 1-4 mark the central positions for the color-color selection of different galaxy types  with the intervals as specified in the text, while the boxes show the template colors as above: 1: passive, 2: transition objects, 3: post-starburst, 4: starburst/starforming.
      } 
         \label{fig_color_color} 
\end{figure}

%
%
%

The availability of deep imaging data in the i, J, and Ks band  allows us to proceed beyond the color-magnitude relation and study cluster galaxies in J$-$Ks versus i$-$Ks color-color-space to improve 
the distinction of different object types. 
Figure\,\ref{fig_color_color} shows 
the color-color space distribution of all detected galaxies with magnitudes 16$\le$Ks$\le$24\,mag and colored symbols as in the CMDs of Fig.\,\ref{fig_CMR},  with the additional information of 
the Ks-magnitude range 
indicated by the symbol shape. 
As visual reference, the colored boxes mark the color-color regions centered on 
the positions of the template spectra discussed in Fig.\,\ref{fig_TemplateFilters}: red for the passive template, green for the post-starbust galaxy, and blue for the starbust SED.   

The lower panel of  Fig.\,\ref{fig_color_color}  illustrates the   color-color net cluster  density contours after background subtraction. 
In this representation, the main cluster galaxy population components of the 30\arcsec \ core region become evident as three very significant high-density 
 regions   plus extensions into the starburst regime in the lower left corner.
To more closely investigate the spatial distribution of the different concentrations in color-color space, we defined color-color cuts to distinguish four different galaxy types\footnote{ This is a purely phenomenological classification based on color alone with a naming inspired by the most simple galaxy templates. Contributions of more complex galaxy types (e.g.,~dust reddened) can be expected, motivating further spectroscopic investigations.} that represent the four components  discussed qualitatively in Sect.\,\ref{s3_CMR}: (1) the red-locus population with colors consistent with a passive SED  (1.84$\le$J$-$Ks$\le$2.6 \& 4.7$\le$i$-$Ks$\le$5.6), 
(2) the `red-sequence transition population' (1.5$\le$J$-$Ks$\le$2.2 \& 3.6$\le$i$-$Ks$\le$4.6), (3) the post-starburst population 
(1.4$\le$J$-$Ks$\le$1.8 \& 2.7$\le$i$-$Ks$<$3.6), and (4) the very blue star-forming/starburst population (0$\le$J$-$Ks$<$1.4 \& 1$\le$i$-$Ks$<$2.9). The centers of these intervals for the color-color selection are marked by the corresponding numbers in the lower panel. 
For simplicity, we use a general terminology of the color-color selected galaxy classes based on the best-matching SED templates of Fig.\,\ref{fig_TemplateFilters}, which may contain contributions of other galaxy types, for example,~dusty starbursts.


%

\subsection{Spatial distribution of color-selected galaxies}
\label{s3_SpatialDistribution}

\begin{figure*}[t]
    \centering
    \includegraphics[width=0.495\linewidth, clip]{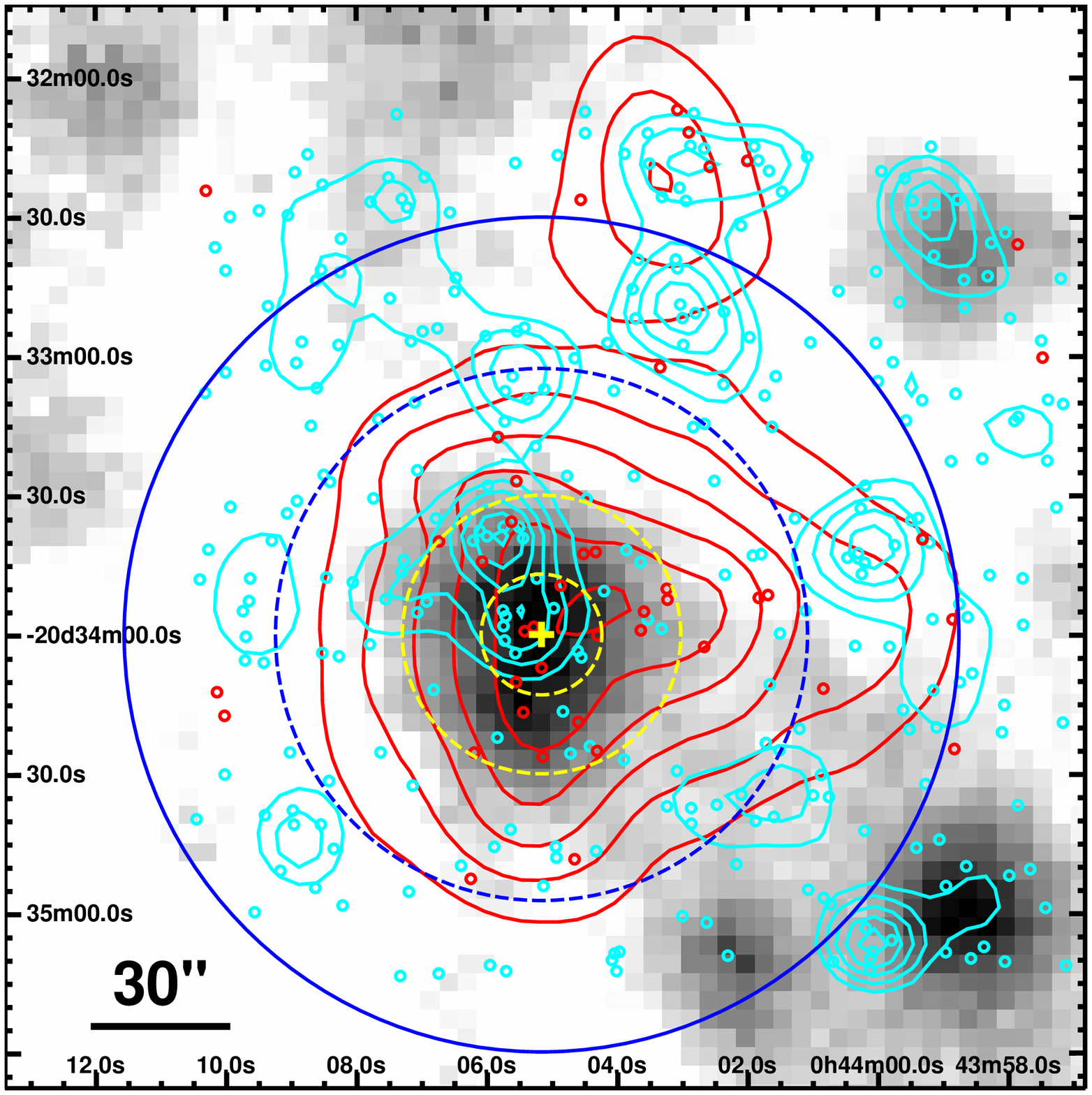}    
    \includegraphics[width=0.495\linewidth, clip]{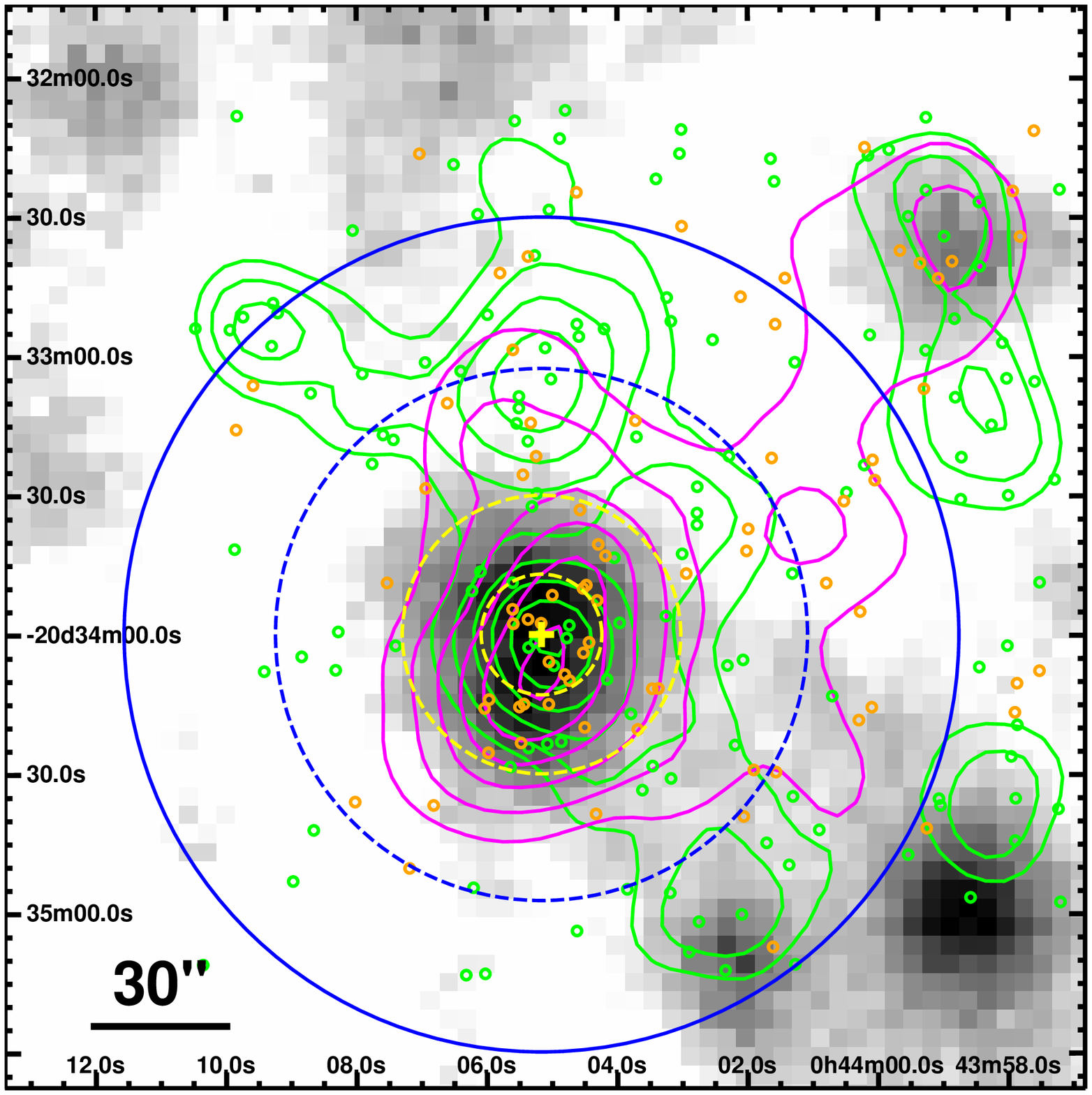} 
    \includegraphics[width=0.495\linewidth, clip]{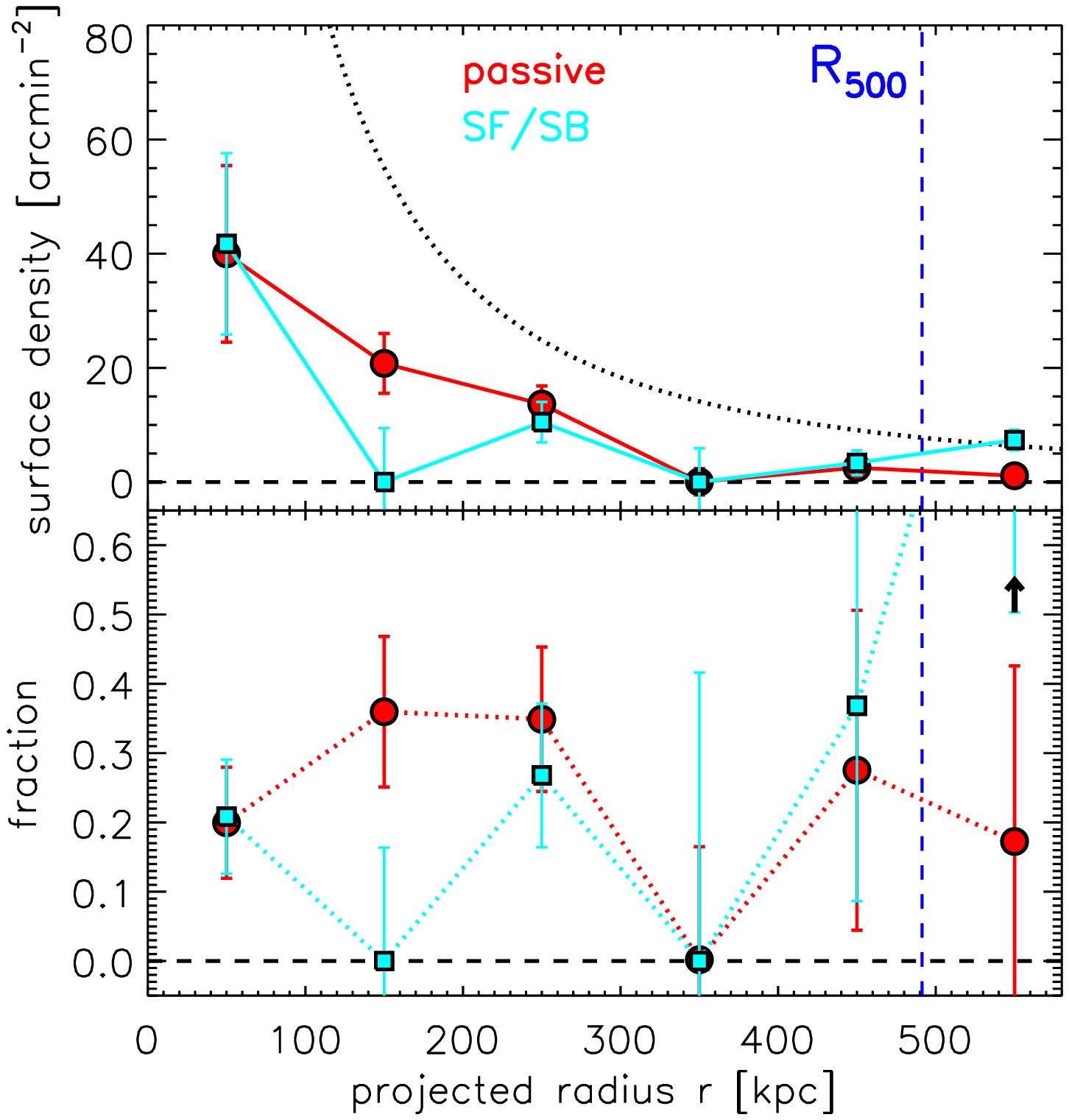}    
    \includegraphics[width=0.495\linewidth, clip]{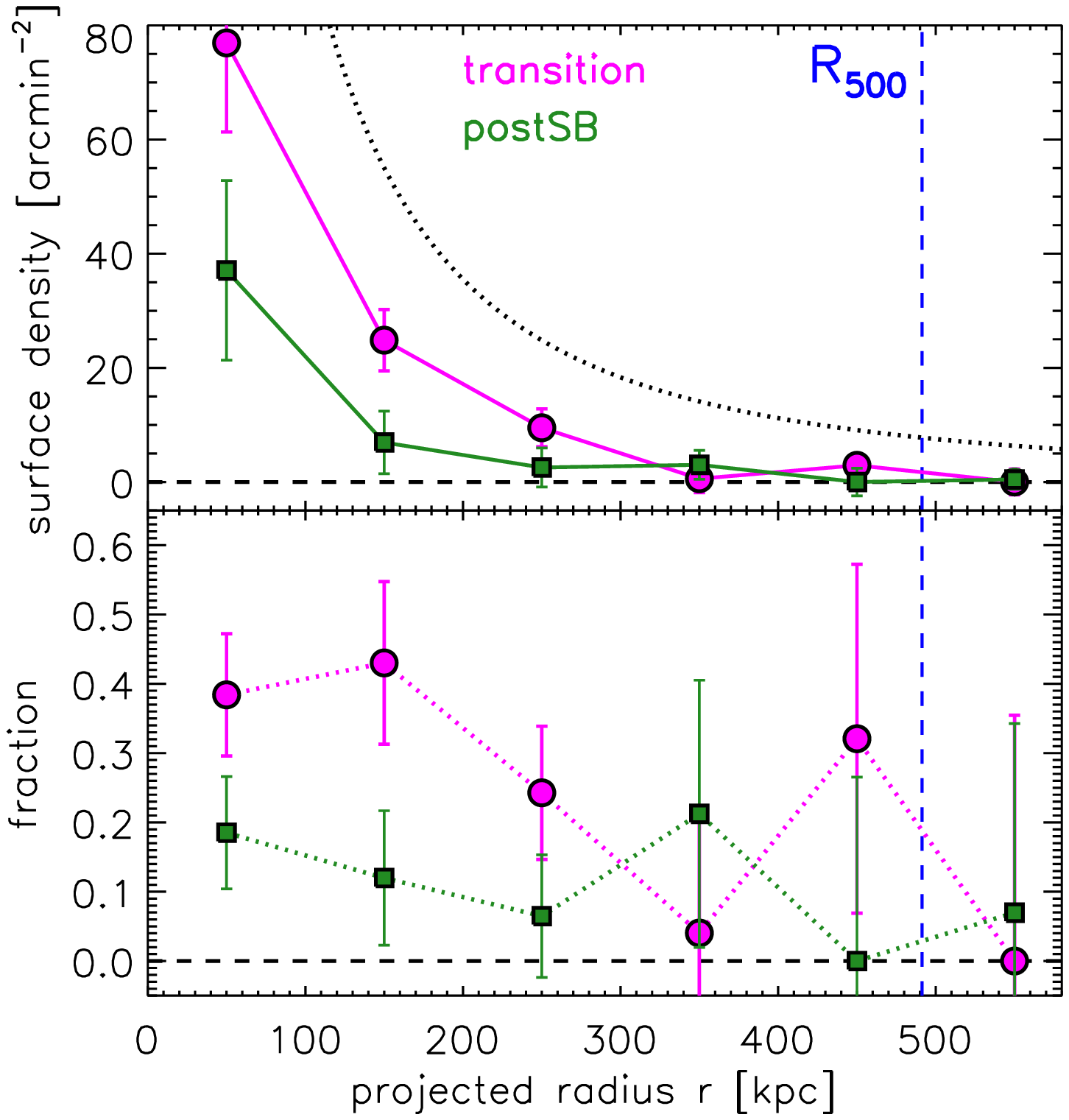}     
      \caption{{\it Top panels}: spatial-excess 
      distributions of color-color-selected galaxies overlaid on the gray-scale XMM-{\it Newton} X-ray image (3.9\arcmin$\times$3.9\arcmin). The 
  blue circles indicate R$_{200}$ (solid) and 
R$_{500}$ (dashed), and the 
dashed yellow circles mark the 13\arcsec/30\arcsec \ analysis radii about the X-ray centroid position (yellow cross).
{\it Top-left panel}: logarithmically spaced density contours (red) of galaxies  with colors consistent with a passive SSP SED (red circles) in J$-$Ks versus i$-$Ks color space and for objects with colors similar to the starburst template (cyan contours and circles). {\it Top-right panel}: same field showing the spatial distribution of red-sequence transition galaxies (magenta contours and orange circles) and galaxies with post-starburst colors (green contours and circles). 
{\it Bottom panels}: corresponding azimuthally averaged and background-subtracted surface-density profiles (upper half) and galaxy  fractions (bottom half) as function of projected cluster-centric radius for the same four color-color selected galaxy types as above. The black dotted line indicates the total best-fit galaxy surface-density profile as derived in Fig.\,\ref{fig_NFW_SD} as reference.
The black upward arrow in the outermost bin of the starburst category (cyan) in the lower left panel marks the 1$\sigma$ lower limit for this class. } 
         \label{fig_Spatial_GalOD}
\end{figure*}

With the definition of these dominant four subclasses of the distant cluster galaxy population, we can now investigate their spatial distribution within the cluster region and beyond in more detail. 
The top panels of Fig.\,\ref{fig_Spatial_GalOD} show the spatial locations of the four different color-color selected galaxy types, indicated by small colored circles, in the cluster region environment of  XDCP\,J0044.0-2033 with the observed  XMM-{\it Newton} X-ray emission used as gray-scale background image.  Logarithmically spaced surface density contours were derived based on the smoothed spatial distribution of objects for each class, which are shown with the corresponding colors.   
The contour levels 
reflect the full dynamic range of the overdensities above the background levels of the individual classes, which are increasing significantly from the red-locus category (1) with negligible background to the blue starburst category (4) with large foreground/background contributions (see top panel of Fig.\,\ref{fig_color_color}).

The surface density contours of the red-locus galaxy population span significance levels of 10-120\,$\sigma$ above the background and are located mostly within the cluster 
R$_{500}$ region, in general correspondence to the 
extended X-ray emission. The star-forming/starbust galaxy population (cyan,  2.5-10\,$\sigma$) 
exhibits a 
5\,$\sigma$ overdensity in the central 100\,kpc core region and appears to be mostly concentrated in clumps farther outside, in particular, at the  
R$_{500}$ radius and beyond. The categories of  red-sequence transition objects (magenta,  4-50\,$\sigma$) and the post-starburst population (green,  2-15\,$\sigma$) in the upper right panel exhibit a 
similar spatial distribution with a compact and sharply peaked core coincident with the regions of maximum X-ray emission.  


The lower panels of Fig.\,\ref{fig_Spatial_GalOD} 
show the azimuthally averaged, background-subtracted radial surface-density profiles of the four color-color selected galaxy categories (top) and the galaxy fractions with respect to the total net cluster population (bottom). At radii beyond R$_{500}$ (rightmost bin) the profiles are consistent with being purely star-forming/starburst  dominated within the given significant uncertainties attributed to the outer low-surface density regions.   
At cluster-centric radii of $\la$300\,kpc, the environmental effects of the high-density cluster regions becomes evident in the 
distribution 
of the overall galaxy population. Within this radius, the surface density of post-starburst systems (category 3), transition objects (2), and red-locus galaxies (1)  rise continuously toward the center, most sharply for the red-sequence transition category (magenta), which dominates the cluster population at r$\la$200\,kpc with a net galaxy fraction of 
$\sim$40\%. Post-starburst galaxies are found out to 
400\,kpc with a slowly varying  fraction of 10-25\%. The fraction of galaxies of the red-locus category (1) reach their maximum of 
 $\sim$35\% in the 100-300\,kpc region, before dropping again to 
 $\sim$20\% in the central 100\,kpc core region, in stark contrast to low-redshift systems. In the 
 very core region, the contribution of star-forming/starburst galaxies with a fraction of about 20\% is similar in number to the red-locus population.  
The lower panel of Fig.\,\ref{fig_Videt_CoreView} shows a V+i+JKs color image of the inner cluster region out to projected distances of $\simeq$300\,kpc with all color-color selected galaxies of the four categories overlaid. 

While the azimuthally averaged profiles of the different galaxy types in the lower panels of Fig.\,\ref{fig_Spatial_GalOD} are robust in a 
statistical sense out to the displayed projected radial distance (r$<$600\,kpc), the individual clumps beyond the R$_{500}$ radius (dashed circle) suggested by overdensities in the upper panels may not all be part of the surrounding cluster environment. Owing to the increasing foreground/background dominance at these radii, in particular for the bluer galaxy types, some contamination by unrelated foreground structures is to be expected. The suggested main matter-accretion axis onto the cluster coming in from the northern direction as indicated by overdensity contours of all galaxy types, however, is spectroscopically backed up by three secure member galaxies. The currently sparse spectroscopic sampling along the NE to SW axis does not allow any definitive statement on the cluster association of potentially infalling individual group structures  beyond R$_{500}$. 
At this point, only the overdensity in the NW (upper right) corner of the FoV featuring weak extended X-ray emission is a spectroscopically confirmed foreground group at  $z_{\mathrm{group}}\!=\!0.721$, while all other possible substructures have yet to be spectroscopically identified.      

\subsection{Cluster core and formation of the brightest cluster galaxy}
\label{s3_BCGobs}

As a final application, we 
focus on the central 100\,kpc  core region of  XDCP\,J0044.0-2033 
to investigate the current situation and probable future evolution of the formation of the brightest cluster galaxy.  As mentioned earlier, we consider the most central secure spectroscopic member with ID\,1 (see Table\,\ref{tab_specmembers} and Fig.\,\ref{fig_Videt_CoreView}) as the current BCG. This system is a Ks*-1 galaxy located at a projected distance 
 from the X-ray centroid of 70\,kpc with a low 
radial restframe velocity offset from 
 the systemic cluster redshift of $\la$100\,km/s and no signs of \OII-emission in its optical spectrum. Because its
 color is bluer than that of the red-locus population, it is classified as a red-sequence transition object 
 observed at an epoch of 
 strong merging activity with a number of satellite galaxies, as discussed in Sect.\,\ref{s3_morphologies} and shown in Figs.\,\ref{fig_PostageStamps}\,and
\ref{fig_Videt_CoreView}.

The cluster-centric offset, merging activity, and fairly blue color are in stark contrast to the more massive dominant central BCG galaxies in other bona fide distant clusters observed at 
lower redshifts $z\!\simeq\!1.2$-1.4 \citep[e.g.,][]{Collins2009a,Rettura2010a,Strazzullo2010a}, that is,~at cosmic times 0.5-1.5\,Gyr later than XDCP\,J0044.0-2033.
However, this difference in lookback time is longer than the short dynamical friction timescales  of 
$\la$100\,Myrs for ID\,1  and other massive galaxies in the 100\,kpc core region 
 as a result of the dynamic interaction with the underlying dark matter halo of the cluster \citep[e.g.,][]{Hausman1978a,Binney1987a}. 
 This implies
that a very significant amount of evolution is expected to occur on these timescales with respect to  
the  BCG 
mass assembly, structural and color evolution, cluster-centric offset position, and the general reshaping of the densest cluster core  region.

The 100\,kpc core region already contains an enclosed total stellar mass of about $M*_{100\,\mathrm{kpc}}\!\simeq\!1.5\!\times\!10^{12}\,\mathrm{M_{\sun}}$ 
(Sect.\,\ref{s3_MIRproperties}) and a net cluster excess of about two dozen galaxies (Sect.\,\ref{s3_Profiles}) of all color-color-selected galaxy categories (Sect.\,\ref{s3_SpatialDistribution}). All general ingredients for the build-up of a classical 
dominant and very massive brightest cluster galaxy are thus already present in the cluster core and are awaiting to be remixed and reshaped within 
the short dynamical and galaxy evolution timescales in this ultradense region.  
This process of {\it galactic cannibalism} \citep{Hausman1978a} seems to be observed at work here and can be expected to induce a rapid evolution of the BCG properties and its structural appearance.


On timescales of gigayears, the massive cluster galaxies at projected offsets of 150-200\,kpc might also 
end up in the 
cluster core through dynamical friction and be relevant for the late-time shaping and further evolution of the central brightest galaxy at  $z\!\la\!1$. This includes the spectroscopic member ID\,3a at a current projected offset of 145\,kpc and the equally bright, but still spectroscopically unconfirmed, candidate cluster galaxy (labeled c2) 
located at a projected cluster-centric distance of 180\,kpc to the northwest of the core. Both these galaxies are redder than ID\,1 and appear to be the central galaxies of an environment rich in satellite galaxies, meaning that~they are possibly part of accreted groups.
Both galaxies exhibit a total Ks magnitude within 0.1\,mag of the ID\,1 luminosity, implying that the current magnitude gap $\Delta m_{12}$ between the first- and second-ranked galaxy is negligible. To establish the typical observed luminosity gap of
$\Delta m_{12,{\mathrm{med}}}\!\simeq\!0.7$\,mag  \citep{Smith2010a} between the dominant BCGs and the second-ranked galaxies in low-redshift massive clusters, one could either invoke the discussed expected fast stellar mass growth of the most massive object  (ID\,1) in the core and/or the coalescence with at least one of the other two bright systems at  $z\!\la\!1$.




\section{Global view of XDCP\,J0044.0-2033}
\label{c4_GlobalView}

In this section, we summarize the current knowledge on XDCP\,J0044.0-2033 with the aim to tie together the 
results of Sect.\,\ref{c3_Results} and  provide a global characterization of the cluster and its immediate large-scale structure environment.  


\subsection{Dynamical state of XDCP\,J0044.0-2033}
\label{s4_DynamicalState}

Based on the currently available data from the X-ray to the mid-infrared regime \citep[][this work]{Santos2011a,Fassbender2011c},  XDCP\,J0044.0-2033 can be established to be in a quite advanced state of dynamical relaxation, considering that the age of the Universe at the time of observation is only about 4\,Gyr. First, the system's extended X-ray emission detected with XMM-{\it Newton} (Fig.\,\ref{fig_OpticalOverview_X0044}) is compact and to first order symmetrical, with only a slight elongation along the North-South axis. Second, the redshift distribution of the 
spectroscopically confirmed 
member galaxies 
do not show outliers in velocity space that would indicate dynamical assembly activity along the line of sight.  And third, the spatial distribution of the overall galaxy population of  XDCP\,J0044.0-2033 is also compact and highly peaked, with a galaxy density maximum in good spatial concordance with the X-ray centroid position (Sect.\,\ref{s3_Profiles}).

Based on this evidence, we can rule out ongoing cluster-scale major-merger activity,  although 
mass-accretion activity with group-scale substructures are still viable at the current resolution of the data.   
A high degree of cluster virialization implies 
that 
the X-ray luminosity-based mass estimate derived from the
M-L$_X$ scaling relation \citep{Reichert2011a}  
provides a robust total mass estimate, establishing XDCP\,J0044.0-2033 as a massive bona fide galaxy cluster at $z\!=\!1.58$.   
Consequently, the inferred total stellar fraction (Sect.\,\ref{s3_MIRproperties}) of  f$_{*,500} =(3.3\pm1.4)$\% can also be considered as robust within the specified uncertainties.

Owing to its selection as an extended X-ray source \citep{Fassbender2011c},  that is,~independent of its optical/IR properties, XDCP\,J0044.0-2033 provides an unbiased view of the overall galaxy population of a massive bona fide cluster at a lookback time of 9.5\,Gyr. By extrapolating the predicted cluster mass growth histories forward in cosmic time \citep[e.g.,][]{Fakhouri2010a}, XDCP\,J0044.0-2033 is expected to be a precursor system of the local very massive cluster population with
$M_{200}(z\!=\!0)\!\sim\!2\!\times\!10^{15}\,\mathrm{M_{\sun}}$.

\subsection{Global view of the galaxy population in the densest environment at  $z\!=\!1.58$}
\label{s4_GlobalGalPopulation}

The central 30\arcsec \ (250\,kpc) core region of XDCP\,J0044.0-2033 constitutes the densest galaxy environment currently known at $z\!\simeq\!1.6$. Visually, this core environment is also richer in galaxies than   IDCS\,J1426.5+3508 at   $z\!=\!1.75$  \citep{Stanford2012a}, implying that it is likely the densest galaxy concentration known in the $z\!>\!1.5$ Universe. 
This unique $z\!\simeq\!1.6$ cluster galaxy overdensity  enabled applying statistical background-subtraction techniques to 
 robustly determine the enclosed stellar mass profile (Sect.\,\ref{s3_MIRproperties}), the J- and Ks-band galaxy NIR luminosity functions    (Sect.\,\ref{s3_NIR_LF}), background-subtracted color-magnitude relations   (Sect.\,\ref{s3_CMR}), and red-galaxy counts    (Sect.\,\ref{s3_FaintEnd_RS}), as well as color-color selected spatial distributions and radial profiles of different galaxy types      (Sect.\,\ref{s3_SpatialDistribution}).

The most luminous (Ks$\la$Ks*) and most massive cluster galaxies  in this ultradense core environment observed at a lookback time of 9.5\,Gyr are still found in an active state of development  compared with the more evolved galaxies in similarly massive clusters at  $z\!\la\!1.4$.
Evidence for these marked galaxy differences at the observed epoch was seen based on morphological appearance with clear signs of active merging and mass assembly (Sect.\,\ref{s3_morphologies}) and their deviating colors with respect to the red-locus population 
 (Sects.\,\ref{s3_CMR}\,and\,\ref{s3_color_color}).
In particular, the currently brightest cluster galaxy is 
in a distinct early-evolution stage compared with the  most dominant low-$z$ BCGs in terms of total NIR luminosity and therefore stellar mass, cluster-centric offset, and magnitude gap to the second ranked galaxy 
(Sect.\,\ref{s3_BCGobs}), but fully consistent with the trends found for high-$z$ cluster samples  \citep[e.g.,][]{Fassbender2011c}.

\noindent
At intermediate NIR luminosities with Ks*$\la$Ks$\la$Ks*+2, the cluster galaxies in the core span  
a wide range of 
colors. 
A large population of galaxies was found with colors consistent with being in the transition stage between a recent star formation quenching period and their migration phase toward the red sequence  (Sect.\,\ref{s3_color_color}). Although a red-locus population of galaxies with colors 
expected for old passively evolving systems  is 
present in color-magnitude space at these intermediate luminosities, it is not straightforward to define a classical red sequence of early-type galaxies in XDCP\,J0044.0-2033 because the red
locus still contains a significant fraction of galaxies with detected \OII-emission lines and/or deformed morphologies (Sect.\,\ref{s3_CMR}).  However, 
we observe a very prominent suppression of red-locus galaxies beyond a magnitude of  Ks$\ga$20.5, corresponding to about Ks*+1.6 (Sect.\,\ref{s3_FaintEnd_RS}). This truncation magnitude is 
remarkably sharp and 
likely to be associated with the mass-quenching scale at  $z\!\sim\!1.6$ in this dense environment. The observations point toward a preferred Ks magnitude scale between Ks*+1 and Ks*+1.5 for the upward migration of core cluster galaxies from the blue cloud toward the red sequence in color-magnitude space.  


At faint NIR magnitudes (Ks$\ga$Ks*+2) the cluster galaxy population is found to be dominated by 
blue sources 
that exhibit colors consistent with star-forming/star-bursting systems. In the very cluster core, this starburst population consists of about 20\%. Outside the core, star-bursting cluster galaxies within the  R$_{200}$   volume appear to be clustered in clumps and are not smoothly distributed, especially around the   R$_{500}$ radius and beyond, where this population becomes the predominant galaxy type (Sect.\,\ref{s3_SpatialDistribution}).

\begin{figure}[h]
  \centering
  \includegraphics[width=\linewidth, clip]{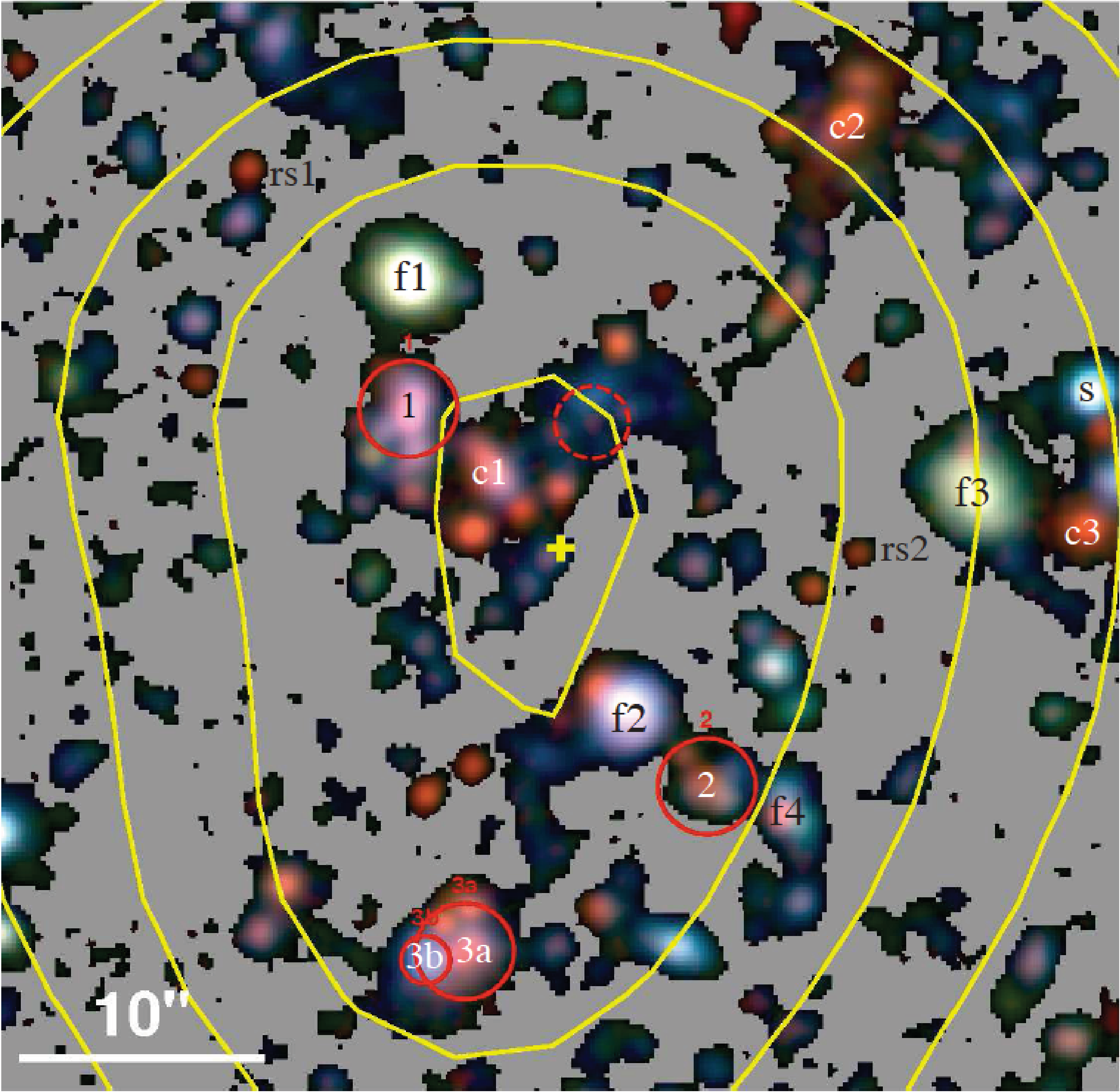} 
  \includegraphics[width=\linewidth, clip]{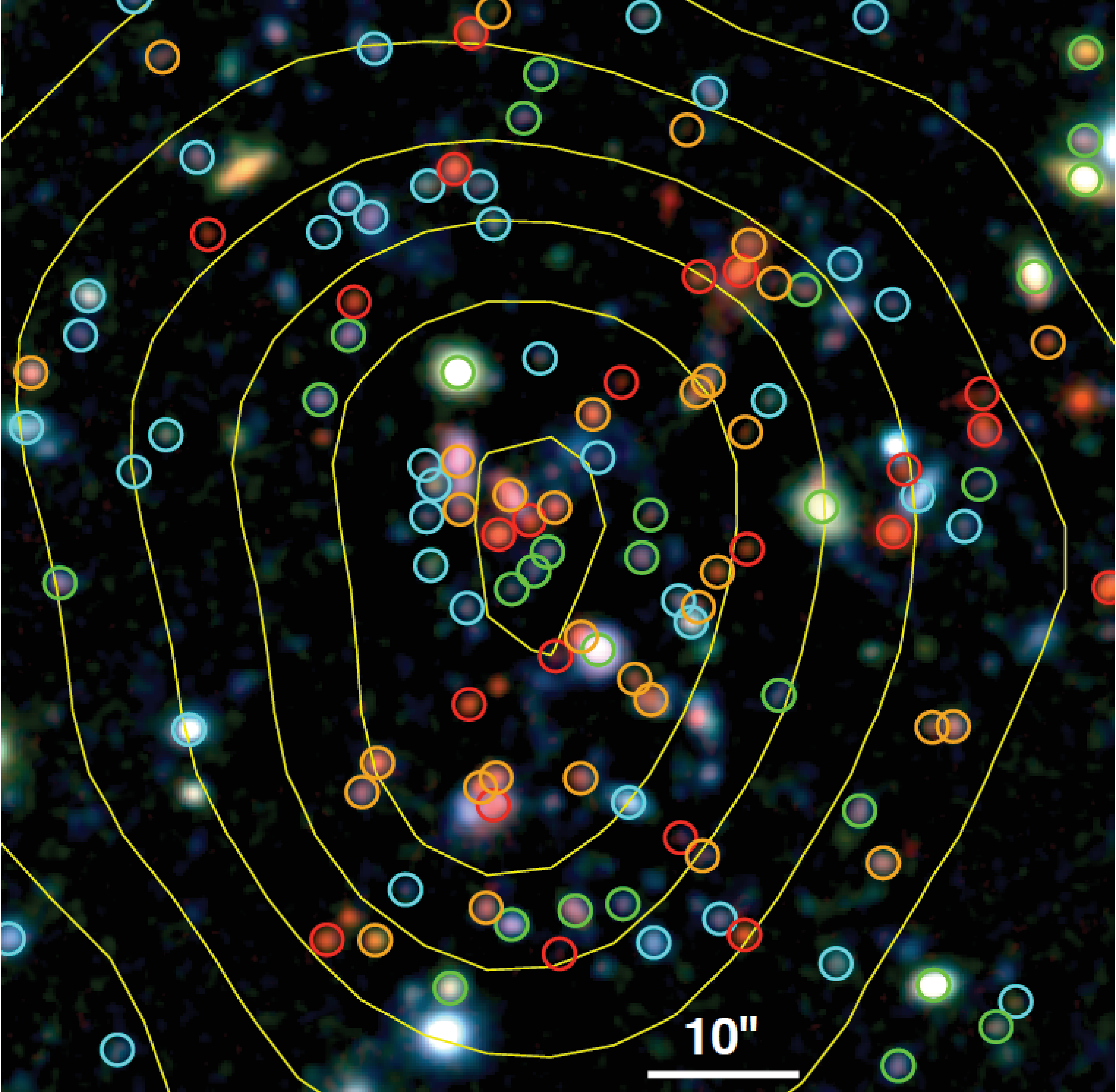}    
      \caption{Core view of XDCP\,J0044.0-2033. {\it Top panel}:  45\arcsec$\times$\,45\arcsec \  color composite showing the Subaru V- (blue channel) and i-band (green) image plus the combined J+Ks image (red) smoothed with a 0.5\arcsec \ kernel. The black background was rescaled to gray scale for contrast enhancement. The yellow cross indicates the location of the X-ray centroid, whereas contours and symbols for spectroscopic members are as in Fig.\,\ref{fig_OpticalOverview_X0044}. 
The brightest cluster member candidates without available spectroscopic confirmation are labelled `c1-c3'. Obvious low-$z$ foreground galaxies are marked `f1-f3', `f4' is a spectroscopically confirmed interloper ($z\!=\!0.592$),  and a foreground star is indicated by `s'.  `rs1'  (Ks$\simeq$20.4)  and `rs2' (Ks$\simeq$21.2) mark red-locus galaxies at and beyond the found red galaxy truncation magnitude of  Ks$\simeq$20.5.
{\it Bottom panel}: 70\arcsec$\times$\,70\arcsec \ (600$\times$600\,kpc) color composite based on the same data as above. Small colored circles indicate the color-color selected galaxies as 
defined in Fig.\,\ref{fig_Spatial_GalOD}. 
      } 
         \label{fig_Videt_CoreView}
\end{figure}

\clearpage

 

\subsection{Environmental effects on the cluster galaxy population}
\label{s4_EnvironmentalEffects}

From the spatial distribution and background-subtracted radial profiles for the different color-color selected galaxy types, the influence of the cluster environment in shaping the galaxy population becomes clear (Sect.\,\ref{s3_SpatialDistribution}). 
Within projected cluster-centric radii of $r\!\la\!300$\,kpc, the fraction of red-sequence transition objects rapidly increases to 
$\sim$40\%, accompanied by an increase of the post-starburst fraction to approximately  20\%. 
These two transitional populations  
dominate the total cluster galaxy budget  within the  regions of the 
extended X-ray emission or, correspondingly,  at net galaxy surface densities beyond 
$\sim$20\,arcmin$^{-2}$.  
The star formation quenching process(es) for these transitional galaxies  in this high-density region must have acted within 
about the last 1.5\,Gyr since the epoch of observation.

As expected, the fraction of red-locus objects also increases at  $r\!\la\!300$\,kpc to a level of 
$\sim$35\%. However, in the central  
100\,kpc, this fraction again drops to 
$\sim$20\%. While the statistical significance of this drop for a single cluster is naturally limited, it might imply that a good fraction of the reddest cluster galaxies are actually accreted as such, for example,~as central galaxies of evolved infalling groups. The galaxies with IDs 3a and c2 in Fig.\,\ref{fig_Videt_CoreView} are strong candidates for this latter scenario.





\section{Discussion}
\label{c5_Discussion}

After presenting the results on the galaxy population properties of XDCP\,J0044.0-2033 in the last two sections, we now compare our findings with previous observational results of high-$z$ cluster environments (Sect.\,\ref{s5_ClusterComparison}),  with
 predictions from galaxy evolution models  (Sect.\,\ref{s5_ModelComparison}), with 
 field galaxies at similar redshifts (Sect.\,\ref{s5_FieldComparison}), and finally attempt to draw 
  a consistent galaxy evolution picture 
  for the densest high-$z$ environments in Sect.\,\ref{s5_EmergingPicture}.


\subsection{Comparison with other  high-$z$  galaxy clusters}
\label{s5_ClusterComparison}

In Sect.\,\ref{s3_NIR_LF}, we measured the total J- and Ks-band {\it luminosity functions} for the core galaxy population at r$\la$250\,kpc. Previous studies of  the cluster galaxy luminosity function have consistently found the faint-end slope to  be close to  $\alpha\!\simeq\!-1$ from low redshifts \citep[$z\!<\!0.1$,][]{Lin2004b} to intermediate-$z$ \citep[$z\!\la\!0.6$,][]{DePropris2013a}, and up to  
$z\!\la\!1.3$ \citep{Strazzullo2006a,Mancone2012a}.  The three-parameter LF fits for
XDCP\,J0044.0-2033 agree with this fiducial faint-end slope. With a fixed  $\alpha\!\simeq\!-1$, we find characteristic magnitudes J* and Ks* that  fully match the predictions based on a passive luminosity evolution model with a formation redshift of   $z_{\mathrm{f}}\!=\!3$, which is commonly found in the studies at lower redshifts as
well.

However, these findings do not imply that the bright end of the LF is already fully in place at  $z\!\sim\!1.6$. On the contrary, 
we showed that the brightest cluster galaxies are only about one magnitude brighter than the characteristic magnitude m* and that close-up views reveal clear signs of ongoing merging and {\it mass-assembly activity}  (Sect.\,\ref{s3_morphologies}). 
Similar conclusions on an active cluster galaxy-mass assembly epoch at $1.3\!<\!z\!\la\!2$ were reached by \citet{Mancone2010a} based on a high-$z$ cluster sample study and for the specific  $z\!=\!1.62$  group environment of XMM-LSS\,J02182-05102
by \citet{Lotz2013a} based on morphological information. 
Recent claims that ruled out any active mass-assembly in bright cluster galaxies at  $z\!<\!1.8$ by  \citet{Andreon2013a} evidently do
 not agree with our direct observations and might be attributable to blending effects in the MIR {\it Spitzer} data, as shown in Fig.\,\ref{fig_Spitzer_View_100kpc}.

%


 For the observed {\it evolution of brightest cluster galaxies} it has now been established that BCGs grow significantly at   $z\!\la\!1.6$ in NIR luminosity and stellar mass  \citep[e.g.,][]{Lin2013a,Lidmann2012a,RF2007PhD}
 driven mainly by mass accretion via merging events  \citep{Lidman2013a,Burke2013a}.
These results are in line with the direct observations of the major BCG assembly phase of XDCP\,J0044.0-2033  (Sect.\,\ref{s3_BCGobs}), which was also reported for a lower mass $z\!\sim\!1.5$ system by  \citet{Nastasi2011a}. New observational results at the highest accessible cluster redshifts agree qualitatively well with BCG growth predictions from simulations  \citep[e.g.,][]{DeLucia2007a,Ruszkowsk2009a}, which is at odds with previous claims of little evolution in BCGs since $z\!\sim\!1.5$ \citep[e.g.,][]{Collins2009a,Stott2010a}.




Significant {\it central star formation activity} down to the very cluster core has been reported for most known systems
 at $z\!\ga\!1.5$ \citep[e.g.,][]{Hilton2010a,Hayashi2010a,Tran2010a,Fassbender2011a,Tadaki2012a}. The inferred fraction of blue starburst/star-forming galaxies of about 20\% in the very core of XDCP\,J0044.0-2033 (Sect.\,\ref{s3_SpatialDistribution}) therefore is 
 a confirmation of the previously observed trend, although extended to a more massive system in this case. 
The highest mass cluster with an observed 
 high star formation activity is the recently discovered system SPT-CL\,J2040-4451 at 
$z\!=\!1.48$ with a mass of $M_{200}\!\simeq\!6\!\times\!10^{14}\,\mathrm{M_{\sun}}$ \citep{Bayliss2013a}. On the other hand, XMMU\,J2235.3-2557 at $z\!=\!1.39$  with  $M_{200}\!\simeq\!7\!\times\!10^{14}\,\mathrm{M_{\sun}}$  \citep{Rosati2009a}  is the highest redshift cluster with a fully quenched core region within 200\,kpc  \citep{Strazzullo2010a,Bauer2011a,Gruetzbauch2012a}. 
Total cluster mass $M_{200}$ and redshift $z$ are clearly two key parameters that determine the level of star formation activity in the densest core regions. However, at lookback times of about 9.5\,Gyr or more, the 
assembly histories of the individual clusters must also play an increasingly important role.  
An early collapse time without  major recent mass  accretion might then explain the low-mass X-ray group at  $z\!=\!1.61$ with predominantly quiescent galaxies reported by  \citet{Tanaka2013a} and similar observations for JKCS\,041 at $z\!=\!1.80$ \citep{Andreon2014a,Newman2014a}, while most 
systems at similar redshifts exhibit strong SF activity.

%
%
%
%
%

Recent observational studies of the {\it quiescent  red-galaxy population and red sequence} in distant cluster environments have provided an increasing amount of evidence of significant changes in the old evolved galaxy population in the  $z\!>\!1.4$ regime. 
Although a significant population of massive, quiescent, early-type galaxies in dense environments has been identified  out to $z\!\simeq\!2$ \citep{Strazzullo2013a,Gobat2013a}, the results of \citet{Snyder2012a} 
 indicate that most of the stellar mass on the red sequence is 
 assembled at  $z\!<\!2$. This 
 is in line with the findings 
 by \citet{Rudnick2012a}, 
who investigated 
in detail the quiescent 
population of  the system XMM-LSS\,J02182-05102 at $z\!=\!1.62$ and pointed out the importance of merging for the 
 red-sequence build-up. 
However, the partial red sequences that are already in place in cluster environments at $z\!>\!1.4$ appear to be all broadly consistent in color and slope with simple expectations from galaxy evolution models where the bulk of stars formed around  $z\!\sim\!3$
\citep[e.g.,][]{Snyder2012a,Hilton2009a,Muzzin2013a,Tanaka2013a,Willis2013a}.
All this agrees well with our 
results, which showed that there is an active build-up phase of the red sequence of XDCP\
J0044.0-2033, while at the same time the locus of the existing red galaxy population is consistent with a simple  SSP with formation redshift $z_f\!=\!3$.







Moreover, the 
evidence for a sharp {\it truncation of the red galaxy population} at magnitudes $\simeq$Ks*+1.6 in Sect.\,\ref{s3_FaintEnd_RS} confirms 
the 
reported trends 
that the faint end of the red sequence is increasingly underpopulated with increasing redshift  
\citep[e.g.,][]{DeLucia2004a,Koyama2007a,Tanaka2007a,Stott2009a,Rudnick2009a,Fassbender2011b,Lerchster2011a,Rudnick2012a,Tanaka2013a}. Assuming a stellar mass of $\simeq$$10^{11}\,\mathrm{M_{\sun}}$ for a galaxy with characteristic magnitude Ks*  \citep{Strazzullo2006a}, the observed truncation limit in XDCP\,J0044.0-2033 can be approximated to be on the order of $M_{\mathrm{*,trunc}}\!\simeq\!0.25\times M_*(Ks*)\!\sim\!2.5\!\times\!10^{10}\,\mathrm{M_{\sun}}$. This corresponds well with the 
red galaxy cut-off scales reported by  \citet{Tanaka2013a} 
and  \citet{Rudnick2012a}  for low-mass group environments at  similar  $z\!\simeq\!1.6$. 




\citet{Muzzin2012a} have investigated in detail the {\it environmental  effects}  at $z\!\sim\!1.0$ based on a sample of rich MIR selected clusters. They found 
that the population of post-starburst galaxies is increased by a factor of three in cluster environments, based on which, they argued for 
a rapid environmental-quenching timescale at $z\!\sim\!1$ or a dominant role of the mass-quenching process at  $z\!>\!1$. The 
high fractions of post-starburst and red-sequence transition objects in XDCP\,J0044.0-2033 (Sects.\,\ref{s3_SpatialDistribution}\,\&\,\ref{s4_EnvironmentalEffects}), which comprise more than half of the core galaxy population, extend 
this trend toward higher redshift.

\subsection{Comparison with model predictions}
\label{s5_ModelComparison}

\begin{figure}[t]
    \centering
     \includegraphics[width=0.98\linewidth, clip]{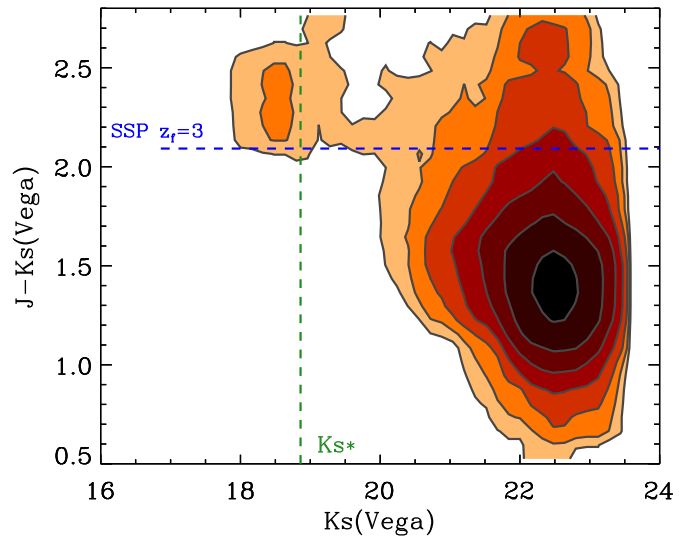}
      \caption{Predicted densities of galaxies at $z\!\simeq\!1.6$ 
 in the J$-$Ks versus Ks observer-frame CMD for cluster environments with $M_{200}\!\ge\!2\!\times\!10^{14}\,\mathrm{M_{\sun}}$ 
 based on the semi-analytic models of   \citet{Menci2008a}. The different contours indicate an increase by factors of two in CMD density from light to dark colors. Dashed lines have the same meaning as in the observed version of the diagram shown in Fig.\,\ref{fig_CMRdensity}. 
      } 
         \label{fig_CMR_models} 
\end{figure}

We now 
compare predictions based on cosmological simulations and semi-analytic models (SAMs) with our observations of galaxy population properties in the densest $z\!\simeq\!1.6$ environment. 
Figure\,\ref{fig_CMR_models}  shows the predicted galaxy densities in the J$-$Ks versus Ks CMD based on an extended version of the semi-analytic galaxy formation models by \citet{Menci2008a} that 
also include stellar mass stripping processes and an improved treatment of the metallicity-dependent cooling function. The model CMD is constructed based on the averaged galaxy populations of all objects within halos with a total mass of $M_{200}\!\ge\!2\!\times\!10^{14}\,\mathrm{M_{\sun}}$ in the redshift slice $1.5\!\le\!z\!\le\!1.6$
The figure is to be compared with the observational analog in Fig.\,\ref{fig_CMRdensity} (top panel), with the difference that the rightmost parts of the observed diagram are affected by incompleteness  and that the observed contours follow linear steps instead of the logarithmic ones 
of Fig.\,\ref{fig_CMR_models}. 



\begin{figure}[t]
    \centering
     \includegraphics[width=0.98\linewidth, clip]{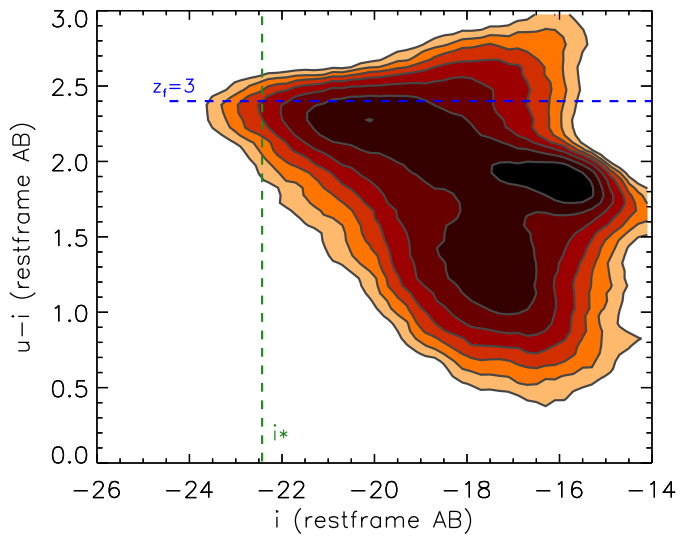}
      \caption{Model predictions for the density distribution of cluster galaxies at  $z\!\simeq\!1.4$ in the restframe u$-$i versus i CMD based on \citet{Bertone2007a}. 
      Contour levels increase by factors of 1.5 in CMD density from light to dark colors, dashed lines are the same as in Fig.\,\ref{fig_CMRdensity}. 
      } 
         \label{fig_CMR_models_Bertone} 
\end{figure}

The 
peak in the predicted color-magnitude distribution in Fig.\,\ref{fig_CMR_models} at Ks$\simeq$22.5 and J$-$Ks$\simeq$1.3 exactly  coincides with the observed  population of low-mass blue star-forming cluster galaxies in Sect.\,\ref{s3_CMR}. After accounting for the approximate factor of two incompleteness corrections in this part of the observed CMD, we 
conclude that the  color-magnitude location of the  bulk of the faint blue galaxies agrees with the observations. The model galaxies also agree reasonably well with our results on the bright end of the luminosity function, where the observed magnitudes of the most luminous objects match well, while the predicted colors appear 
slightly 
redder than the measured ones. 

The most significant discrepancy between the predictions of the
model reported by  \citet{Menci2008a} and the observations occurs in the intermediate-magnitude range 
19$\la$Ks$\la$20.5, where the model is almost devoid of objects owing to fast galaxy evolutionary processes in the model that drive objects toward the bright end on short timescales. 
For XDCP\,J0044.0-2033, on the other hand, we find this region to be densely populated 
by a mix of red-locus galaxies and recently quenched galaxies that transition toward the red sequence. Overall, we 
conclude that the semi-analytic galaxy evolution model of \citet{Menci2008a} agrees well for the starting point of the cluster galaxies in the blue cloud and the end point of massive red galaxies, 
while the model seems discrepant and does not produce the galaxies found in the transition region  in color-magnitude space. 

Next, we compare the  observations with the implementation of \citet{Bertone2007a} of the {\em Munich galaxy formation model}, 
which is based on the Millennium Simulation \citep{Springel2005a}. Figure\,\ref{fig_CMR_models_Bertone} shows the restframe  u$-$i versus i 
color-magnitude diagram for all galaxies in cluster environments with  $M_{\mathrm{vir}}\!\ge\!10^{14}\,h^{-1}\mathrm{M_{\sun}}$ at the  $z\!\simeq\!1.385$ simulation output snapshot, which
was made about half a gigayear later in cosmic time than the observations of Fig.\,\ref{fig_CMRdensity}. This  u$-$i restframe color represents  the intrinsic galaxy color, which lies in between the observed  i$-$Ks and J$-$Ks measurements,  as can be seen in Fig.\,\ref{fig_TemplateFilters}.

The   model predictions of \citet{Bertone2007a}  in Fig.\,\ref{fig_CMR_models_Bertone} show the bright end of the red sequence already 
in place. 
Colors and magnitudes relative to the model lines  are 
fairly consistent  with the  XDCP\,J0044.0-2033 observations  for the brightest (m$<$m*) cluster galaxies and, moreover, the highly populated part of red galaxies at intermediate magnitudes  (m*$\la$m$\la$m*+2) agrees well with the observed  J$-$Ks CMD. Qualitatively, the drop toward bluer colors, that
is,~the observed red-locus truncation, is  also reproduced by the model, 
although it 
occurs at fainter magnitudes (m$\ga$m*+3.5) in the model. The bulk population of faint blue galaxies at 
 m$\simeq$m*+4, on the other hand, is again fully consistent with the observational findings. 

We also checked the predicted  $z\!\simeq\!1.6$ cluster galaxy populations 
based on the N-body plus hydrodynamical simulations of {\it rezoomed} clusters in  \citet{Romeo2005a,Romeo2008a}, which include in particular supernova feedback, metal-dependent cooling, and
non-instantaneous chemical recycling. 
For this model, the distant cluster galaxy population shows a distribution in color magnitude space that is in between the two models discussed above.     


One common feature for 
the galaxy evolution models of  \citet{Bertone2007a}, \citet{Romeo2008a}, and   \citet{Menci2008a} is a significant population of very red galaxies at faint magnitudes (m$\ga$m*+3) in the upper right part of the CMDs. Observationally, such galaxies are notoriously hard to detect because of increasing incompleteness and photometric errors. However, up to the observational limit of the available data this predicted abundant very red, faint cluster galaxy population seems to be suppressed in the measurements.  


Overall, the tested galaxy evolution models for distant cluster environments seem to clearly reproduce the observed faint blue bulk population of star-forming galaxies, and the CMD location of the brightest red galaxies is reproduced reasonably
well. At intermediate magnitudes, where most of the evolutionary action takes place, the models differ significantly, but seem to commonly underestimate the recently quenched transitioning population of galaxies at  m*$\la$m$\la$m*+1.5, as suggested by the presented observations.

Our observed red-galaxy truncation scale of  $M_{\mathrm{*,trunc}}\!\sim\!2.5\!\times\!10^{10}\,\mathrm{M_{\sun}}$, however, is consistently predicted by  \citet{Gabor2012a} based on their hydrodynamical cosmological simulation. Here, galaxies passing a critical mass threshold of $\sim\!3\!\times\!10^{10}\,\mathrm{M_{\sun}}$ enter the mass-quenching regime and are subsequently  moving upward in the CMD, reaching the red sequence on a timescale of 1-2Gyr. Once on the red sequence, the galaxies continue to grow significantly in stellar mass via minor-merging processes, predicting the most massive galaxies to be bluer in color owing to the increasing contribution of infalling low-metallicity, low-mass galaxies. This halo-quenching model thus qualitatively reproduces many of the discussed features of our observed cluster galaxy population in color-magnitude space.


\subsection{Comparison with field observations}
\label{s5_FieldComparison}

\citet{Peng2010a} and \citet{Peng2012a} have shown that {\it mass-quenching} and  {\it environment-quenching} are distinct and separable processes out to at least   $z\!\simeq\!1$, in the sense that the former acts quasi-independently of the environment and the latter varies little with the halo or stellar mass of the individual galaxy.
Previously, \citet{Bundy2006a} found evidence for an evolving critical mass-quenching scale above which star formation is suppressed and additional indications that this downsizing process is  accelerated in very dense environments, meaning that~the mass scale above which feedback processes halt star formation increases toward higher redshifts, while it decreases in denser environments.    

In a recent field study at a median redshift of $z_{\mathrm{med}}\!\sim\!1.8$, Sommariva et~al.~(2014)\nocite{Sommariva2013a} identified a stellar mass threshold for the population of passive galaxies. Here the identified peak of the passive galaxy population as a function of Ks magnitude  corresponds well with the identified truncation magnitude of this work, while the sharp decline in number density sets in at a scale about one magnitude fainter.  
Another finding in common of the field galaxy study by Sommariva et~al.~(2014) 
and the observations in this work is the discrepancy
between the data and theoretical predictions regarding the presence of red galaxies at faint magnitudes, as was discussed in Sect.\,\ref{s5_ModelComparison}.

\subsection{Emerging picture of galaxy evolution in the densest $z\!>\!1.5$ cluster environments}
\label{s5_EmergingPicture}



After this comprehensive look at the galaxy population properties of XDCP\,J0044.0-2033, we now attempt to combine all observational results into a more general scenario of the galaxy evolution process in the densest high-$z$ cluster environments. 
As a starting point, we consider the suggested cluster galaxy evolution toy models by \citet{Muzzin2012a} to explain the distribution of the post-starburst population at $z\!\sim\!1$ and by \citet{Rudnick2012a}, which describes the build-up of stellar mass on the red sequence since  $z\!\simeq\!1.6,$  and 
extend them as necessary to incorporate the results of this work. 

As remarked before, at a lookback time of 9.5\,Gyr a number of relevant timescales related to the cluster formation and galaxy evolution process are similar to each other and the lifetime of the cluster itself, which gives rise to the observation of new phenomena in the redshift regime at $z\!>\!1.5$. A closer look at these timescales can therefore provide new insights and help in combining many of the recent observational results.   

If we assume an early cluster formation epoch for the systems under consideration of  $z\!\sim\!3$, the cluster lifetime at $z\!\simeq\!1.6$  is at most t$_{\mathrm{cl,life}}$$\sim$2\,Gyr, while the accretion and relaxation timescales are  t$_{\mathrm{cl,accr}}$$\sim$t$_{\mathrm{cl,relax}}$$\sim$0.5-1\,Gyr. Because of the stochastic accretion process of infalling matter and structures from the surrounding environment, it is immediately evident that we can expect to observe significant cluster-to-cluster variations  in the properties of the galaxy and matter content at $z\!\simeq\!1.6$, depending on the specific formation histories of the individual systems. 

For the galaxy population properties in $z\!\simeq\!1.6$ clusters, and in particular, for the formation of the most massive galaxies and the red sequence, we have to consider at least  three additional timescales: (i) the environment-quenching timescale  t$_{\mathrm{quench,env}}$,  (ii) the galaxy mass-quenching timescale t$_{\mathrm{quench,mass}}$, and (iii) the galaxy mass-assembly timescale t$_{\mathrm{gal,as}}$, that is,~the time needed to double the stellar mass through merging processes. 

Galaxies accreted by the cluster from the large-scale structure surroundings are in their majority blue star-forming systems, but can also enter as pre-processed galaxies as part of group environments. Once part of the main cluster halo, the  environmental quenching starts to suppress star formation on the  timescale  t$_{\mathrm{quench,env}}$ for all galaxies, that is,~it causes a parallel shift upward in the CMD. This  environmental quenching timescale for XDCP\,J0044.0-2033 must be longer than the typical residence time of a galaxy in the cluster since accretion (i.e.,~$>$\,1Gyr) to explain the lack of faint red galaxies and the presence of central SF galaxies, and short enough to add a significant number of galaxies to the red sequence in the time span from the cluster formation epoch to $z\!\sim\!1$ (i.e.,~$\la$\,4Gyr). Based on the observed CMD densities at faint magnitudes and intermediate colors, this timescale for XDCP\,J0044.0-2033 can be estimated to be on the order of t$_{\mathrm{quench,env}}$$\sim$2-3\,Gyr, which is~similar to the cluster age t$_{\mathrm{cl,life}}$. 

The mass-quenching timescale t$_{\mathrm{quench,mass}}$, on the other hand, must be shorter and sets in after galaxies pass the identified stellar mass threshold of $log(M_{*,{\mathrm{trunc}}}/M_{\sun})\!\simeq\!10.4,$ visible as the truncation magnitude of the red-galaxy population in XDCP\,J0044.0-2033. The mass quenching then drives galaxies with magnitudes brighter than Ks*+1.6 upward in the CMD toward the red sequence on a timescale on the order of  t$_{\mathrm{quench,mass}}$$\sim$1\,Gyr.

The critical third timescale is now the  galaxy mass-assembly time t$_{\mathrm{gal,as}}$, which is in general longer than the mass-quenching timescale t$_{\mathrm{quench,mass}}$, which in most cases results in an evolutionary path along the red sequence through dry-merging activity \citep[e.g.,][]{Faber2007a}. However, this galaxy mass-assembly timescale strongly depends
on~the local galaxy density, the relative velocity structure of neighboring galaxies, and in particular, on the local dynamic friction timescale  t$_{\mathrm{DF}}$.
As shown throughout this paper, the central cluster region of XDCP\,J0044.0-2033 is an extreme environment in these aspects, featuring very high galaxy densities (e.g.,~Sect.\,\ref{s3_Profiles}), a spectroscopic example of low radial velocities for galaxies with IDs\,3a\,and\,3b
(Sect.\,\ref{s2_Spectroscopy}), and a short dynamic-friction timescale in the core (Sect.\,\ref{s3_BCGobs}). In this favorable environment, the mass-assembly rate is highly accelerated and can become shorter than the mass-quenching timescale for a significant fraction of the cluster galaxies, that is,~t$_{\mathrm{gal,as}}$$\la$t$_{\mathrm{quench,mass}}$ or t$_{\mathrm{gal,as}}$$\la$1\,Gyr. 

This accelerated mass-assembly activity at $z\!\simeq\!1.6$  has now two important consequences for the observed galaxy population properties as objects are driven leftward toward brighter magnitudes in the CMD. First, many galaxies in the cluster core are driven across the mass-quenching threshold  $M_{*,{\mathrm{trunc}}}$ on a short timescale, where the star formation suppression is accelerated and the rapid transition toward the red sequence begins. Second, as some galaxies continue to rapidly grow in stellar mass through continuing merging activity, the transition toward brighter magnitudes in the CMD can be as fast or faster than the upward movement, with the observational consequence that a good fraction of the most massive galaxies are still found in their transition state with colors bluer than the expected red sequence, as seen, for example,~in Figs.\,\ref{fig_CMRdensity}\,and\,\ref{fig_color_color}.

This scenario would imply that the immediate influence of the cluster environment  at $z\!\simeq\!1.6$ is mainly characterized by the accelerated short  galaxy mass-assembly timescale of t$_{\mathrm{gal,as}}$$\la$1\,Gyr. As a consequence of this, a large portion of cluster galaxies becomes massive  enough within a short time span to pass the mass-quenching threshold and exhibit an increased transitional speed toward the red sequence on the timescale  t$_{\mathrm{quench,mass}}$$\sim$1\,Gyr. The direct environmental quenching mechanism, on the other hand, must be subdominant and suppresses star formation only on timescales  of t$_{\mathrm{quench,env}}$$\sim$2-3\,Gyr. All three relevant timescales that consistently match the observations of XDCP\,J0044.0-2033 are likely to evolve with decreasing redshift and vary with the total cluster mass  $M_{200}(z)$.

The proposed scenario, which rests upon a very rapid 
galaxy mass-assembly timescale t$_{\mathrm{gal,as}}$ in high-$z$ cluster cores,
is fully consistent with \citet{Rudnick2012a}, who derived a necessary high merger rate at $z\!\simeq\!1.6$ to explain the red-sequence build-up, and \citet{Muzzin2012a}, who speculated on the dominance of mass quenching over environmental quenching at  $z\!>\!1$.
Moreover, \citet{Strazzullo2013a} also tentatively identified the mass-quenching process as the likely  dominant mechanism in a $z\!\simeq\!2$ group environment. From the theoretical and modeling side, \citet{DeLucia2012a} identified the mass-quenching-dominated regime for galaxies with  $log(M_{*}/M_{\sun})\!>\!10.5$, which
quantitatively agrees very well with our estimated threshold mass.

In summary, we conclude that in the massive cluster environment of XDCP\,J0044.0-2033 at $z\!\simeq\!1.58$ we finally directly observe the `fireworks' of massive cluster galaxy assembly at play.

\section{Summary and conclusions}
\label{c6_Conclusions}


We have presented a comprehensive study of the galaxy population properties of the massive galaxy cluster XDCP\,J0044.0-2033 at $z\!=\!1.58$ based on deep HAWK-I imaging data in the J- and Ks bands, complemented by Subaru imaging in V and i, {\it Spitzer} observations at 4.5\,\microns, and new spectroscopic observations with VLT/FORS\,2. 
This X-ray luminous cluster 
constitutes the densest environment at $z\!\simeq\!1.6$ currently known, which allowed applying statistical background-subtraction techniques to robustly measure and derive various properties of the cluster-associated galaxy population. 

Our main results are summarized as follows:
 
\begin{enumerate}
\item Based on the deep HAWK-I near-infrared imaging data, we detected a  net excess of about 90 cluster-associated galaxies, of which approximately 75\% are concentrated within the central 30\arcsec \, (250\,kpc) about the centroid position of the extended X-ray emission.   

\item The radial galaxy surface-density profile of the cluster is consistent with a centrally peaked NFW profile with a high concentration parameter of $c_{200}\!\simeq\!10$. The radial 4.5\,\microns \ surface brightness profile, on the other hand, is best described by a less compact projected NFW form with  $c_{200}\!\simeq\!4.4$, because the stellar mass peak is offset from the X-ray centroid position.

\item Based on the {\it Spitzer}  4.5\,\microns \ imaging data, we measured a total enclosed stellar mass within the projected R$_{500}$  (R$_{200}$) radii of  $M*_{500}\!\simeq\!(6.3 \pm 1.6)\!\times\!10^{12}\,\mathrm{M_{\sun}}$  ($M*_{200}\!\simeq\!(9.0 \pm 2.4)\!\times\!10^{12}\,\mathrm{M_{\sun}}$) and a resulting stellar mass fraction of  f$_{*,500}=M_{*,500}/M_{500}  = (3.3\pm1.4)$\%, consistent with local values.


\item We measured the total J- and Ks band  galaxy luminosity functions in the cluster core region within a projected radius of 250\,kpc  and found characteristic apparent (Vega) magnitudes at the cluster redshift of  J*$=$20.90$\pm$0.50 and  Ks*$=$18.71$\pm$0.66 for a fixed faint-end slope of  $\alpha\!=\!-1$, in good agreement with predictions from simple stellar population models for a formation redshift of $z_{\mathrm{f}}\!=\!3$.

\item Although the overall shape of the galaxy luminosity function seems to be in place at $z\!\simeq\!1.6$, this does not imply that the bright end of the galaxy population is fully evolved. On the contrary, a closer look at the morphologies and color distributions of the secure spectroscopic members reveals a very active ongoing mass-assembly epoch for the most massive galaxies with clearly visible merger signatures.  


\item The J$-$Ks and  i$-$Ks versus Ks  color-magnitude diagrams of the cluster core region revealed a more complex distribution of cluster galaxies in color-magnitude space than lower redshift systems.  Although a fairly rich red-locus population is present in the magnitude range Ks*$\la$Ks$\la$Ks*+1.6 at a color consistent with predictions of $z_{\mathrm{f}}\!=\!3$ SSP models, a significant number of these galaxies still exhibit \OII \ line emission and/or deformed morphologies and thus are not readily identifiable as early-type systems. Moreover, most galaxies at the bright end of the luminosity function at Ks$\la$Ks* feature colors that are too blue to be associated with a red sequence.

\item  The CMDs additionally revealed a large population of galaxies that undergoes color transitioning toward the red sequence, in particular in the magnitude range  Ks*+1$\la$Ks$\la$Ks*+1.5. Even in the cluster core region, we also readily identified the bulk population of faint blue star-forming cluster galaxies at magnitudes around  Ks$\simeq$Ks*+3.5.

\item Number counts along the red-locus galaxy population revealed  a sharp truncation magnitude at Ks(Vega)$\simeq$20.5$\simeq$Ks*+1.6, above which very few red galaxies are present in the cluster up to the detection limit of the data. The faint end of the red-locus population at  Ks(Vega)$\ga$Ks*+1.6 therefore is not yet fully developed either. 

\item We used  the J$-$Ks versus  i$-$Ks plane to identify and select different cluster galaxy types in color-color space, distinguishing four significant populations in the cluster core:
(1) the red-locus population with colors consistent with a passive SED,  (2) a `red-sequence transition population' just blueward of the red locus, (3) a post-starburst population whose colors are consistent with quenched star formation 1\,Gyr ago, and (4) the very blue star-forming/starburst population. In the very densest cluster core region at $r\!\le\!100$\,kpc we identified relative fractions of these galaxy types that comprise about 20\% red-locus galaxies (type 1), 40\% red-sequence transition objects (type 2), 20\% post-starburst galaxies (type 3), and 20\% star-forming/starburst systems (type 4).

\item Based on the color-color selection of the four different galaxy types, we also investigated their spatial distribution and radial trends with cluster-centric distance. In terms of absolute number of detected galaxies the first three galaxy types show monotonically rising profiles toward the cluster center, while the star-forming/starburst population appears to be found mostly in clumpy structures, in particular, at and outside the  R$_{500}$ radius, where this blue population becomes dominant. The relative fractions, on the other hand, indicate a rising fraction of post-starburst galaxies toward the center (type 3), a plateau fraction of   40\%  for the  red-sequence transition galaxies reached within  $r\!\le\!200$\,kpc (type 2), and the red-locus populations that reaches peak fractions of about 35\% in the radial range 100-300\,kpc and then drops again toward the most central region.    


\item We investigated the central cluster core at $r\!\le\!100$\,kpc with respect to the most likely formation and evolution scenario of the brightest cluster galaxy. The current cluster BCG is identified to be located at a projected cluster-centric distance of 70\,kpc from the X-ray centroid with a magnitude of   Ks*-1.  Its color is blueward of the red-locus population, which places it in the transition-object class (type 2), and it features a highly deformed morphology, which indicates ongoing  merging activity along several directions. 
Based on a measured total stellar mass within  $r\!\le\!100$\,kpc of of about $M*_{100\,\mathrm{kpc}}\!\simeq\!1.5\!\times\!10^{12}\,\mathrm{M_{\sun}}$ and estimated very short dynamical friction timescales of $\la$100\,Myr, we argued for a very rapid evolution timescale 
of the current BCG toward increasing stellar masses, redder colors, and smaller cluster-centric offsets, as observed in the majority of cluster cores at lower redshifts. 

\item We compared our observational results with model predictions from semi-analytic galaxy evolution models as well as N-body/hydrodynamical simulations and found that the predictions for the locations of the bulk population of the blue star-forming galaxies (type 4) and the brightest cluster galaxies in  color-magnitude space are consistent with the observations, while the predictions for the intermediate transition regions are generally deviant. In particular, we found that the galaxy evolution model of  \citet{Menci2008a} predicts a transition region that is underpopulated compared with the observations because of a mass evolution toward the bright end that is too fast, while the      
  model of \citet{Bertone2007a} predicts the bulk of transitioning galaxies at magnitudes significantly fainter than observed. 
The qualitatively best match to our results is given by the model predictions of  \citet{Gabor2012a}, which feature an efficient mass-quenching mechanism above a threshold in stellar mass of $\sim\!3\!\times\!10^{10}\,\mathrm{M_{\sun}}$.

\item Our observational results on the galaxy population properties in the very dense core region of XDCP\,J0044.0-2033  at $z\!\simeq\!1.6$  point toward a scenario in which environmental quenching is a subdominant process  that acts on a timescale on the order of  t$_{\mathrm{quench,env}}$$\sim$2-3\,Gyr, while galaxy mass-quenching with a timescale of t$_{\mathrm{quench,mass}}$$\sim$1\,Gyr is the more important process to shape the observed galaxy population properties. We see evidence that galaxy evolution in this superdense environment is highly accelerated owing to a rapid mass-assembly timescale through merging processes of  t$_{\mathrm{gal,as}}$$\la$1\,Gyr, which is for many galaxies similar to or shorter than the mass-quenching timescale. This scenario naturally explains the significant fraction of very bright cluster galaxies that are not yet part of the red sequence, the sharp truncation magnitude of the red-locus population, and the magnitude scale where most galaxies are transitioning toward the red sequence. This would imply that the  dominant role of the $z\!\simeq\!1.6$ cluster environment in shaping the observed galaxy population is to drive a large number of core cluster galaxies across the identified mass-quenching threshold of  $log(M_{*,{\mathrm{trunc}}}/M_{\sun})\!\simeq\!10.4$, while the effect of direct environmental star formation quenching processes are only becoming of increasing relevance at later cosmic epochs.

\end{enumerate}



\noindent
Owing to its richness in galaxies and the observed ongoing galaxy transformation phenomena, the cluster XDCP\,J0044.0-2033 at $z\!\simeq\!1.58$ is an excellent testbed to observationally probe in detail  galaxy formation processes at the highest matter-density peak at a lookback time of 9.5\,Gyr. Upcoming results based on the most recent observations of the cluster galaxies, for example, with~VLT/KMOS and the very deep X-ray coverage with  {\it Chandra,} could lead the way to improved insights into the physical mechanisms that shape the observed galaxy population at this epoch.


\begin{acknowledgements}
The research leading to these results has received funding from the European Union Seventh Framework Programme (FP7/2007-2013) under grant agreement \num\,267251 `Astronomy Fellowships in Italy' (AstroFIt).
This research was supported by the DFG  under grant BO 702/16-3.
This work is based in part on observations made with the Spitzer Space Telescope, which is operated by the Jet Propulsion Laboratory, California Institute of Technology under a contract with NASA.

\end{acknowledgements}

\bibliographystyle{aa} 

\begin{thebibliography}{94}
\expandafter\ifx\csname natexlab\endcsname\relax\def\natexlab#1{#1}\fi

\bibitem[{{Andreon}(2013)}]{Andreon2013a}
{Andreon}, S. 2013, \aap, 554, A79

\bibitem[{{Andreon} {et~al.}(2014){Andreon}, {Newman}, {Trinchieri},
  {Raichoor}, {Ellis}, \& {Treu}}]{Andreon2014a}
{Andreon}, S., {Newman}, A.~B., {Trinchieri}, G., {et~al.} 2014, \aap, 565,
  A120

\bibitem[{{Bauer} {et~al.}(2011){Bauer}, {Gr{\"u}tzbauch}, {J{\o}rgensen},
  {Varela}, \& {Bergmann}}]{Bauer2011a}
{Bauer}, A.~E., {Gr{\"u}tzbauch}, R., {J{\o}rgensen}, I., {Varela}, J., \&
  {Bergmann}, M. 2011, \mnras, 411, 2009

\bibitem[{{Bayliss} {et~al.}(2013){Bayliss}, {Ashby}, {Ruel}, {Brodwin},
  {Aird}, {Bautz}, {Benson}, {Bleem}, {Bocquet}, {Carlstrom}, {Chang}, {Cho},
  {Clocchiatti}, {Crawford}, {Crites}, {Desai}, {Dobbs}, {Dudley}, {Foley},
  {Forman}, {George}, {Gettings}, {Gladders}, {Gonzalez}, {de Haan},
  {Halverson}, {High}, {Holder}, {Holzapfel}, {Hoover}, {Hrubes}, {Jones},
  {Joy}, {Keisler}, {Knox}, {Lee}, {Leitch}, {Liu}, {Lueker}, {Luong-Van},
  {Mantz}, {Marrone}, {Mawatari}, {McDonald}, {McMahon}, {Mehl}, {Meyer},
  {Miller}, {Mocanu}, {Mohr}, {Montroy}, {Murray}, {Padin}, {Plagge}, {Pryke},
  {Reichardt}, {Rest}, {Ruhl}, {Saliwanchik}, {Saro}, {Sayre}, {Schaffer},
  {Shirokoff}, {Song}, {Stalder}, {Suhada}, {Spieler}, {Stanford},
  {Staniszewski}, {Stark}, {Story}, {Stubbs}, {van Engelen}, {Vanderlinde},
  {Vieira}, {Vikhlinin}, {Williamson}, {Zahn}, \& {Zenteno}}]{Bayliss2013a}
{Bayliss}, M.~B., {Ashby}, M.~L.~N., {Ruel}, J., {et~al.} 2013, arXiv:1307.2903

\bibitem[{{Bertin} \& {Arnouts}(1996)}]{Bertin1996a}
{Bertin}, E. \& {Arnouts}, S. 1996, \aaps, 117, 393

\bibitem[{{Bertin} {et~al.}(2002){Bertin}, {Mellier}, {Radovich}, {Missonnier},
  {Didelon}, \& {Morin}}]{Bertin2002a}
{Bertin}, E., {Mellier}, Y., {Radovich}, M., {et~al.} 2002, in Astronomical
  Society of the Pacific Conference Series, Vol. 281, Astronomical Data
  Analysis Software and Systems XI, ed. D.~A. {Bohlender}, D.~{Durand}, \&
  T.~H. {Handley}, 228

\bibitem[{{Bertone} {et~al.}(2007){Bertone}, {De Lucia}, \&
  {Thomas}}]{Bertone2007a}
{Bertone}, S., {De Lucia}, G., \& {Thomas}, P.~A. 2007, \mnras, 379, 1143

\bibitem[{{Binney} \& {Tremaine}(1987)}]{Binney1987a}
{Binney}, J. \& {Tremaine}, S. 1987, {Galactic dynamics} (Princeton University
  Press)

\bibitem[{{Brodwin} {et~al.}(2006){Brodwin}, {Brown}, {Ashby}, {Bian}, {Brand},
  {Dey}, {Eisenhardt}, {Eisenstein}, {Gonzalez}, {Huang}, {Jannuzi},
  {Kochanek}, {McKenzie}, {Murray}, {Pahre}, {Smith}, {Soifer}, {Stanford},
  {Stern}, \& {Elston}}]{Brodwin2006a}
{Brodwin}, M., {Brown}, M.~J.~I., {Ashby}, M.~L.~N., {et~al.} 2006, \apj, 651,
  791

\bibitem[{{Budzynski} {et~al.}(2012){Budzynski}, {Koposov}, {McCarthy},
  {McGee}, \& {Belokurov}}]{Budzynski2012a}
{Budzynski}, J.~M., {Koposov}, S.~E., {McCarthy}, I.~G., {McGee}, S.~L., \&
  {Belokurov}, V. 2012, \mnras, 423, 104

\bibitem[{{Bundy} {et~al.}(2006){Bundy}, {Ellis}, {Conselice}, {Taylor},
  {Cooper}, {Willmer}, {Weiner}, {Coil}, {Noeske}, \&
  {Eisenhardt}}]{Bundy2006a}
{Bundy}, K., {Ellis}, R.~S., {Conselice}, C.~J., {et~al.} 2006, \apj, 651, 120

\bibitem[{{Burke} \& {Collins}(2013)}]{Burke2013a}
{Burke}, C. \& {Collins}, C.~A. 2013, \mnras, 434, 2856

\bibitem[{{Collins} {et~al.}(2009){Collins}, {Stott}, {Hilton}, {Kay},
  {Stanford}, {Davidson}, {Hosmer}, {Hoyle}, {Liddle}, {Lloyd-Davies}, {Mann},
  {Mehrtens}, {Miller}, {Nichol}, {Romer}, {Sahl{\'e}n}, {Viana}, \&
  {West}}]{Collins2009a}
{Collins}, C.~A., {Stott}, J.~P., {Hilton}, M., {et~al.} 2009, \nat, 458, 603

\bibitem[{{De Lucia} \& {Blaizot}(2007)}]{DeLucia2007a}
{De Lucia}, G. \& {Blaizot}, J. 2007, \mnras, 375, 2

\bibitem[{{De Lucia} {et~al.}(2004){De Lucia}, {Poggianti},
  {Arag{\'o}n-Salamanca}, {Clowe}, {Halliday}, {Jablonka}, {Milvang-Jensen},
  {Pell{\'o}}, {Poirier}, {Rudnick}, {Saglia}, {Simard}, \&
  {White}}]{DeLucia2004a}
{De Lucia}, G., {Poggianti}, B.~M., {Arag{\'o}n-Salamanca}, A., {et~al.} 2004,
  \apjl, 610, L77

\bibitem[{{De Lucia} {et~al.}(2012){De Lucia}, {Weinmann}, {Poggianti},
  {Arag{\'o}n-Salamanca}, \& {Zaritsky}}]{DeLucia2012a}
{De Lucia}, G., {Weinmann}, S., {Poggianti}, B.~M., {Arag{\'o}n-Salamanca}, A.,
  \& {Zaritsky}, D. 2012, \mnras, 423, 1277

\bibitem[{{De Propris} {et~al.}(2013){De Propris}, {Phillipps}, \&
  {Bremer}}]{DePropris2013a}
{De Propris}, R., {Phillipps}, S., \& {Bremer}, M.~N. 2013, \mnras, 434, 3469

\bibitem[{{Faber} {et~al.}(2007){Faber}, {Willmer}, {Wolf}, {Koo}, {Weiner},
  {Newman}, {Im}, {Coil}, {Conroy}, {Cooper}, {Davis}, {Finkbeiner}, {Gerke},
  {Gebhardt}, {Groth}, {Guhathakurta}, {Harker}, {Kaiser}, {Kassin},
  {Kleinheinrich}, {Konidaris}, {Kron}, {Lin}, {Luppino}, {Madgwick},
  {Meisenheimer}, {Noeske}, {Phillips}, {Sarajedini}, {Schiavon}, {Simard},
  {Szalay}, {Vogt}, \& {Yan}}]{Faber2007a}
{Faber}, S.~M., {Willmer}, C.~N.~A., {Wolf}, C., {et~al.} 2007, \apj, 665, 265

\bibitem[{{Fakhouri} {et~al.}(2010){Fakhouri}, {Ma}, \&
  {Boylan-Kolchin}}]{Fakhouri2010a}
{Fakhouri}, O., {Ma}, C.-P., \& {Boylan-Kolchin}, M. 2010, \mnras, 406, 2267

\bibitem[{{Fassbender}(2007)}]{RF2007PhD}
{Fassbender}, R. 2007, Phd thesis, Ludwig-Maximilians-Unversit{\"a}t
  M{\"u}nchen, astro-ph/0806.0861

\bibitem[{{Fassbender} {et~al.}(2011{\natexlab{a}}){Fassbender},
  {B{\"o}hringer}, {Nastasi}, {{\v S}uhada}, {M{\"u}hlegger}, {de Hoon},
  {Kohnert}, {Lamer}, {Mohr}, {Pierini}, {Pratt}, {Quintana}, {Rosati},
  {Santos}, \& {Schwope}}]{Fassbender2011c}
{Fassbender}, R., {B{\"o}hringer}, H., {Nastasi}, A., {et~al.}
  2011{\natexlab{a}}, New Journal of Physics, 13, 125014

\bibitem[{{Fassbender} {et~al.}(2011{\natexlab{b}}){Fassbender},
  {B{\"o}hringer}, {Santos}, {Pratt}, {{\v S}uhada}, {Kohnert}, {Lerchster},
  {Rovilos}, {Pierini}, {Chon}, {Schwope}, {Lamer}, {M{\"u}hlegger}, {Rosati},
  {Quintana}, {Nastasi}, {de Hoon}, {Seitz}, \& {Mohr}}]{Fassbender2011b}
{Fassbender}, R., {B{\"o}hringer}, H., {Santos}, J.~S., {et~al.}
  2011{\natexlab{b}}, \aap, 527, A78

\bibitem[{{Fassbender} {et~al.}(2011{\natexlab{c}}){Fassbender}, {Nastasi},
  {B{\"o}hringer}, {{\v S}uhada}, {Santos}, {Rosati}, {Pierini},
  {M{\"u}hlegger}, {Quintana}, {Schwope}, {Lamer}, {de Hoon}, {Kohnert},
  {Pratt}, \& {Mohr}}]{Fassbender2011a}
{Fassbender}, R., {Nastasi}, A., {B{\"o}hringer}, H., {et~al.}
  2011{\natexlab{c}}, \aap, 527, L10

\bibitem[{{Fioc} \& {Rocca-Volmerange}(1997)}]{Fioc1997a}
{Fioc}, M. \& {Rocca-Volmerange}, B. 1997, \aap, 326, 950

\bibitem[{{Gabor} \& {Dav{\'e}}(2012)}]{Gabor2012a}
{Gabor}, J.~M. \& {Dav{\'e}}, R. 2012, \mnras, 427, 1816

\bibitem[{{Galametz} {et~al.}(2013){Galametz}, {Grazian}, {Fontana},
  {Ferguson}, {Ashby}, {Barro}, {Castellano}, {Dahlen}, {Donley}, {Faber},
  {Grogin}, {Guo}, {Huang}, {Kocevski}, {Koekemoer}, {Lee}, {McGrath}, {Peth},
  {Willner}, {Almaini}, {Cooper}, {Cooray}, {Conselice}, {Dickinson}, {Dunlop},
  {Fazio}, {Foucaud}, {Gardner}, {Giavalisco}, {Hathi}, {Hartley}, {Koo},
  {Lai}, {de Mello}, {McLure}, {Lucas}, {Paris}, {Pentericci}, {Santini},
  {Simpson}, {Sommariva}, {Targett}, {Weiner}, {Wuyts}, \& {the CANDELS
  team}}]{Galametz2013a}
{Galametz}, A., {Grazian}, A., {Fontana}, A., {et~al.} 2013, \apjs, 206, 10

\bibitem[{{Gehrels}(1986)}]{Gehrels1986a}
{Gehrels}, N. 1986, \apj, 303, 336

\bibitem[{{Giodini} {et~al.}(2009){Giodini}, {Pierini}, {Finoguenov}, {Pratt},
  {Boehringer}, {Leauthaud}, {Guzzo}, {Aussel}, {Bolzonella}, {Capak}, {Elvis},
  {Hasinger}, {Ilbert}, {Kartaltepe}, {Koekemoer}, {Lilly}, {Massey},
  {McCracken}, {Rhodes}, {Salvato}, {Sanders}, {Scoville}, {Sasaki}, {Smolcic},
  {Taniguchi}, {Thompson}, \& {the COSMOS Collaboration}}]{Giodini2009a}
{Giodini}, S., {Pierini}, D., {Finoguenov}, A., {et~al.} 2009, \apj, 703, 982

\bibitem[{{Gobat} {et~al.}(2011){Gobat}, {Daddi}, {Onodera}, {Finoguenov},
  {Renzini}, {Arimoto}, {Bouwens}, {Brusa}, {Chary}, {Cimatti}, {Dickinson},
  {Kong}, \& {Mignoli}}]{Gobat2011a}
{Gobat}, R., {Daddi}, E., {Onodera}, M., {et~al.} 2011, \aap, 526, A133

\bibitem[{{Gobat} {et~al.}(2013){Gobat}, {Strazzullo}, {Daddi}, {Onodera},
  {Carollo}, {Renzini}, {Finoguenov}, {Cimatti}, {Scarlata}, \&
  {Arimoto}}]{Gobat2013a}
{Gobat}, R., {Strazzullo}, V., {Daddi}, E., {et~al.} 2013, \apj, 776, 9

\bibitem[{{Gr{\"u}tzbauch} {et~al.}(2012){Gr{\"u}tzbauch}, {Bauer},
  {J{\o}rgensen}, \& {Varela}}]{Gruetzbauch2012a}
{Gr{\"u}tzbauch}, R., {Bauer}, A.~E., {J{\o}rgensen}, I., \& {Varela}, J. 2012,
  \mnras, 423, 3652

\bibitem[{{Hausman} \& {Ostriker}(1978)}]{Hausman1978a}
{Hausman}, M.~A. \& {Ostriker}, J.~P. 1978, \apj, 224, 320

\bibitem[{{Hayashi} {et~al.}(2010){Hayashi}, {Kodama}, {Koyama}, {Tanaka},
  {Shimasaku}, \& {Okamura}}]{Hayashi2010a}
{Hayashi}, M., {Kodama}, T., {Koyama}, Y., {et~al.} 2010, \mnras, 402, 1980

\bibitem[{{Hilton} {et~al.}(2010){Hilton}, {Lloyd-Davies}, {Stanford}, {Stott},
  {Collins}, {Romer}, {Hosmer}, {Hoyle}, {Kay}, {Liddle}, {Mehrtens}, {Miller},
  {Sahl{\'e}n}, \& {Viana}}]{Hilton2010a}
{Hilton}, M., {Lloyd-Davies}, E., {Stanford}, S.~A., {et~al.} 2010, \apj, 718,
  133

\bibitem[{{Hilton} {et~al.}(2009){Hilton}, {Stanford}, {Stott}, {Collins},
  {Hoyle}, {Davidson}, {Hosmer}, {Kay}, {Liddle}, {Lloyd-Davies}, {Mann},
  {Mehrtens}, {Miller}, {Nichol}, {Romer}, {Sabirli}, {Sahl{\'e}n}, {Viana},
  {West}, {Barbary}, {Dawson}, {Meyers}, {Perlmutter}, {Rubin}, \&
  {Suzuki}}]{Hilton2009a}
{Hilton}, M., {Stanford}, S.~A., {Stott}, J.~P., {et~al.} 2009, \apj, 697, 436

\bibitem[{{Kissler-Patig} {et~al.}(2008){Kissler-Patig}, {Pirard}, {Casali},
  {Moorwood}, {Ageorges}, {Alves de Oliveira}, {Baksai}, {Bedin}, {Bendek},
  {Biereichel}, {Delabre}, {Dorn}, {Esteves}, {Finger}, {Gojak}, {Huster},
  {Jung}, {Kiekebush}, {Klein}, {Koch}, {Lizon}, {Mehrgan}, {Petr-Gotzens},
  {Pritchard}, {Selman}, \& {Stegmeier}}]{Kissler2008a}
{Kissler-Patig}, M., {Pirard}, J.-F., {Casali}, M., {et~al.} 2008, \aap, 491,
  941

\bibitem[{{Kotulla} {et~al.}(2009){Kotulla}, {Fritze}, {Weilbacher}, \&
  {Anders}}]{Kotulla2009a}
{Kotulla}, R., {Fritze}, U., {Weilbacher}, P., \& {Anders}, P. 2009, \mnras,
  396, 462

\bibitem[{{Koyama} {et~al.}(2007){Koyama}, {Kodama}, {Tanaka}, {Shimasaku}, \&
  {Okamura}}]{Koyama2007a}
{Koyama}, Y., {Kodama}, T., {Tanaka}, M., {Shimasaku}, K., \& {Okamura}, S.
  2007, \mnras, 382, 1719

\bibitem[{{Landolt}(1992)}]{Landolt1992a}
{Landolt}, A.~U. 1992, \aj, 104, 340

\bibitem[{{Lerchster} {et~al.}(2011){Lerchster}, {Seitz}, {Brimioulle},
  {Fassbender}, {Rovilos}, {B{\"o}hringer}, {Pierini}, {Kilbinger},
  {Finoguenov}, {Quintana}, \& {Bender}}]{Lerchster2011a}
{Lerchster}, M., {Seitz}, S., {Brimioulle}, F., {et~al.} 2011, \mnras, 411,
  2667

\bibitem[{{Lidman} {et~al.}(2013){Lidman}, {Iacobuta}, {Bauer}, {Barrientos},
  {Cerulo}, {Couch}, {Delaye}, {Demarco}, {Ellingson}, {Faloon}, {Gilbank},
  {Huertas-Company}, {Mei}, {Meyers}, {Muzzin}, {Noble}, {Nantais}, {Rettura},
  {Rosati}, {S{\'a}nchez-Janssen}, {Strazzullo}, {Webb}, {Wilson}, {Yan}, \&
  {Yee}}]{Lidman2013a}
{Lidman}, C., {Iacobuta}, G., {Bauer}, A.~E., {et~al.} 2013, \mnras, 433, 825

\bibitem[{{Lidman} {et~al.}(2008){Lidman}, {Rosati}, {Tanaka}, {Strazzullo},
  {Demarco}, {Mullis}, {Ageorges}, {Kissler-Patig}, {Petr-Gotzens}, \&
  {Selman}}]{Lidman2008a}
{Lidman}, C., {Rosati}, P., {Tanaka}, M., {et~al.} 2008, \aap, 489, 981

\bibitem[{{Lidman} {et~al.}(2012){Lidman}, {Suherli}, {Muzzin}, {Wilson},
  {Demarco}, {Brough}, {Rettura}, {Cox}, {DeGroot}, {Yee}, {Gilbank},
  {Hoekstra}, {Balogh}, {Ellingson}, {Hicks}, {Nantais}, {Noble}, {Lacy},
  {Surace}, \& {Webb}}]{Lidmann2012a}
{Lidman}, C., {Suherli}, J., {Muzzin}, A., {et~al.} 2012, \mnras, 427, 550

\bibitem[{{Lin} {et~al.}(2013){Lin}, {Brodwin}, {Gonzalez}, {Bode},
  {Eisenhardt}, {Stanford}, \& {Vikhlinin}}]{Lin2013a}
{Lin}, Y.-T., {Brodwin}, M., {Gonzalez}, A.~H., {et~al.} 2013, \apj, 771, 61

\bibitem[{{Lin} {et~al.}(2004){Lin}, {Mohr}, \& {Stanford}}]{Lin2004b}
{Lin}, Y.-T., {Mohr}, J.~J., \& {Stanford}, S.~A. 2004, \apj, 610, 745

\bibitem[{{{\L}okas} \& {Mamon}(2001)}]{Lokas2001a}
{{\L}okas}, E.~L. \& {Mamon}, G.~A. 2001, \mnras, 321, 155

\bibitem[{{Lotz} {et~al.}(2013){Lotz}, {Papovich}, {Faber}, {Ferguson},
  {Grogin}, {Guo}, {Kocevski}, {Koekemoer}, {Lee}, {McIntosh}, {Momcheva},
  {Rudnick}, {Saintonge}, {Tran}, {van der Wel}, \& {Willmer}}]{Lotz2013a}
{Lotz}, J.~M., {Papovich}, C., {Faber}, S.~M., {et~al.} 2013, \apj, 773, 154

\bibitem[{{Maihara} {et~al.}(2001){Maihara}, {Iwamuro}, {Tanabe}, {Taguchi},
  {Hata}, {Oya}, {Kashikawa}, {Iye}, {Miyazaki}, {Karoji}, {Yoshida}, {Totani},
  {Yoshii}, {Okamura}, {Shimasaku}, {Saito}, {Ando}, {Goto}, {Hayashi},
  {Kaifu}, {Kobayashi}, {Kosugi}, {Motohara}, {Nishimura}, {Noumaru},
  {Ogasawara}, {Sasaki}, {Sekiguchi}, {Takata}, {Terada}, {Yamashita}, {Usuda},
  \& {Tokunaga}}]{Maihara2001a}
{Maihara}, T., {Iwamuro}, F., {Tanabe}, H., {et~al.} 2001, \pasj, 53, 25

\bibitem[{{Mancone} {et~al.}(2012){Mancone}, {Baker}, {Gonzalez}, {Ashby},
  {Stanford}, {Brodwin}, {Eisenhardt}, {Snyder}, {Stern}, \&
  {Wright}}]{Mancone2012a}
{Mancone}, C.~L., {Baker}, T., {Gonzalez}, A.~H., {et~al.} 2012, \apj, 761, 141

\bibitem[{{Mancone} {et~al.}(2010){Mancone}, {Gonzalez}, {Brodwin}, {Stanford},
  {Eisenhardt}, {Stern}, \& {Jones}}]{Mancone2010a}
{Mancone}, C.~L., {Gonzalez}, A.~H., {Brodwin}, M., {et~al.} 2010, \apj, 720,
  284

\bibitem[{{Mei} {et~al.}(2009){Mei}, {Holden}, {Blakeslee}, {Ford}, {Franx},
  {Homeier}, {Illingworth}, {Jee}, {Overzier}, {Postman}, {Rosati}, {Van der
  Wel}, \& {Bartlett}}]{Mei2009a}
{Mei}, S., {Holden}, B.~P., {Blakeslee}, J.~P., {et~al.} 2009, \apj, 690, 42

\bibitem[{{Menci} {et~al.}(2008){Menci}, {Rosati}, {Gobat}, {Strazzullo},
  {Rettura}, {Mei}, \& {Demarco}}]{Menci2008a}
{Menci}, N., {Rosati}, P., {Gobat}, R., {et~al.} 2008, \apj, 685, 863

\bibitem[{{Muzzin} {et~al.}(2013){Muzzin}, {Wilson}, {Demarco}, {Lidman},
  {Nantais}, {Hoekstra}, {Yee}, \& {Rettura}}]{Muzzin2013a}
{Muzzin}, A., {Wilson}, G., {Demarco}, R., {et~al.} 2013, \apj, 767, 39

\bibitem[{{Muzzin} {et~al.}(2012){Muzzin}, {Wilson}, {Yee}, {Gilbank},
  {Hoekstra}, {Demarco}, {Balogh}, {van Dokkum}, {Franx}, {Ellingson}, {Hicks},
  {Nantais}, {Noble}, {Lacy}, {Lidman}, {Rettura}, {Surace}, \&
  {Webb}}]{Muzzin2012a}
{Muzzin}, A., {Wilson}, G., {Yee}, H.~K.~C., {et~al.} 2012, \apj, 746, 188

\bibitem[{{Nastasi} {et~al.}(2011){Nastasi}, {Fassbender}, {B{\"o}hringer},
  {{\v S}uhada}, {Rosati}, {Pierini}, {Verdugo}, {Santos}, {Schwope}, {de
  Hoon}, {Kohnert}, {Lamer}, {M{\"u}hlegger}, \& {Quintana}}]{Nastasi2011a}
{Nastasi}, A., {Fassbender}, R., {B{\"o}hringer}, H., {et~al.} 2011, \aap, 532,
  L6

\bibitem[{{Nastasi} {et~al.}(2013){Nastasi}, {Scodeggio}, {Fassbender},
  {B{\"o}hringer}, {Pierini}, {Verdugo}, {Garilli}, \&
  {Franzetti}}]{Nastasi2013a}
{Nastasi}, A., {Scodeggio}, M., {Fassbender}, R., {et~al.} 2013, \aap, 550, A9

\bibitem[{{Newman} {et~al.}(2014){Newman}, {Ellis}, {Andreon}, {Treu},
  {Raichoor}, \& {Trinchieri}}]{Newman2014a}
{Newman}, A.~B., {Ellis}, R.~S., {Andreon}, S., {et~al.} 2014, \apj, 788, 51

\bibitem[{{Ouchi} {et~al.}(2004){Ouchi}, {Shimasaku}, {Okamura}, {Furusawa},
  {Kashikawa}, {Ota}, {Doi}, {Hamabe}, {Kimura}, {Komiyama}, {Miyazaki},
  {Miyazaki}, {Nakata}, {Sekiguchi}, {Yagi}, \& {Yasuda}}]{Ouchi2004a}
{Ouchi}, M., {Shimasaku}, K., {Okamura}, S., {et~al.} 2004, \apj, 611, 660

\bibitem[{{Papovich} {et~al.}(2012){Papovich}, {Bassett}, {Lotz}, {van der
  Wel}, {Tran}, {Finkelstein}, {Bell}, {Conselice}, {Dekel}, {Dunlop}, {Guo},
  {Faber}, {Farrah}, {Ferguson}, {Finkelstein}, {H{\"a}ussler}, {Kocevski},
  {Koekemoer}, {Koo}, {McGrath}, {McLure}, {McIntosh}, {Momcheva}, {Newman},
  {Rudnick}, {Weiner}, {Willmer}, \& {Wuyts}}]{Papovich2012a}
{Papovich}, C., {Bassett}, R., {Lotz}, J.~M., {et~al.} 2012, \apj, 750, 93

\bibitem[{{Papovich} {et~al.}(2010){Papovich}, {Momcheva}, {Willmer},
  {Finkelstein}, {Finkelstein}, {Tran}, {Brodwin}, {Dunlop}, {Farrah}, {Khan},
  {Lotz}, {McCarthy}, {McLure}, {Rieke}, {Rudnick}, {Sivanandam}, {Pacaud}, \&
  {Pierre}}]{Papovich2010a}
{Papovich}, C., {Momcheva}, I., {Willmer}, C.~N.~A., {et~al.} 2010, \apj, 716,
  1503

\bibitem[{{Peng} {et~al.}(2010){Peng}, {Lilly}, {Kova{\v c}}, {Bolzonella},
  {Pozzetti}, {Renzini}, {Zamorani}, {Ilbert}, {Knobel}, {Iovino}, {Maier},
  {Cucciati}, {Tasca}, {Carollo}, {Silverman}, {Kampczyk}, {de Ravel},
  {Sanders}, {Scoville}, {Contini}, {Mainieri}, {Scodeggio}, {Kneib}, {Le
  F{\`e}vre}, {Bardelli}, {Bongiorno}, {Caputi}, {Coppa}, {de la Torre},
  {Franzetti}, {Garilli}, {Lamareille}, {Le Borgne}, {Le Brun}, {Mignoli},
  {Perez Montero}, {Pello}, {Ricciardelli}, {Tanaka}, {Tresse}, {Vergani},
  {Welikala}, {Zucca}, {Oesch}, {Abbas}, {Barnes}, {Bordoloi}, {Bottini},
  {Cappi}, {Cassata}, {Cimatti}, {Fumana}, {Hasinger}, {Koekemoer},
  {Leauthaud}, {Maccagni}, {Marinoni}, {McCracken}, {Memeo}, {Meneux}, {Nair},
  {Porciani}, {Presotto}, \& {Scaramella}}]{Peng2010a}
{Peng}, Y.-j., {Lilly}, S.~J., {Kova{\v c}}, K., {et~al.} 2010, \apj, 721, 193

\bibitem[{{Peng} {et~al.}(2012){Peng}, {Lilly}, {Renzini}, \&
  {Carollo}}]{Peng2012a}
{Peng}, Y.-j., {Lilly}, S.~J., {Renzini}, A., \& {Carollo}, M. 2012, \apj, 757,
  4

\bibitem[{{Pierini} {et~al.}(2004){Pierini}, {Maraston}, {Bender}, \&
  {Witt}}]{Pierini2004a}
{Pierini}, D., {Maraston}, C., {Bender}, R., \& {Witt}, A.~N. 2004, \mnras,
  347, 1

\bibitem[{{Pierini} {et~al.}(2005){Pierini}, {Maraston}, {Gordon}, \&
  {Witt}}]{Pierini2005a}
{Pierini}, D., {Maraston}, C., {Gordon}, K.~D., \& {Witt}, A.~N. 2005, \mnras,
  363, 131

\bibitem[{{Reichert} {et~al.}(2011){Reichert}, {B{\"o}hringer}, {Fassbender},
  \& {M{\"u}hlegger}}]{Reichert2011a}
{Reichert}, A., {B{\"o}hringer}, H., {Fassbender}, R., \& {M{\"u}hlegger}, M.
  2011, \aap, 535, A4

\bibitem[{{Rettura} {et~al.}(2010){Rettura}, {Rosati}, {Nonino}, {Fosbury},
  {Gobat}, {Menci}, {Strazzullo}, {Mei}, {Demarco}, \& {Ford}}]{Rettura2010a}
{Rettura}, A., {Rosati}, P., {Nonino}, M., {et~al.} 2010, \apj, 709, 512

\bibitem[{{Romeo} {et~al.}(2008){Romeo}, {Napolitano}, {Covone},
  {Sommer-Larsen}, {Antonuccio-Delogu}, \& {Capaccioli}}]{Romeo2008a}
{Romeo}, A.~D., {Napolitano}, N.~R., {Covone}, G., {et~al.} 2008, \mnras, 389,
  13

\bibitem[{{Romeo} {et~al.}(2005){Romeo}, {Portinari}, \&
  {Sommer-Larsen}}]{Romeo2005a}
{Romeo}, A.~D., {Portinari}, L., \& {Sommer-Larsen}, J. 2005, \mnras, 361, 983

\bibitem[{{Rosati} {et~al.}(2009){Rosati}, {Tozzi}, {Gobat}, {Santos},
  {Nonino}, {Demarco}, {Lidman}, {Mullis}, {Strazzullo}, {B{\"o}hringer},
  {Fassbender}, {Dawson}, {Tanaka}, {Jee}, {Ford}, {Lamer}, \&
  {Schwope}}]{Rosati2009a}
{Rosati}, P., {Tozzi}, P., {Gobat}, R., {et~al.} 2009, \aap, 508, 583

\bibitem[{{Rudnick} {et~al.}(2009){Rudnick}, {von der Linden}, {Pell{\'o}},
  {Arag{\'o}n-Salamanca}, {Marchesini}, {Clowe}, {De Lucia}, {Halliday},
  {Jablonka}, {Milvang-Jensen}, {Poggianti}, {Saglia}, {Simard}, {White}, \&
  {Zaritsky}}]{Rudnick2009a}
{Rudnick}, G., {von der Linden}, A., {Pell{\'o}}, R., {et~al.} 2009, \apj, 700,
  1559

\bibitem[{{Rudnick} {et~al.}(2012){Rudnick}, {Tran}, {Papovich}, {Momcheva}, \&
  {Willmer}}]{Rudnick2012a}
{Rudnick}, G.~H., {Tran}, K.-V., {Papovich}, C., {Momcheva}, I., \& {Willmer},
  C. 2012, \apj, 755, 14

\bibitem[{{Ruszkowski} \& {Springel}(2009)}]{Ruszkowsk2009a}
{Ruszkowski}, M. \& {Springel}, V. 2009, \apj, 696, 1094

\bibitem[{{Santos} {et~al.}(2011){Santos}, {Fassbender}, {Nastasi},
  {B{\"o}hringer}, {Rosati}, {{\v S}uhada}, {Pierini}, {Nonino},
  {M{\"u}hlegger}, {Quintana}, {Schwope}, {Lamer}, {de Hoon}, \&
  {Strazzullo}}]{Santos2011a}
{Santos}, J.~S., {Fassbender}, R., {Nastasi}, A., {et~al.} 2011, \aap, 531, L15

\bibitem[{{Schechter}(1976)}]{Schechter1976a}
{Schechter}, P. 1976, \apj, 203, 297

\bibitem[{{Skrutskie} {et~al.}(2006){Skrutskie}, {Cutri}, {Stiening},
  {Weinberg}, {Schneider}, {Carpenter}, {Beichman}, {Capps}, {Chester},
  {Elias}, {Huchra}, {Liebert}, {Lonsdale}, {Monet}, {Price}, {Seitzer},
  {Jarrett}, {Kirkpatrick}, {Gizis}, {Howard}, {Evans}, {Fowler}, {Fullmer},
  {Hurt}, {Light}, {Kopan}, {Marsh}, {McCallon}, {Tam}, {Van Dyk}, \&
  {Wheelock}}]{Skrutskie2006a}
{Skrutskie}, M.~F., {Cutri}, R.~M., {Stiening}, R., {et~al.} 2006, \aj, 131,
  1163

\bibitem[{{Smith} {et~al.}(2010){Smith}, {Khosroshahi}, {Dariush}, {Sanderson},
  {Ponman}, {Stott}, {Haines}, {Egami}, \& {Stark}}]{Smith2010a}
{Smith}, G.~P., {Khosroshahi}, H.~G., {Dariush}, A., {et~al.} 2010, \mnras,
  409, 169

\bibitem[{{Snyder} {et~al.}(2012){Snyder}, {Brodwin}, {Mancone}, {Zeimann},
  {Stanford}, {Gonzalez}, {Stern}, {Eisenhardt}, {Brown}, {Dey}, {Jannuzi}, \&
  {Perlmutter}}]{Snyder2012a}
{Snyder}, G.~F., {Brodwin}, M., {Mancone}, C.~M., {et~al.} 2012, \apj, 756, 114

\bibitem[{{Sommariva}{et~al.}(2014)}]{Sommariva2013a}
{Sommariva}, V., et~al. 2014, submitted

\bibitem[{{Springel} {et~al.}(2005){Springel}, {White}, {Jenkins}, {Frenk},
  {Yoshida}, {Gao}, {Navarro}, {Thacker}, {Croton}, {Helly}, {Peacock}, {Cole},
  {Thomas}, {Couchman}, {Evrard}, {Colberg}, \& {Pearce}}]{Springel2005a}
{Springel}, V., {White}, S.~D.~M., {Jenkins}, A., {et~al.} 2005, \nat, 435, 629

\bibitem[{{Stanford} {et~al.}(2012){Stanford}, {Brodwin}, {Gonzalez},
  {Zeimann}, {Stern}, {Dey}, {Eisenhardt}, {Snyder}, \&
  {Mancone}}]{Stanford2012a}
{Stanford}, S.~A., {Brodwin}, M., {Gonzalez}, A.~H., {et~al.} 2012, \apj, 753,
  164

\bibitem[{{Stott} {et~al.}(2010){Stott}, {Collins}, {Sahl{\'e}n}, {Hilton},
  {Lloyd-Davies}, {Capozzi}, {Hosmer}, {Liddle}, {Mehrtens}, {Miller}, {Romer},
  {Stanford}, {Viana}, {Davidson}, {Hoyle}, {Kay}, \& {Nichol}}]{Stott2010a}
{Stott}, J.~P., {Collins}, C.~A., {Sahl{\'e}n}, M., {et~al.} 2010, \apj, 718,
  23

\bibitem[{{Stott} {et~al.}(2009){Stott}, {Pimbblet}, {Edge}, {Smith}, \&
  {Wardlow}}]{Stott2009a}
{Stott}, J.~P., {Pimbblet}, K.~A., {Edge}, A.~C., {Smith}, G.~P., \& {Wardlow},
  J.~L. 2009, \mnras, 394, 2098

\bibitem[{{Strazzullo} {et~al.}(2013){Strazzullo}, {Gobat}, {Daddi}, {Onodera},
  {Carollo}, {Dickinson}, {Renzini}, {Arimoto}, {Cimatti}, {Finoguenov}, \&
  {Chary}}]{Strazzullo2013a}
{Strazzullo}, V., {Gobat}, R., {Daddi}, E., {et~al.} 2013, \apj, 772, 118

\bibitem[{{Strazzullo} {et~al.}(2010){Strazzullo}, {Rosati}, {Pannella},
  {Gobat}, {Santos}, {Nonino}, {Demarco}, {Lidman}, {Tanaka}, {Mullis},
  {Nu{\~n}ez}, {Rettura}, {Jee}, {B{\"o}hringer}, {Bender}, {Bouwens},
  {Dawson}, {Fassbender}, {Franx}, {Perlmutter}, \&
  {Postman}}]{Strazzullo2010a}
{Strazzullo}, V., {Rosati}, P., {Pannella}, M., {et~al.} 2010, \aap, 524, A17

\bibitem[{{Strazzullo} {et~al.}(2006){Strazzullo}, {Rosati}, {Stanford},
  {Lidman}, {Nonino}, {Demarco}, {Eisenhardt}, {Ettori}, {Mainieri}, \&
  {Toft}}]{Strazzullo2006a}
{Strazzullo}, V., {Rosati}, P., {Stanford}, S.~A., {et~al.} 2006, \aap, 450,
  909

\bibitem[{{Tadaki} {et~al.}(2012){Tadaki}, {Kodama}, {Ota}, {Hayashi},
  {Koyama}, {Papovich}, {Brodwin}, {Tanaka}, \& {Iye}}]{Tadaki2012a}
{Tadaki}, K.-i., {Kodama}, T., {Ota}, K., {et~al.} 2012, \mnras, 423, 2617

\bibitem[{{Tanaka} {et~al.}(2013){Tanaka}, {Finoguenov}, {Mirkazemi}, {Wilman},
  {Mulchaey}, {Ueda}, {Xue}, {Brandt}, \& {Cappelluti}}]{Tanaka2013a}
{Tanaka}, M., {Finoguenov}, A., {Mirkazemi}, M., {et~al.} 2013, \pasj, 65, 17

\bibitem[{{Tanaka} {et~al.}(2007){Tanaka}, {Kodama}, {Kajisawa}, {Bower},
  {Demarco}, {Finoguenov}, {Lidman}, \& {Rosati}}]{Tanaka2007a}
{Tanaka}, M., {Kodama}, T., {Kajisawa}, M., {et~al.} 2007, \mnras, 377, 1206

\bibitem[{{Tozzi} {et~al.}(2013){Tozzi}, {Santos}, {Nonino}, {Rosati},
  {Borgani}, {Sartoris}, {Altieri}, \& {Sanchez-Portal}}]{Tozzi2013a}
{Tozzi}, P., {Santos}, J.~S., {Nonino}, M., {et~al.} 2013, \aap, 551, A45

\bibitem[{{Tran} {et~al.}(2010){Tran}, {Papovich}, {Saintonge}, {Brodwin},
  {Dunlop}, {Farrah}, {Finkelstein}, {Finkelstein}, {Lotz}, {McLure},
  {Momcheva}, \& {Willmer}}]{Tran2010a}
{Tran}, K.-V.~H., {Papovich}, C., {Saintonge}, A., {et~al.} 2010, \apjl, 719,
  L126

\bibitem[{{Willis} {et~al.}(2013){Willis}, {Clerc}, {Bremer}, {Pierre},
  {Adami}, {Ilbert}, {Maughan}, {Maurogordato}, {Pacaud}, {Valtchanov},
  {Chiappetti}, {Thanjavur}, {Gwyn}, {Stanway}, \& {Winkworth}}]{Willis2013a}
{Willis}, J.~P., {Clerc}, N., {Bremer}, M.~N., {et~al.} 2013, \mnras, 430, 134

\bibitem[{{Windhorst} {et~al.}(2011){Windhorst}, {Cohen}, {Hathi}, {McCarthy},
  {Ryan}, {Yan}, {Baldry}, {Driver}, {Frogel}, {Hill}, {Kelvin}, {Koekemoer},
  {Mechtley}, {O'Connell}, {Robotham}, {Rutkowski}, {Seibert}, {Straughn},
  {Tuffs}, {Balick}, {Bond}, {Bushouse}, {Calzetti}, {Crockett}, {Disney},
  {Dopita}, {Hall}, {Holtzman}, {Kaviraj}, {Kimble}, {MacKenty}, {Mutchler},
  {Paresce}, {Saha}, {Silk}, {Trauger}, {Walker}, {Whitmore}, \&
  {Young}}]{Windhorst2011a}
{Windhorst}, R.~A., {Cohen}, S.~H., {Hathi}, N.~P., {et~al.} 2011, \apjs, 193,
  27

\bibitem[{{Zeimann} {et~al.}(2012){Zeimann}, {Stanford}, {Brodwin}, {Gonzalez},
  {Snyder}, {Stern}, {Eisenhardt}, {Mancone}, \& {Dey}}]{Zeimann2012a}
{Zeimann}, G.~R., {Stanford}, S.~A., {Brodwin}, M., {et~al.} 2012, \apj, 756,
  115

\bibitem[{{Zhang} {et~al.}(2008){Zhang}, {Finoguenov}, {B{\"o}hringer},
  {Kneib}, {Smith}, {Kneissl}, {Okabe}, \& {Dahle}}]{Zhang2008a}
{Zhang}, Y., {Finoguenov}, A., {B{\"o}hringer}, H., {et~al.} 2008, \aap, 482,
  451

\end{thebibliography}


\end{document}